\documentclass[nohyper,12pt,letterpaper]{JHEP3}
\usepackage{amsmath}
\usepackage{amsthm}
\usepackage{amsfonts}

\DeclareMathOperator{\mult}{mult} 
\DeclareMathOperator{\hgt}{ht} 
\DeclareMathOperator{\Ker}{Ker} 
\DeclareMathOperator{\sgn}{sgn} 
\DeclareMathOperator{\supp}{supp} 
\DeclareMathOperator{\Aut}{Aut} 

\date{\today}

\title{$E_{10}$ Orbifolds}

\author{
Jeffrey Brown${}^1$, Surya Ganguli,${}^2$ 
Ori J. Ganor${}^2$ and Craig Helfgott${}^3$\\ 

${}^1$Department of Mathematics,
             University of California, Berkeley, CA 94720 \\ \\
${}^2$Department of Physics,
             University of California, Berkeley, CA 94720 \\
  and\\
Theoretical Physics Group,
            Lawrence Berkeley National Laboratory,\\
Berkeley, CA 94720\\ \\
${}^3$Department of Physics,
             Raymond and Beverly Sackler Faculty of Exact Science\\
            Tel-Aviv University, Israel \\

Emails:\\ \email{jbrown@Math.Berkeley.edu},
       \email{sganguli@socrates.berkeley.edu} \\
       \email{origa@socrates.berkeley.edu},
       \email{helfgott@socrates.berkeley.edu}
}

\abstract{
We study $Z_2$ orbifolds of M-theory in terms of $E_{10}.$
We find a simple relation between the $Z_2$ action on $E_{10}$ and
the imaginary root that corresponds [hep-th/0401053] to the
``twisted sector'' branes.
We discuss the connection between the Kac-Moody algebra $DE_{10}$
and the ``untwisted'' sector, and we demonstrate how 
$DE_{18}$ can describe both the untwisted and twisted sectors simultaneously.
}


\received{September 2, 2004}

\preprint{\hepth{0409037}\\ UCB-PTH-04/24\\ LBNL-56196}
\begin{document}

\newtheorem{thm}{Theorem}[section]
\newtheorem{lem}[thm]{Lemma}
\newtheorem{prop}[thm]{Proposition}
\newtheorem{claim}[thm]{Claim}
\newtheorem{conj}[thm]{Conjecture}

\theoremstyle{definition}
\newtheorem{defn}{Definition}[section]

\numberwithin{equation}{section}

\newcommand{\thmref}[1]{Theorem~\ref{#1}}
\newcommand{\secref}[1]{\S\ref{#1}}
\newcommand{\lemref}[1]{Lemma~\ref{#1}}
\newcommand{\clmref}[1]{Claim~\ref{#1}}

\newcommand{\propref}[1]{Proposition~\ref{#1}}
\newcommand{\figref}[1]{Figure~\ref{#1}}
\newcommand{\appref}[1]{Appendix~\ref{#1}}

\def\be{\begin{equation}} 
\def\ee{\end{equation}} 
\def\bear{\begin{eqnarray}} 
\def\eear{\end{eqnarray}} 
\def\nn{\nonumber}

\newcommand\belabel[1]{\begin{equation}\label{#1}}
 
\newcommand\bra[1]{{\langle {#1}|}}  
\newcommand\ket[1]{{|{#1}\rangle}}  
 
\def\defineas{{:=}} 
 
\def\a{\alpha} 
\def\b{\beta} 
\def\g{\gamma} 
\def\u{\mu} 
\def\r{\rho} 
\def\th{{\theta}} 
\def\lam{{\lambda}} 

\def\bth{{\overline{\theta}}} 
\def\blam{{\overline{\lambda}}} 
\def\bpsi{{\overline{\psi}}} 
\def\bsig{{\overline{\sigma}}} 
\def\Dslash{{\relax{\not\kern-.18em \partial}}} 
\def\Spin{{{\mbox{\rm Spin}}}} 
\def\SL{{{\mbox{\rm SL}}}} 
\def\GL{{{\mbox{\rm GL}}}} 
\def\rt{{\rightarrow}}  
\def\cc{{\mbox{c.c.}}} 

\newcommand\SUSY[1]{{${\cal N}={#1}$}}  
\newcommand\px[1]{{\partial_{#1}}} 
\newcommand\qx[1]{{\partial^{#1}}} 

\newcommand\ppx[1]{{\frac{\partial}{\partial {#1}}}} 
\newcommand\pspxs[1]{{\frac{\partial^2}{\partial {#1}^2}}} 
\newcommand\pspxpx[2]{{\frac{\partial^2}{\partial {#1}\partial {#2}}}}


\newcommand{\field}[1]{\mathbb{#1}}
\newcommand{\ring}[1]{\mathbb{#1}}
\newcommand{\C}{\field{C}}
\newcommand{\R}{\field{R}}
\newcommand{\Q}{\field{Q}}
\newcommand{\F}{\field{F}}
\newcommand{\A}{\field{A}}
\newcommand{\Z}{\ring{Z}}
\newcommand{\N}{\ring{N}}


\providecommand{\abs}[1]{{\lvert#1\rvert}}
\providecommand{\norm}[1]{{\lVert#1\rVert}}
\providecommand{\divides}{{\vert}}
\providecommand{\suchthat}{{:\quad}}

\def\bz{{\overline{z}}}
\def\gYM{{g_{\textit{YM}}}} 
\def\gst{{g_s}} 
\def\Mst{{M_s}} 
\newcommand\rep[1]{{\bf {#1}}} 
\def\Brane{{\cal B}} 
\def\Ac{{S}} 
\def\MinkAc{{\tilde{S}}} 
\def\Lag{{L}} 

\newcommand\inner[2]{{\langle {#1}, {#2} \rangle}} 
\newcommand\kform[2]{{({#1}|{#2})}} 
\def\NatNum{{\bf N}} 

\def\rtB{{\Theta}} 
\def\ct{{\tilde{\tau}}} 

\def\const{{{\mbox{\rm const}}\,}}

\def\Car{{\mathfrak h}} 
\newcommand\CarOf[1]{{\Car({{#1}})}} 
\def\DualCar{{\Car^*}} 
\newcommand\DualCarOf[1]{{\DualCar({{#1}})}} 
\def\AlgG{{\mathfrak g}} 
\def\Rts{{\Delta}} 
\newcommand\RtsOf[1]{{\Rts({{#1}})}} 
\def\RtLat{{Q}} 
\newcommand\RtLatOf[1]{{\RtLat({{#1}})}} 
\def\Weyl{{W}} 
\newcommand\WeylOf[1]{{\Weyl({{#1}})}} 
\def\WgtLat{{P}} 
\newcommand\WgtLatOf[1]{{\WgtLat({{#1}})}} 
\def\Fwgt{{\Lambda}} 
\def\DENFwgt{{\tilde{\Lambda}'}} 

\def\ReRoots{{\Delta_{\textit{Re}}}} 
\newcommand\ReRootsOf[1]{{\ReRoots({{#1}})}} 
\def\ImRoots{{\Delta_{\textit{Im}}}} 
\newcommand\ImRootsOf[1]{{\ImRoots({{#1}})}} 

\def\PosRoots{{\Delta^{+}}} 
\newcommand\PosRootsOf[1]{{\PosRoots({{#1}})}} 
\def\PosReRoots{{\Delta^{+}_{\textit{Re}}}} 
\newcommand\PosReRootsOf[1]{{\PosReRoots({{#1}})}} 
\def\PosImRoots{{\Delta^{+}_{\textit{Im}}}} 
\newcommand\PosImRootsOf[1]{{\PosImRoots({{#1}})}} 

\def\NegRoots{{\Delta^{-}}} 
\newcommand\NegRootsOf[1]{{\NegRoots({{#1}})}} 
\def\NegReRoots{{\Delta^{-}_{\textit{Re}}}} 
\newcommand\NegReRootsOf[1]{{\NegReRoots({{#1}})}} 
\def\NegImRoots{{\Delta^{-}_{\textit{Im}}}} 
\newcommand\NegImRootsOf[1]{{\NegImRoots({{#1}})}} 

\newcommand\PosRootsRep[1]{{\Delta^{+}_{\rep{{#1}}}}} 

\def\PosWgts{{P^{+}}} 
\newcommand\PosWgtsOf[1]{{\PosWgts({{#1}})}} 
\def\NegWgts{{P^{-}}} 
\newcommand\NegWgtsOf[1]{{\NegWgts({{#1}})}} 

\newcommand\RootsAtLevel[1]{{\Delta_{[{#1}]}}} 
\newcommand\RootsOfAtLevel[2]{{\RootsAtLevel{{#2}}({{#1}})}} 

\newcommand\pcoeff[1]{{p^{({#1})}}} 

\def\AlgInv{{\AlgG^{({\textit{inv}})}}} 
\newcommand\AlgInvOf[1]{{\AlgInv({{#1}})}} 

\def\RtsInv{{\Rts^{({\textit{inv}})}}} 
\def\RtLatInv{{\RtLat^{({\textit{inv}})}}} 
\def\WeylInv{{\Weyl^{({\textit{inv}})}}} 
\def\WgtsInv{{\Wgts^{({\textit{inv}})}}} 
\def\PosRtsInv{{\Rts_+^{({\textit{inv}})}}} 
\def\NegRtsInv{{\Rts_-^{({\textit{inv}})}}} 

\newcommand\ModuleAtLevelOf[2]{{({#2})_{[{#1}]}}} 

\def\cO{{\cal O}} 

\def\Fl{{\cal C}} 
\def\ModSp{{\cal M}} 
\def\vp{{\vec{p}}} 
\def\vh{{\vec{h}}} 
\def\veps{{\vec{\epsilon}}} 

\def\whD{{\widehat{D}}} 

\numberwithin{equation}{section}

\def\Horava{{Ho\v{r}ava\ }}

\def\AlgNil{{\mathfrak n}} 
\def\AlgAbl{{\mathfrak a}} 
\def\AlgKom{{\mathfrak k}} 
\newcommand\AlgKomOf[1]{{\AlgKom({{#1}})}} 

\def\Id{{I}} 
\def\wa{{\tilde{\a}}} 

\def\Ac{{I}} 
\def\Mass{{m}} 
\def\bv{\b^\vee} 

\def\Alggl{{\mathfrak gl}} 
\def\Affgl{{\widehat{\Alggl}}} 
\def\Algsl{{\mathfrak sl}} 
\def\Affsl{{\widehat{\Algsl}}} 

\def\Algso{{\mathfrak so}} 
\def\Affso{{\widehat{\Algso}}} 
\newcommand\Eb[2]{{E^{({#1},{#2})}}} 
\newcommand\JEn[3]{{J_{{#3}}^{({#1},{#2})}}} 

\def\OrbRt{{\tau}} 
\newcommand\OrbChg[1]{{q({#1})}} 

\def\Xrt{{\xi}} 
\def\Yrt{{\chi}} 
\def\fDEE{{\upsilon}} 
\def\wfDEE{{\tilde{\upsilon}}} 
\def\fOH{{\sigma}} 
\def\pHO{{\varpi}} 
\def\eSODE{{\varrho}} 
\def\rtAlgG{{\tilde{\AlgG}}} 

\def\gv{\g^\vee} 
\def\wdelta{{\tilde{\delta}}} 
\def\wLambda{{\widetilde{\Lambda}}} 

\newcommand\coDE[2]{\tilde{C}^{({#1})}_{#2}} 
\newcommand\coE[2]{C^{({#1})'}_{#2}} 
\newcommand\coEXD[2]{C^{({#1})}_{#2}} 
\newcommand\DcoEXDE[2]{\Delta C^{({#1})}_{#2}} 

\def\Commutant{{\AlgG^{({\textit{com}})}}} 

\def\Yrtv{{\Yrt^\vee}} 
\def\rtAlgG{{\tilde{\AlgG}}} 

\def\rtlam{{\tilde{\lam}}}
\def\tn{{\tilde{n}}} 
\def\half{{\tfrac{1}{2}}} 

\def\wwLambda{{\tilde{\tilde{\Lambda}}}} 

\def\FRR{{\cal F}^{\textit{RR}}} 
\def\vph{{\varphi}} 

\def\Action{{I}} 
\def\dPS{{\mathbb{B}}} 
\newcommand{\CP}{\mathbb{CP}} 

\centerline{\bf DISCLAIMER}
This document was prepared as an account of work sponsored by the United States Government. While this document is believed to contain correct information, neither the United States Government nor any agency thereof, nor The Regents of the University of California, nor any of their employees, makes any warranty, express or implied, or assumes any legal responsibility for the accuracy, completeness, or usefulness of any information, apparatus, product, or process disclosed, or represents that its use would not infringe privately owned rights. Reference herein to any specific commercial product, process, or service by its trade name, trademark, manufacturer, or otherwise, does not necessarily constitute or imply its endorsement, recommendation, or favoring by the United States Government or any agency thereof, or The Regents of the University of California. The views and opinions of authors expressed herein do not necessarily state or reflect those of the United States Government or any agency thereof or The Regents of the University of California.

\newpage
\tableofcontents

\section{Introduction}\label{sec:intro}
\paragraph{}
In this paper we extend a previous work \cite{Brown:2004jb} on the 
connection between M-theory and $E_{10}.$
There is a very intriguing connection between the infinite dimensional
Lie group $E_{10}$ and M-theory on $T^{10}$ \cite{Julia}\cite{Nicolai:kx}.
Generally speaking, fluxes of M-theory on $T^{10}$ correspond
to certain generators, or more precisely {\it roots}, of $E_{10}.$
Yet, not all generators of $E_{10}$ are accounted for by fluxes.
It was proposed in \cite{Brown:2004jb} that Kaluza-Klein particles
and all the objects that can be obtained
from them by U-duality (M2-branes, M5-branes, etc.)
can be associated with other generators of $E_{10},$ which are not
associated with fluxes.
The interpretation of all generators is not yet complete,
and there are an infinite number of $E_{10}$ generators that
are associated with neither fluxes nor particles.
Nevertheless, the physical interpretation of the $E_{10}$ generators 
that we have so far is sufficient to start analyzing 
more complicated backgrounds of M-theory.

In general, starting with an M-theory background $X$
and a discrete symmetry group $\Gamma,$ 
we can often construct a new consistent background
$X/\Gamma$ -- the $\Gamma$-orbifold of $X.$
In this procedure, we keep the $\Gamma$-invariant ``modes'' and discard the
``modes'' that are not invariant. In addition, consistency requirements
such as anomaly cancellation and charge conservation often necessitate
the insertion of extra degrees of freedom -- the ``twisted'' sector.
In perturbative string theory,
there is a well defined procedure to find the twisted sectors \cite{Dixon:1985jw}.
Recent developments in this field appear in 
\cite{Halpern:2004ud}
(which are the latest in a series of papers starting with
\cite{Borisov:1997nc}).
Unfortunately, at the moment there is no general principle to determine
the contents of the twisted sector in M-theory; a case-by-case analysis,
which sometimes relies on a clever guess, is required.

In this paper we will take $X$ to be $T^{10},$ and $\Gamma$ will be some $\Z_2$ action,
which needs to be specified.
One of the most famous examples of such a $T^{10}/\Z_2$ orbifold
is $T^9\times (S^1/\Z_2),$ for which the twisted sector at low-energy was found by
\Horava and Witten \cite{Horava:1995qa} to be an
$E_8\times E_8$ 9+1D Super-Yang-Mills theory.

Often, the twisted-sector contains an extra finite number of branes.
For example the \Horava-Witten orbifold can be viewed as type-IIA string theory
compactified on $S^1/\Z_2$ ($\times T^8,$ in our case) with 16 extra
($T^8$-filling) D8-branes \cite{Polchinski:1995df}.
This setting is known as type-IA.
The compactification of M-theory on $T^5/\Z_2,$ as another example,
was studied in \cite{Dasgupta:1995zm}\cite{Witten:1995em}
 and argued in \cite{Witten:1995em} to contain $16$ M5-branes.
Similarly, M-theory on $T^8/\Z_2$ contains
$16$ M2-branes
\cite{Sen:1996na}-\cite{Dasgupta:1997cd},
and M-theory on $T^9/\Z_2$ contains $16$ units
of Kaluza-Klein momentum 
\cite{Dasgupta:1995zm}\cite{Sen:1996zq}-\cite{Kumar:1996mj}.
All of these examples are, in fact, dual to each other 
\cite{Dasgupta:1995zm}\cite{Sen:1996zq}.

In this paper,
we will re-examine the orbifolds $T^{10}/\Z_2$ and redefine them in terms of $E_{10}.$
We will extend the $\Z_2$ action to an automorphism of 
the root-lattice of $E_{10}$ and then to an automorphism of the 
whole $E_{10}$ Lie algebra.
We will show that some of the features of the twisted-sector
have a simple $E_{10}$-description.
In particular, there is an algebraic connection between 
the $\Z_2$ action on the Cartan subalgebra of $E_{10}$ and the 
$E_{10}$-root that, according to \cite{Brown:2004jb}, is associated with the
branes of the twisted sector.
Thus, the details of the $\Z_2$ action
algebraically determine the types of branes in the twisted sector.

Naturally,
we will refer to the subalgebra of $E_{10}$ that is fixed by (invariant under)
the $\Z_2$ automorphism as the {\it untwisted sector}.
We will also refer to the $\Z_2$-invariant sublattice of the 
$E_{10}$ root lattice as the {\it untwisted sector of the root lattice}.
In \appref{app:EquivRt}
we will demonstrate that for each of the $\Z_2$ orbifolds that
we study, the untwisted sector of the root lattice
is isomorphic to the root lattice of the infinite dimensional
Kac-Moody algebra $DE_{10}.$
We will also show that the untwisted sector of $E_{10}$ contains
a $DE_{10}$ subalgebra, but is actually larger than $DE_{10}.$


The orbifolds that we will study are also dual to heterotic string theory
compactified on $T^9.$
This compactification has been studied in \cite{Motl:1999cy}.
There is  reason to suspect that the Kac-Moody algebra $DE_{18}$ appears in this setting.
Its exact role is yet unknown, but it is probably similar to the role of $E_{10}$ in
toroidal M-theory compactification.
$DE_{18}$ is a natural extension of the affine Lie algebra
$\Affso(24,8)$ that appears when heterotic string theory
is compactified on $T^8$ 
(see 
\cite{Nicolai:1987vy}-\cite{Schwarz:1995td}
and references therein).

$DE_{10}$ is obviously a subalgebra of $DE_{10+k}$ for any $k\ge 0.$
The Lie algebra $DE_{10+k}$ actually has a 
$DE_{10}\oplus D_k$ subalgebra, where $D_k$ is isomorphic to the
classical Lie algebra $\Algso(2k).$
The case $2k=16$ is physically special because
the number $16$ that appears here is, as we shall show,
directly related to the appearance of $16$ branes in the twisted sector
of the orbifolds above. 
(These could, for example, be the $16$ M5-branes of
the twisted sector of $T^5/\Z_2.$)
This extra $\Algso(16)$ subalgebra will allow us to group the
generators of $DE_{18}$ in $\Algso(16)$-representations.
We will propose that if a generator belongs
to a nontrivial $\Algso(16)$-representation it is related to
the twisted sector, and the details of the representation
determine the relationship between the generator and the $16$ 
branes of the twisted sector.
For example, a $DE_{18}$ generator belonging to the fundamental
representation is associated with a single twisted-sector brane.
We will present a few examples to support this proposal.

We have to mention that there is yet another infinite dimensional
Kac-Moody Lie algebra that is believed to be associated with heterotic string
theory compactified on $T^9.$
This is the non-simply-laced
$BE_{10},$ and it was shown in \cite{Damour:2000hv}
that it is associated to the cosmological evolution of
the toroidally compactified heterotic universe.
We will not discuss $BE_{10}$ in this paper, since it does not seem
to contain the information needed to distinguish between the $16$
twisted sectors.

As this work was close to completion, another paper 
that has some overlap with our work appeared \cite{Kleinschmidt:2004dy}.
In this paper the orbifold invariant subalgebra of $E_{10}$ is also identified
and it is also noted that it contains a proper $DE_{10}$ subalgebra.
This is demonstrated by decomposing under the finite $D_9$ subalgebra.
We had arrived at the same conclusion, however,
by decomposing under the affine $\whD_8$ subalgebra.
$\Z_2$-orbifolds of $E_{10}$ and other Kac-Moody algebras have
been also recently discussed in \cite{Mkrtchyan:2004ah},
with similar conclusions.
Other recent ideas related to orbifolds of infinite dimensional
``hidden symmetries'' of M-theory appear in 
\cite{Chaudhuri:2004zh}.

The paper is organized as follows.
In \secref{sec:Prelim} we briefly review $E_{10}$ and its relation
to M-theory on $T^{10}.$
In \secref{sec:Orbifolds} we discuss the $\Z_2$ orbifolds
and demonstrate the algebraic connection between the 
associated $\Z_2$ action on the $E_{10}$ root lattice
and the root that is associated with the twisted-sector branes.
In \secref{sec:Untwisted} we demonstrate that the $\Z_2$-invariant
part of the $E_{10}$ root lattice is isomorphic to the $DE_{10}$ root lattice,
and we show that $DE_{10}$ is a proper subalgebra of the $\Z_2$ invariant
part of $E_{10}.$
In \secref{sec:DE18} we review $DE_{18},$
and in \secref{sec:Twisted} we study the algebraic properties
of the twisted sector -- the generators of $DE_{18}$ that
fall into nontrivial $\Algso(16)$ representations.
In \secref{sec:PhysicalInt} we connect the physical 
twisted sector of the orbifold to its algebraic description
in terms of $DE_{18}.$
Most of our discussion is presented for the $T^5/\Z_2$ orbifold,
but in \secref{sec:TypeIA} we write down, for completeness,
the relevant formulas for the dual type-IA string theory.
In \secref{sec:T4T4}
we present a preliminary discussion on a more complicated orbifold,
$(T^4/\Z_2)\times (T^4/\Z_2),$
which is dual to M-theory on certain Calabi-Yau $3$-folds
\cite{Sethi:1996es}.
We conclude with a discussion in \secref{sec:concl}.
In appendices \ref{app:proof}-\ref{app:ProofProp} we fill-in some technical 
details about the definitions of the various subalgebras.
In \appref{app:EquivRt} we explicitly show that the $\Z_2$
invariant part of the root lattice of $E_{10}$ is equivalent
to the root lattice of $DE_{10}$ and also to the sublattice
of $DE_{18}$ that is orthogonal to the roots of
$\Algso(16)\subset DE_{18}.$
In appendix \ref{app:DenomFormula} we present a generating 
function for the multiplicities of the $\Z_2$ invariant roots of $E_{10}$
that can be compared with a variant of
the ``denominator formula'' for the multiplicities of roots of $DE_{10}.$

\section{Preliminaries}\label{sec:Prelim}
\paragraph{}
We assume basic familiarity with Kac-Moody algebras.
For a thorough treatment see \cite{KacBook}.

Given a Kac-Moody Lie algebra $X,$ we use the following notation:
\vskip 10pt
\begin{tabular}{cl}
$\CarOf{X}=$ & Cartan subalgebra (CSA) of $X$ \\ 
$\RtsOf{X}=$ & roots of $X$ \\ 
$\RtLatOf{X}=$ & root lattice of $X$ \\ 
$\WgtLatOf{X}=$ & weight lattice of $X$ \\ 
$\WeylOf{X}=$ & Weyl group of $X$ \\ 
$\AlgG(X)_\a=$ & root space (for any root $\a\in\RtsOf{X}$) \\
$\kform{\a}{\b}=$ & The Killing form [for any two roots $\a,\b\in\CarOf{X}^*$] \\
$\inner{\a}{h}=$ & The inner product 
[for any root $\a\in\CarOf{X}^*$ and CSA element $h\in\CarOf{X}$] \\
$L(\Lambda)=$ & Irreducible representation of $X$ 
            with a dominant highest-weight $\Lambda\in\WgtLatOf{X}.$ \\
\end{tabular}
\vskip 10pt
We have the root space decomposition
\belabel{eqn:RtSpaces}
X = \bigoplus_{\a\in\RtsOf{X}}\AlgG(X)_\a,
\qquad
\AlgG(X)_\a\defineas
\{x\in X\suchthat [h,x]=\inner{\a}{h}x\quad\forall h\in\CarOf{X}\}.
\ee

\subsection{$E_{10}$}
\paragraph{}
The Dynkin diagram of $E_{10}$ is given in \figref{fig:DynkinE10}.

\vskip 10pt
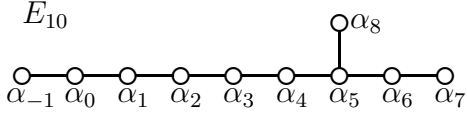
\begin{figure}[h]
\begin{picture}(370,80)
%
%
\thicklines
\multiput(20,20)(20,0){9}{\circle{6}}
\multiput(23,20)(20,0){8}{\line(1,0){14}}

\put(140,23){\line(0,1){14}}
\put(140,40){\circle{6}}
\put( 14,10){$\a_{-1}$}
\put( 36,10){$\a_0$}
\put( 56,10){$\a_1$}
\put( 76,10){$\a_2$}
\put( 96,10){$\a_3$}
\put(116,10){$\a_4$}
\put(136,10){$\a_5$}
\put(156,10){$\a_6$}
\put(176,10){$\a_7$}

\put(144,38){$\a_8$}
\put(20,40){$E_{10}$}
\end{picture}
\caption{The Dynkin diagram of $E_{10}.$}
\label{fig:DynkinE10}
\end{figure}
\vskip 10pt
It is possible to find a basis of $\CarOf{X}^*$ in which
the root-lattice $\RtLatOf{E_{10}}$ is given by
\belabel{eqn:RootLatRep}
\{(n_1, \dots, n_{10})\suchthat \sum_{i=1}^{10} n_i\in 3\Z,\quad n_i\in\Z
\quad (i=1,\dots,10)\}.
\ee
In this basis a root $\a$ is given by a series of $10$ integers,
\belabel{eqn:annn}
\a = (n_1, n_2, n_3, n_4, n_5, n_6, n_7, n_8, n_9, n_{10}),
\ee
and the square of the root is
$$
\a^2 = \sum_1^{10} n_i^2 -\frac{1}{9}\Bigl(\sum_1^{10} n_i\Bigr)^2.
$$
The simple roots can be written in this basis as
\bear
\a_{-1} &\defineas& (1,-1,0,0,0,0,0,0,0,0),\nn\\
\a_0    &\defineas& (0,1,-1,0,0,0,0,0,0,0),\nn\\
\a_1    &\defineas& (0,0,1,-1,0,0,0,0,0,0),\nn\\
\a_2    &\defineas& (0,0,0,1,-1,0,0,0,0,0),\nn\\
\a_3    &\defineas& (0,0,0,0,1,-1,0,0,0,0),\nn\\
\a_4    &\defineas& (0,0,0,0,0,1,-1,0,0,0),\nn\\
\a_5    &\defineas& (0,0,0,0,0,0,1,-1,0,0),\nn\\
\a_6    &\defineas& (0,0,0,0,0,0,0,1,-1,0),\nn\\
\a_7    &\defineas& (0,0,0,0,0,0,0,0,1,-1),\nn\\
\a_8    &\defineas& (0,0,0,0,0,0,0,1,1, 1),\nn\\
&& \label{eqn:AlphaExplicit}
\eear
For the rest of the paper,
whenever
we express an $E_{10}$ root $\a$ as a series of $10$ integers,
it will be understood to be in the basis \eqref{eqn:AlphaExplicit}.

We define the {\it fundamental weights} 
$\Fwgt_i\in\CarOf{E_{10}}^*$ ($i=-1,\dots,8$) as
the solutions to the linear equations
$$
\kform{\Fwgt_i}{\a_j}=\delta_{ij},\qquad i,j=-1,\dots,8.
$$

For future reference, we need the sum of the fundamental weights of $E_{10}.$
This is the weight that enters into character formulas and is given by
$$
\rho =
-30\a_{-1} -61\a_0 -93\a_1 -126\a_2 -160\a_3 -195\a_4 -231\a_5
-153\a_6 -76\a_7 -115\a_8.
$$
We can also write it as
\belabel{eqn:rhoE10}
\rho = -(30, 31, 32, 33, 34, 35, 36, 37, 38, 39).
\ee

\subsection{M-theory on $T^{10}$}
\paragraph{}
In the basis dual to \eqref{eqn:AlphaExplicit},
an element of $\CarOf{E_{10}}$ can be written as
$$
\vh = (h_1, h_2, h_3, h_4, h_5, h_6, h_7, h_8, h_9, h_{10}),
$$
so that
$$
\inner{\vh}{\a} = \sum_{i=1}^{10} n_i h_i.
$$

For M-theory applications, we will set $h_i=\log (M_p R_i)$
($i=1,\dots,10$), where $R_i$ are called the {\it compactification radii},
and $M_p$ is the eleven-dimensional Planck mass.
$2\pi R_i$ are the sizes of the directions of the $T^{10}$
on which M-theory is compactified.
Thus, the connection with M-theory is given by the following identification,
\belabel{eqn:vhRadii}
\vh = (
\log[M_p R_1],
\log[M_p R_2],
\dots,
\log[M_p R_{10}]).
\ee
A classical approximation is valid when $M_p R_i\gg 1$ for all $i=1,\dots,10.$
In a dynamical setting, $R_i$ and therefore $\vh$
vary as a function of time, but we will not need to discuss
time dependence in the scope of this work.

\subsection{Roots and Fluxes}
\paragraph{}
A few roots of $E_{10}$ can be associated with zero modes
of 10+1D supergravity fields.
The bosonic fields of 10+1D supergravity are the metric and 3-form.
We denote their (spatial) components by $g_{IJ}$ ($1\le I\le J\le 10$)
and $C_{IJK}$ ($1\le I<J<K\le 10$).
We denote the supergravity 
variable associated with the root $\a,$ if it is established,
by $\Fl_\a.$
For the simple root $\a_8,$ for example,
we have \textit{$\Fl_{\a_8} = C_{89\,10}$},
according to equation \eqref{eqn:AlphaExplicit}.
The other simple roots $\a_i$ ($i=-1,\dots,7$),
listed in equation \eqref{eqn:AlphaExplicit},
are associated with ratios $g_{(i+2),(i+3)}/g_{(i+3),(i+3)}$ of metric components.

Similarly, the field $C_{IJK}$ corresponds to the positive root \eqref{eqn:annn}
with $n_I=n_J=n_K=1,$ and $n_i=0$ for $i\neq I,J,K.$
The ratio of metric components $g_{IJ}/g_{JJ}$ 
corresponds to \eqref{eqn:annn} with $n_I=1,$ $n_J=-1$ and $n_i=0$ for $i\neq I,J.$

There is a simple way to see which variable is associated to a given root $\a.$
Using the identification \eqref{eqn:vhRadii}, the expression
$2\pi\exp\inner{\vh}{\a}$ can be written as a monomial in the radii
$R_1,\dots,R_{10}.$
This monomial can be interpreted as the real part of
an action of an instanton. The imaginary part of the
action is the variable $\Fl_\a.$
For example, $2\pi\exp\inner{\vh}{\a_{-1}}=2\pi R_1/R_2$
is the real part of the action of a Kaluza-Klein instanton
whose imaginary part is $2\pi i g_{12}/g_{22}.$
Similarly, $2\pi\exp\inner{\vh}{\a_8}=2\pi M_p^3 R_8 R_9 R_{10}$
is the real part of the action of an M2-brane instanton.
Its imaginary part is $(2\pi)^3 i C_{89\,10}.$
(See \cite{Obers:1998fb} for more details.)

Any {\it real root}, i.e. a root $\a$ that satisfies
$\kform{\a}{\a}\equiv\a^2=2,$ 
is Weyl-dual to any one of the simple roots. In other words,
for any $i=-1,\dots,8,$ 
there exists an element $w$ in the Weyl-group of $E_{10},$
$w\in\WeylOf{E_{10}},$ such that $w(\a)=\a_i$
\cite{KacBook}.
In M-theory, $\WeylOf{E_{10}}$ is a subgroup of the U-duality group
\cite{Elitzur:1997zn}\cite{Banks:1998vs},
which is $E_{10}(\Z)$ \cite{Hull:1995mz}.
Thus, any real root $\a$ can be associated with an instanton formally
\cite{Obers:1998fb}\cite{Ganor:1999ui}.
If we assume
\belabel{eqn:WeakAsympt}
R_1\gg R_2 \gg\cdots\gg R_{10} \gg \frac{1}{M_p^3 R_8 R_9},
\ee
then whenever $\a$ is a positive root
$2\pi\exp\inner{\vh}{\a}$ is large, and thus can be 
interpreted as an instanton action of some object that
is formally U-dual to a Kaluza-Klein instanton (or M2-brane).
The requirement \eqref{eqn:WeakAsympt} is always satisfied in
the classical limit of eleven-dimensional supergravity,
\belabel{eqn:Asymptotic}
R_1\gg R_2 \gg\cdots\gg R_{10}\gg M_p^{-1}.
\ee

Imaginary roots,
which satisfy $\a^2\le 0,$
are not Weyl-equivalent (U-dual) to real roots
and therefore cannot be interpreted as instantons.
It was proposed in \cite{Brown:2004jb} that at least a subset of them
can be associated with Minkowski brane charges.
This subset is the set of
positive prime isotropic roots -- positive roots
that satisfy $\a^2=0$ and cannot be written as a nontrivial integer
multiple of another root.

To find the brane charge that corresponds to a given positive
prime isotropic root $\a,$ we need to examine the monomial
\belabel{eqn:MassRoot}
\Mass(\a)= \frac{e^{\inner{\vh}{\a}}}{M_p^9 R_1\cdots R_{10}}
\ee
and identify it with the energy of a Minkowski brane.
The denominator of \eqref{eqn:MassRoot} comes naturally
if we realize that the Weyl-invariant numerator $\exp\inner{\vh}{\a}$
measures the energy in units dual to {\it conformal time}
\cite{Brown:2004jb},
which is usually defined by the differential equation
\belabel{eqn:ct}
d\ct = \frac{dt}{2\pi M_p^9 R_1(t) \cdots R_{10}(t)},
\ee
where the explicit dependence of radii on time $t$ was restored.

We will now present two examples of positive prime isotropic roots
and their associated brane charges.
For the first example take
$$
\a = (0,1,1,1,1,1,1,1,1,1).
$$
We get $\Mass(\a)=1/R_1$ which identifies it with a Kaluza-Klein particle.
For the second example take
$$
\a = (2,2,1,1,1,1,1,1,1,1).
$$
We get $\Mass(\a) = M_p^3 R_1 R_2,$ which identifies it with
an M2-brane.
All positive prime isotropic roots of $E_{10}$ are Weyl-equivalent,
and therefore any such root can be identified with a formal U-dual
of a Minkowski brane.
More details and more arguments in favor of the correspondence
between brane charges and imaginary roots
can be found in \cite{Brown:2004jb}.

In \cite{Damour:2002cu} a different program
for identifying imaginary roots of $E_{10}$
with supergravity fields was suggested.
There, imaginary roots were associated with multiple derivatives
of supergravity fields.
In this paper we will only discuss imaginary roots in the
context of their corresponding brane charges,
but it would be interesting to extend our discussion of $E_{10}$ orbifolds
also to the multiple-derivative fields of \cite{Damour:2002cu}.

Also, we have to mention a different scheme 
\cite{West:2001as}-\cite{Keurentjes:2004bv}
in which brane charges are related to real-roots of $E_{11}.$
In this paper we will restrict ourselves to $E_{10}.$
We will also discuss only bosonic excitations.
A more complete description probably requires an extension
of $E_{10}$ to a superalgebra.
A supersymmetric version of $E_{11}$ was recently described in
\cite{Miemiec:2004iv}.

Finally, we mention a recent interesting idea about yet another connection
between $E_9$ and M-theory 
\cite{HoravaE9}-\cite{Bergman:2004ne}.
In this context $E_9$ is conjectured to be a symmetry of a
certain ``topological'' sector of M-theory.
It appears that this $E_9$ has a different real structure than
the one in our context.

\section{The Orbifolds}\label{sec:Orbifolds}
\paragraph{}
We will now begin to study the $\Z_2$-orbifolds.
The following definition will be useful.
\begin{defn}\label{defn:RootsMod2}
We say that two roots $\a,\b\in\Rts$ are {\it equivalent mod $2$} if
$$
\kform{\a}{\eta}\equiv\kform{\b}{\eta}\pmod 2
$$
for every $\eta\in\RtLat$ (where $\RtLat$ is the root lattice).
\end{defn}
For each orbifold we will first identify the $\Z_2$-charge of each supergravity
field $C_{IJK}$ and $g_{IJ}.$ This will allow us to identify the $\Z_2$-charge
of an arbitrary variable $\Fl_\a$ associated with the positive root $\a.$
Next, we will find a relation between the exceptional branes and the $\Z_2$ action.
According to \cite{Brown:2004jb},
the exceptional branes, like any other Minkowski brane, 
are associated with an imaginary root $\OrbRt.$
We will show that the $\Z_2$-charge of any $\Fl_\a$ is determined by $\OrbRt$;
the variable $\Fl_\a$ is $\Z_2$-even ($\Z_2$-odd)
when $\kform{\a}{\OrbRt}$ is an even (odd) integer.
We will now demonstrate these ideas for various $\Z_2$-orbifolds.

\subsection{$T^5/\Z_2$}\label{subsec:T5Z2}
\paragraph{}
We start with the orbifold $T^5/\Z_2.$
The charge assignments are \cite{Dasgupta:1995zm}\cite{Witten:1995em}
\bear
g_{i j} &\rightarrow& g_{ij},\qquad 1\le i,j\le 5,\nn\\
g_{i I} &\rightarrow& -g_{i I},\qquad 1\le i\le 5,\quad 6\le I\le 10\nn\\
g_{I J} &\rightarrow& g_{I J},\qquad 6\le I,J\le 10\nn\\
C_{i j k} &\rightarrow& -C_{i j k},\qquad 1\le i < j < k \le 5,\nn\\
C_{i j K} &\rightarrow& C_{i j K},\qquad 1\le i < j\le 5,\quad 6\le K\le 10\nn\\
C_{i J K} &\rightarrow& -C_{i J K},\qquad 1\le i\le 5,\quad 6\le J<K\le 10\nn\\
C_{I J K} &\rightarrow& C_{I J K},\qquad 6\le I<J<K\le 10\nn\\
\label{eqn:ChgAssignT5}
\eear
Every $C_{IJK}$ ($1\le I<J<K\le 10$) and $g_{IJ}$ ($1\le I\le J\le 10$)
is associated with
a positive real root.
Using \eqref{eqn:AlphaExplicit},
we can associate a $\Z_2$-charge with every simple root as follows,
$$
\Fl_{\a_i}\rightarrow \Fl_{\a_i}\quad (i\neq 3),\qquad
\Fl_{\a_3}\rightarrow -\Fl_{\a_3}.
$$
We say that the $\Z_2$-charge of $\a_{-1},\dots,\a_2,\a_4,\dots,\a_8$
is even and the $\Z_2$-charge of $\a_3$ is odd. 
It is easy to check that if we require the $\Z_2$-charge to be multiplicative
[$\OrbChg{\a+\b}=\OrbChg{\a}\OrbChg{\b}$], we get all $\Z_2$-charge
assignments \eqref{eqn:ChgAssignT5}.
The $\Z_2$-charge of a generic root
\textit{$\a = \sum_{i=-1}^8 k_i\a_i$} ($0\le k_i\in\Z$), 
is therefore \textit{$(-1)^{k_3}$}.

Now, let us move on to the twisted sector of the orbifold.
It consists of $16$ M5-branes, which are associated with the imaginary
root
\belabel{eqn:OrbRtT5Z2}
\OrbRt_{\textit{M5}} = (2,2,2,2,2,1,1,1,1,1).
\ee
We see that the $\Z_2$-charge associated with a generic root $\b$ is determined
by \textit{$\OrbChg{\b}=(-1)^{\kform{\b}{\OrbRt_{\textit{M5}}}}$}.

\subsection{$T^8/\Z_2$}\label{subsec:T8Z2}
\paragraph{}
Now we repeat the calculation for the orbifold $T^8/\Z_2.$
This time the charge assignments are
\bear
g_{i j} &\rightarrow& g_{ij},\qquad 1\le i,j\le 2,\nn\\
g_{i I} &\rightarrow& -g_{i I},\qquad 1\le i\le 2,\quad 3\le I\le 10\nn\\
g_{I J} &\rightarrow& g_{I J},\qquad 3\le I,J\le 10\nn\\
C_{1 2 K} &\rightarrow& -C_{1 2 K},\qquad 3\le K\le 10\nn\\
C_{i J K} &\rightarrow& C_{i J K},\qquad 1\le i\le 2,\quad 3\le J<K\le 10\nn\\
C_{I J K} &\rightarrow& -C_{I J K},\qquad 3\le I<J<K\le 10\nn
\eear
Using \eqref{eqn:AlphaExplicit}, we find that
the $\Z_2$-charge of $\a_{-1},\a_1,\dots,\a_7$
is even and the $\Z_2$-charge of $\a_0,\a_8$ is odd.
The $\Z_2$-charge of a generic root
\textit{$\a = \sum_{i=-1}^8 k_i\a_i$} ($0\le k_i\in\Z$) 
is \textit{$(-1)^{k_0+k_8}$}.

The twisted sector contains 16 M2-branes, which are associated with the imaginary
root
$$
\OrbRt_{\textit{M2}} = (2,2,1,1,1,1,1,1,1,1).
$$
We see that the charge of a generic root $\b$ is again determined
by
\textit{$(-1)^{\kform{\b}{\OrbRt_{\textit{M2}}}}$}.

\subsection{$S^1/\Z_2$}\label{subsec:S1Z2}
\paragraph{}
We will now discuss the \Horava-Witten orbifold $S^1/\Z_2.$
Let the $S^1$ be in the $10^{th}$ direction.
According to \cite{Horava:1995qa},
the $\Z_2$ acts on the bosonic fields of M-theory as follows
\bear
g_{i j} &\rightarrow& g_{ij},\qquad 1\le i,j\le 9,\nn\\
g_{i 10} &\rightarrow& -g_{i 10},\qquad 1\le i\le 9,\nn\\
C_{i j k} &\rightarrow& -C_{i j k},\qquad 1\le i < j < k \le 9,\nn\\
C_{i j 10} &\rightarrow& C_{i j 10},\qquad 1\le i < j \le 9,\nn
\eear
The $\Z_2$ charge of a generic root
\textit{$\a = \sum_{i=-1}^8 k_i\a_i$} ($0\le k_i\in\Z$)
is now defined as \textit{$(-1)^{k_7}$}.

\subsection{Type IA}\label{subsec:typeIA}
\paragraph{}
If we switch the role of the $9^{th}$ and $10^{th}$ directions
in the $S^1/\Z_2$ orbifold of \secref{subsec:S1Z2},
and take the limit $M_p R_{10}\rightarrow 0,$
we get type-IA \cite{Polchinski:1995df}.
The $\Z_2$ acts on the bosonic fields of M-theory as follows
\bear
g_{i j} &\rightarrow& g_{ij},\qquad i,j=1\dots 8,10,\nn\\
g_{i 9} &\rightarrow& -g_{i 9},\qquad i=1\dots 8,10,\nn\\
C_{i j k} &\rightarrow& -C_{i j k},\qquad i,j,k=1\dots8,10,\nn\\
C_{i j 9} &\rightarrow& C_{i j 9},\qquad i,j=1\dots8,10.\nn
\eear
The $\Z_2$-charge of a generic root
\textit{$\a = \sum_{i=-1}^8 k_i\a_i$} ($0\le k_i\in\Z$)
is now defined as \textit{$(-1)^{k_6+k_7}$}.

Now let us analyze the twisted sector.
The twisted sector of type-IA consists of 16 D8-branes
\cite{Horava:1989ga}\cite{Polchinski:1995df}.
The D8-branes span directions $1\dots 8.$
If formally we lift them back to M-theory,
they correspond to the imaginary root
\belabel{eqn:OrbRootD8}
\OrbRt_{\textit{D8}} = (2,2,2,2,2,2,2,2,1,4).
\ee
The associated mass is \cite{Brown:2004jb}
$$
\frac{e^{\inner{\OrbRt_{\textit{D8}}}{\vh}}}{M_p^{9} R_1\cdots R_{10}} = 
M_p^{12} R_1\cdots R_8 R_{10}^3 = \frac{1}{g_s}M_s^9 R_1\cdots R_8,
$$
where $M_s= M_p^{3/2} R_{10}^{1/2}$ is the string scale, and 
$g_s= (M_p R_{10})^{3/2}$ is the string coupling constant.

Note that for any root
\textit{$\b = \sum_{i=-1}^8 k_i\a_i$},
we have
$\kform{\b}{\OrbRt_{\textit{D8}}} = k_6 -3k_7,$
and therefore the $\Z_2$-charge of any 
root can be written as
$$
\OrbChg{\b} = (-1)^{\kform{\b}{\OrbRt_{\textit{D8}}}}.
$$
Of course, any other root of the form $\OrbRt_{\textit{D8}}+2\a'',$
with $\a''$ in the root lattice, will have the same property.
Had we chosen another direction, instead of the $10^{th},$ 
on which to reduce to type-IA, say the $8^{th},$ we would have
gotten another imaginary root, say
$$
\OrbRt'_{\textit{D8}} = (2,2,2,2,2,2,2,4,1,2).
$$
The mass associated with this root is
$$
\frac{e^{\inner{\OrbRt'_{\textit{D8}}}{\vh}}}{M_p^{9} R_1\cdots R_{10}} = 
M_p^{12} R_1\cdots R_7 R_8^3 R_{10}.
$$
It is interesting to check what the mass of the object associated
with $\OrbRt'_{\textit{D8}}$ is in string variables
corresponding  to the original reduction
along the $10^{th}$ direction. We get \cite{Elitzur:1997zn}
$$
M_p^{12} R_1\cdots R_7 R_8^3 R_{10} = 
\frac{M_s^{11}}{g_s^3} R_1\cdots R_7 R_8^3,
$$
where we used
$$
R_{10} = \frac{g_s}{M_s},\qquad
M_p = g_s^{-\frac{1}{3}}M_s.
$$
Both $\OrbRt_{\textit{D8}}$ and 
$\OrbRt'_{\textit{D8}}$ determine the same $\Z_2$-charge
$$
\OrbChg{\b} = (-1)^{\kform{\b}{\OrbRt_{\textit{D8}}}}
 = (-1)^{\kform{\b}{\OrbRt'_{\textit{D8}}}}.
$$
So what distinguishes $\OrbRt'_{\textit{D8}}$?
The answer seems to be that the particular choice of
simple roots \eqref{eqn:AlphaExplicit} implies a particular 
asymptotic region 
\eqref{eqn:WeakAsympt} of $\vh$-space, and
the reduction to type-IA on the $8^{th}$ direction is less favorable
than the reduction on the $10^{th}.$

\subsection{$T^9/\Z_2$}\label{subsec:T9Z2}
\paragraph{}
Now we repeat the calculation for the orbifold $T^9/\Z_2.$
This time the charge assignments are
\bear
g_{1 I} &\rightarrow& -g_{1 I},\qquad 2\le I\le 10\nn\\
g_{I J} &\rightarrow& g_{I J},\qquad 2\le I,J\le 10\nn\\
C_{1 J K} &\rightarrow& -C_{1 J K},\qquad 2\le J<K\le 10\nn\\
C_{I J K} &\rightarrow& C_{I J K},\qquad 2\le I<J<K\le 10\nn
\eear
Using \eqref{eqn:AlphaExplicit}, we find that
the $\Z_2$-charge of $\a_0,\dots,\a_8$
is even and the $\Z_2$-charge of $\a_{-1}$ is odd.
The $\Z_2$ charge of a generic root
\textit{$\a = \sum_{i=-1}^8 k_i\a_i$} ($0\le k_i\in\Z$)
is \textit{$(-1)^{k_{-1}}$}.

The twisted sector contains KK-particles, which are associated with the imaginary
root
$$
\OrbRt_{\textit{KK}} = (0,1,1,1,1,1,1,1,1,1).
$$
We see that the charge of a generic root $\b$ is again determined
by \textit{$(-1)^{\kform{\b}{\OrbRt_{\textit{KK}}}}$}.

\subsection{$T^4/\Z_2$}\label{subsec:T4Z2}
\paragraph{}
For the last example, we take the orbifold $T^4/\Z_2.$
The charge assignments are
\bear
g_{i j} &\rightarrow& g_{ij},\qquad 1\le i,j\le 6,\nn\\
g_{i I} &\rightarrow& -g_{i I},\qquad 1\le i\le 6,\quad 7\le I\le 10\nn\\
g_{I J} &\rightarrow& g_{I J},\qquad 7\le I,J\le 10\nn\\
C_{i j k} &\rightarrow& C_{i j k},\qquad 1\le i < j < k \le 6,\nn\\
C_{i j K} &\rightarrow& -C_{i j K},\qquad 1\le i < j\le 6,\quad 7\le K\le 10\nn\\
C_{i J K} &\rightarrow& C_{i J K},\qquad 1\le i\le 6,\quad 7\le J<K\le 10\nn\\
C_{I J K} &\rightarrow& -C_{I J K},\qquad 7\le I<J<K\le 10\nn
\eear
Using \eqref{eqn:AlphaExplicit}, we find that
the $\Z_2$-charge of $\a_{-1},\dots,\a_3,\a_5,\dots,\a_7$
is even and the $\Z_2$-charge of $\a_4, \a_8$ is odd.
The $\Z_2$-charge of a generic root
\textit{$\a = \sum_{i=-1}^8 k_i\a_i$} ($0\le k_i\in\Z$)
is \textit{$(-1)^{k_4+k_8}$}.

The twisted sectors contain KK-monopoles which are associated with the imaginary
roots
$$
\OrbRt_{\textit{KKM}} = (2,2,2,2,2,2,1,1,1,3),
$$
or any of the permutations
$$
(2,2,2,2,2,2,1,1,3,1),\quad
(2,2,2,2,2,2,1,3,1,1),\quad
(2,2,2,2,2,2,3,1,1,1).
$$
(They are all equal mod $2.$)
We see that the charge of a generic root $\b$ is again determined
by
$$
(-1)^{\kform{\b}{\OrbRt_{\textit{KKM}}}}.
$$

\subsection{Summary}\label{eqn:sumgam}
\paragraph{}
The branes of the twisted sector in a $\Z_2$ orbifold of $T^{10}$
correspond to a positive isotropic imaginary root $\OrbRt$ such that the
$\Z_2$-charge of any root $\b$ is given by
\belabel{eqn:TwistCond}
\OrbChg{\b} = (-1)^{\kform{\b}{\OrbRt}}.
\ee

\section{The Untwisted Sector and $DE_{10}$}\label{sec:Untwisted}
\paragraph{}
We will now study in more detail the subalgebra of $E_{10}$ 
that consists of all elements with even $\Z_2$-charge.

\subsection{Definition of $\AlgInv$}
\paragraph{}
For each of the $\Z_2$-orbifolds above,
we define the ``untwisted'' sector $\AlgInv$ as the Lie subalgebra of $E_{10}$
generated by all the $E_{10}$ generators that correspond to roots with even charge.
In particular, the Cartan subalgebra $\CarOf{E_{10}}\subset E_{10}$
is contained in $\AlgInv,$
since its generators correspond to the root $0.$

In \secref{eqn:sumgam}, we observed that the $\Z_2$-orbifold
is associated with a positive prime isotropic imaginary root $\OrbRt,$
and the $\Z_2$-even roots $\a$ are precisely those with even $\kform{\a}{\OrbRt}.$
We define the subset of even roots
\belabel{eqn:defRtsInv}
\RtsInv\defineas\{\a\in\Rts\suchthat
  \kform{\a}{\OrbRt} \equiv 0\mod 2\},
\qquad \RtsInv\subset\Rts
\ee
and the sublattice
\belabel{eqn:defRtLatInv}
\RtLatInv\defineas\{\a\in\RtLat\suchthat
  \kform{\a}{\OrbRt} \equiv 0\mod 2\},
\qquad
\qquad \RtLatInv\subset\RtLat.
\ee
We proceed to study the subalgebras $\AlgInv.$
Since all positive prime isotropic imaginary roots $\OrbRt$ are
$\Weyl$-equivalent, all $\AlgInv$ algebras are isomorphic.
It is therefore sufficient to concentrate on one particular orbifold,
say $T^5\times (T^5/\Z_2).$

We will demonstrate that $\AlgInv$ contains a proper subalgebra
that is isomorphic to $DE_{10}.$
Here ``proper'' means that $\AlgInv$ is actually bigger than $DE_{10},$
as we will demonstrate in \secref{subsec:subsetDE10}.
These observations have also been made, independently, in \cite{Kleinschmidt:2004dy}.

\subsection{The Kac-Moody Algebra $DE_{10}$}\label{subsec:DE10}
\paragraph{}
We begin with the salient features of the hyperbolic
Kac-Moody algebra $DE_{10}.$ The Dynkin diagram is given in
\figref{fig:DynkinDE10}.

\vskip 10pt
\begin{figure}[h]
\begin{picture}(370,80)
%
%
\thicklines
\multiput(20,20)(20,0){8}{\circle{6}}
\multiput(23,20)(20,0){7}{\line(1,0){14}}

\put(60,23){\line(0,1){14}}
\put(60,40){\circle{6}}
\put(140,23){\line(0,1){14}}
\put(140,40){\circle{6}}
\put( 14,7){$\g_{-1}$}
\put( 36,7){$\g_0$}
\put( 56,7){$\g_1$}
\put( 76,7){$\g_2$}
\put( 96,7){$\g_3$}
\put(116,7){$\g_4$}
\put(136,7){$\g_5$}
\put(156,7){$\g_6$}
\put(64,38){$\g_7$}
\put(144,38){$\g_8$}
\put(20,50){$DE_{10}$}
\end{picture}
\caption{The Dynkin diagram of $DE_{10}.$}
\label{fig:DynkinDE10}
\end{figure}
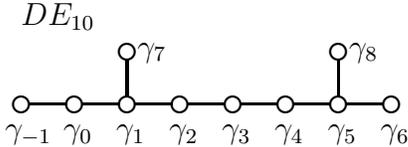
\vskip 10pt

We have the Cartan matrix:
$$
A(DE_{10}) =
\left(\begin{array}{cccccccccc}
2& -1& 0& 0& 0& 0& 0& 0& 0& 0\\ 
-1& 2& -1& 0& 0& 0& 0& 0& 0& 0\\ 
0& -1& 2& -1& 0& 0& 0& 0& -1& 0\\ 
0& 0& -1& 2& -1& 0& 0& 0& 0& 0\\ 
0& 0& 0& -1& 2& -1& 0& 0& 0& 0\\ 
0& 0& 0& 0& -1& 2& -1& 0& 0& 0\\ 
0& 0& 0& 0& 0& -1& 2& -1& 0& -1\\ 
0& 0& 0& 0& 0& 0& -1& 2& 0& 0\\ 
0& 0& -1& 0& 0& 0& 0& 0& 2& 0\\ 
0& 0& 0& 0& 0& 0& -1& 0& 0& 2\\ 
\end{array}\right).
$$
Its inverse is
\belabel{eqn:ADEinverse}
A(DE_{10})^{-1} =
\left(\begin{array}{cccccccccc}
0& -1& -2& -2& -2& -2& -2& -1& -1& -1\\ 
-1& -2& -4& -4& -4& -4& -4& -2& -2& -2\\ 
-2& -4& -6& -6& -6& -6& -6& -3& -3& -3\\ 
-2& -4& -6& -5& -5& -5& -5& -\frac{5}{2}& -3& -\frac{5}{2}\\ 
-2& -4& -6& -5& -4& -4& -4& -2& -3& -2\\ 
-2& -4& -6& -5& -4& -3& -3& -\frac{3}{2}& -3& -\frac{3}{2}\\ 
-2& -4& -6& -5& -4& -3& -2& -1& -3& -1\\ 
-1& -2& -3& -\frac{5}{2}& -2& -\frac{3}{2}& -1& 0& -\frac{3}{2}& -\frac{1}{2}\\ 
-1& -2& -3& -3& -3& -3& -3& -\frac{3}{2}& -1& -\frac{3}{2}\\ 
-1& -2& -3& -\frac{5}{2}& -2& -\frac{3}{2}& -1& -\frac{1}{2}& -\frac{3}{2}& 0\\ 
\end{array}\right).
\ee
The rows of the above matrix are the fundamental weights $\wLambda_i$ 
($i=-1,\dots,8$).
They satisfy
\belabel{eqn:SimpleWgtsDE10}
\kform{\wLambda_i}{\g_j}=\delta_{i,j},\qquad i,j=-1,\dots,10.
\ee
The sum of the simple weights  is
(referring to \figref{fig:DynkinDE10}),
\belabel{eqn:rhoDE10}
\rho(DE_{10}) = \sum_{i=-1}^8\wLambda_i =
-14\g_{-1} -29\g_0 -45\g_1 -40\g_2 -36\g_3 -33\g_4 -31\g_5 -15\g_6  -22\g_7 -15\g_8
\ee
We denote the root space decomposition of $DE_{10}$ by
\belabel{eqn:RtSpacesDE10}
DE_{10} = \bigoplus_{\a\in\RtsOf{DE_{10}}}\AlgG(DE_{10})_\a,
\qquad
\AlgG(DE_{10})_\a\defineas
\{x\in DE_{10}\suchthat [h,x]=\inner{\a}{h}x\quad\forall h\in\CarOf{DE_{10}}\}.
\ee

We will also need the notation of the dual basis
$\gv_{-1},\dots,\gv_8,$ which is a basis of the Cartan subalgebra
$\CarOf{DE_{10}}$ such that
$$
\inner{\gv_i}{\g_j}=A(DE_{10})_{ij},\qquad i,j=-1,\dots,8.
$$
We denote by $\WgtLatOf{DE_{10}}\subset \CarOf{DE_{10}}^*$ the weight lattice
\belabel{eqn:DefWLDE}
\WgtLatOf{DE_{10}}\defineas\{\wLambda\in\CarOf{DE_{10}}^*\suchthat
\kform{\wLambda}{\a}\in\Z
\quad\forall \a\in\RtLatOf{DE_{10}}\}
\equiv\sum_{i=-1}^8\Z\wLambda_i.
\ee

\subsection{The Subgroup $DE_{10}\subset \AlgInv$}\label{subsec:subsetDE10}
\paragraph{}
We will now show that $\AlgInv$ contains a subgroup $DE_{10}.$
For concreteness, we will refer to the
$T^5/\Z_2$ orbifold (discussed in \secref{subsec:T5Z2}).
First, we look for a basis of $\RtLatInv$ [defined in \eqref{eqn:defRtLatInv}]
that is comprised of real roots.
The $\Z_2$-charge is determined by $k_3\pmod 2,$ so all simple roots
$\a_{-1},\dots,\a_8$ except $\a_3$ belong to $\RtLatInv.$
Let us find another real root of $E_{10}$ that completes these $9$ simple roots 
to a basis of $\RtLatInv.$ We need a root with an even coefficient of $k_3,$
and it is not hard to check that 
$$
\Xrt = \a_2 + 2\a_3 + 2\a_4 + 2\a_5 + \a_6 + \a_8
\rightarrow (0,0,0,1,1,0,0,0,0,1)
$$
is the minimal root with this property.
Furthermore,
$$
\RtLatInv = 
\Z\a_{-1} +\Z\a_0 +\Z\a_1 +\Z\a_2 +\Z\a_4 +\Z\a_5 +\Z\a_6 +\Z\a_7 + \Z\a_8 +\Z \Xrt,
$$
and
$$
\RtsInv = \RtLatInv\cap\Rts.
$$

The inner products among the real roots 
$\a_{-1},$ $\a_0,$ $\a_1,$ $\a_2,$ $\a_4,$ $\a_5,$ $\a_6,$ $\a_7,$ $\a_8$ and $\Xrt$
are depicted in a Dynkin-like diagram in \figref{fig:DynkinDE10T5Z2}.
\vskip 10pt
\begin{figure}[h]
\begin{picture}(370,80)
%
%
\thicklines
\multiput(20,20)(20,0){8}{\circle{6}}
\multiput(23,20)(20,0){7}{\line(1,0){14}}

\put(60,23){\line(0,1){14}}
\put(60,40){\circle{6}}
\put(140,23){\line(0,1){14}}
\put(140,40){\circle{6}}
\put( 14,7){$\a_{-1}$}
\put( 36,7){$\a_0$}
\put( 56,7){$\a_1$}
\put( 76,7){$\Xrt$}
\put( 96,7){$\a_7$}
\put(116,7){$\a_6$}
\put(136,7){$\a_5$}
\put(156,7){$\a_4$}
\put(64,38){$\a_2$}
\put(144,38){$\a_8$}
\put(20,50){$DE_{10}$}
\end{picture}
\caption{The Dynkin diagram of the 
$\Z_2$-invariant subalgebra $DE_{10}\subset E_{10}$
that is associated with the $T^5/\Z_2$ orbifold.}
\label{fig:DynkinDE10T5Z2}
\end{figure}
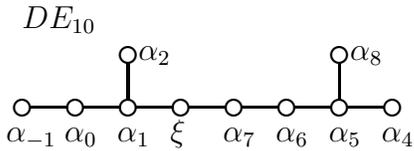
\vskip 10pt
Comparing to \figref{fig:DynkinDE10}, we can define an isometry
$\fDEE:\CarOf{DE_{10}}^*\rightarrow\CarOf{E_{10}}^*$ by
\begin{align}
&\fDEE(\g_{-1})=\a_{-1},\quad
\fDEE(\g_0)=\a_0,\quad
\fDEE(\g_1)=\a_1,\quad
\fDEE(\g_2)=\Xrt,\quad
\fDEE(\g_3)=\a_7,
\nn\\
&\fDEE(\g_4)=\a_6,\quad
\fDEE(\g_5)=\a_5,\quad
\fDEE(\g_6)=\a_4,\quad
\fDEE(\g_7)=\a_2,\quad
\fDEE(\g_8)=\a_8,
\label{eqn:DeffDEE}
\end{align}
Using the nondegenerate bilinear forms $\kform{\cdot}{\cdot}$ on $E_{10}$ and $DE_{10}$
we can identify $\CarOf{DE_{10}}^*$ with $\CarOf{DE_{10}}$
and $\CarOf{E_{10}}^*$ with $\CarOf{E_{10}}.$
The map $\fDEE$ then defines an isomorphism between the Cartan subalgebras
of $E_{10}$ and $DE_{10}.$
We can extend this map to an injective Lie algebra homomorphism
\belabel{eqn:vDE10}
\wfDEE:DE_{10}\rightarrow\AlgInv\subset E_{10}.
\ee
The details can be found in \appref{app:proof}.

\subsection{Comparing $DE_{10}$ to $\AlgInv$}
\label{subsec:CompMul}
\paragraph{}
We have seen in \secref{subsec:subsetDE10}
that $DE_{10}$ is a subalgebra of $\AlgInv.$
We will now demonstrate that it is a {\it proper} subalgebra,
i.e. $\AlgInv$ is bigger.
We note that $DE_{10}$ has an affine $\whD_8$ subalgebra,
and we will study the algebras $DE_{10}$ and $\AlgInv$ as representations
of $\whD_8.$
(Similar observations have been made independently in \cite{Kleinschmidt:2004dy},
but based on a $D_9$ rather than $\whD_8$ subalgebra.)

Both $DE_{10}$ and $\AlgInv$ decompose into irreducible representations
of $\whD_8.$ The complete decomposition is unknown.
It is, however, possible to calculate the number of times that
any representation of $\whD_8$ with affine-level $k=1$ or $2$ appears
in $DE_{10}$ and in $E_{10}.$

Recall that affine $\whD_8\simeq\Affso(16)$ is the Lie algebra 
with generators $J_n^a, K, L_0$ [where $n\in\Z$ and $a=1,\dots,120=\dim\Algso(16)$]
and commutation relations \cite{Bardakci:1970nb}
\belabel{eqn:JJ}
[J_n^a, J_m^b] = {f^{ab}}_c J_{m+n}^c + n \delta_{m,-n}\delta^{ab}K,
\qquad
[K,J_n^a]=[K,L_0]=0,
\qquad
[L_0, J_n^a] = -n J_n^a,
\ee
where ${f^{ab}}_c$ are the structure constants of $\Algso(16)$ in a
basis where the Killing form is $\delta^{ab}.$

Looking at the Dynkin diagram of $DE_{10}$ (\figref{fig:DynkinDE10}), we see that
if we drop the node $\g_{-1},$ we get the Dynkin diagram of $\whD_8$
(\figref{fig:DynkinwhD8}).
\vskip 10pt
\begin{figure}[h]
\begin{picture}(370,80)
%
%
\thicklines
\multiput(40,20)(20,0){7}{\circle{6}}
\multiput(43,20)(20,0){6}{\line(1,0){14}}

\put(60,23){\line(0,1){14}}
\put(60,40){\circle{6}}
\put(140,23){\line(0,1){14}}
\put(140,40){\circle{6}}
\put( 36,7){$\g_0$}
\put( 56,7){$\g_1$}
\put( 76,7){$\g_2$}
\put( 96,7){$\g_3$}
\put(116,7){$\g_4$}
\put(136,7){$\g_5$}
\put(156,7){$\g_6$}
\put(64,38){$\g_7$}
\put(144,38){$\g_8$}
\put(20,50){$\whD_8$}
\end{picture}
\caption{The Dynkin diagram of the $\whD_8$ subalgebra.}
\label{fig:DynkinwhD8}
\end{figure}
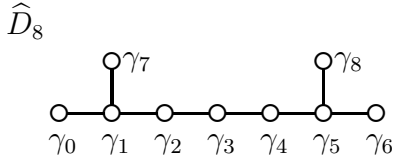
\vskip 10pt
Define the $DE_{10}$-root
$$
\wdelta\defineas\g_0 + 2\g_1 + 2\g_2 + 2\g_3 + 2\g_4 + 2\g_5 + \g_6 + \g_7 + \g_8.
$$
Then the $\whD_8\subset DE_{10}$ subalgebra can be defined as
the sum of the root spaces $\AlgG(DE_{10})_\a$ that correspond to roots
of the form \textit{$\a=\sum_0^8 k_i\g_i.$} This can also be written as,
$$
\whD_8\simeq \bigoplus_{\kform{\a}{\wdelta}=0}\AlgG(DE_{10})_\a.
$$
In particular $\CarOf{DE_{10}}\subset \whD_8.$
The elements $\gv_{-1},\dots,\gv_8,$ which form a basis for
$\CarOf{DE_{10}},$ are identified with the following generators of $\whD_8$
\cite{KacBook}:
$\gv_1,\dots,\gv_8$ are identified with appropriate linear combinations of
the $J_0^a$'s that span the Cartan subalgebra of $D_8;$
the element 
$$
\wdelta^\vee\defineas
\gv_0 + 2\gv_1 + 2\gv_2 + 2\gv_3 + 2\gv_4 + 2\gv_5 + \gv_6 + \gv_7 + \gv_8,
$$
is identified with $K,$ and 
$\gv_{-1}$ is identified with $L_0.$

Furthermore, for any positive integer $k,$ the subspace of all level-$k$ root-spaces,
\belabel{eqn:ModuleDE10}
\ModuleAtLevelOf{k}{DE_{10}}\defineas
\bigoplus_{\kform{\a}{\wdelta}=k}\AlgG(DE_{10})_\a,
\ee
is a representation of $\whD_8$ of affine-level $k$
(i.e. the central element $K$ of $\whD_8$ is constant 
on $\ModuleAtLevelOf{k}{DE_{10}}$ with value $k$).
The weights of this representation are bounded from above,
since only negative roots $\a$ appear in \eqref{eqn:ModuleDE10}.
We need the decomposition of $\ModuleAtLevelOf{k}{DE_{10}}$
into irreducible representations of $\whD_8.$
There is a standard method to find this decomposition 
\cite{KMW}-\cite{FF}
that is also nicely explained in
\cite{Bauer:1996ca}.

First, recall that an irreducible 
highest-weight representation of $\whD_8\simeq\Algso(16)$
is defined by the highest-weight $\wLambda.$
This weight $\wLambda$ is a $10$-dimensional vector and can naturally be identified
with an element of $\CarOf{DE_{10}}^*,$ since
$\CarOf{DE_{10}}=\CarOf{\whD_8}.$
$\wLambda$ is a linear combination of the fundamental
weights $\wLambda_{-1},\dots,\wLambda_8$ \eqref{eqn:SimpleWgtsDE10}.
This representation is denoted by $L(\wLambda).$
It turns out (\S12 of \cite{KacBook}) that 
\belabel{eqn:DE10Level1}
\ModuleAtLevelOf{1}{DE_{10}}\simeq L(-\g_{-1}) \equiv L(\wLambda_0+2\wdelta).
\ee
Here the representation $L(\wLambda_0+2\wdelta)$ can be realized as
the Hilbert space of $16$ chiral Majorana fermions in 1+1D with anti-periodic
boundary conditions \cite{Bardakci:1970nb}.

The decomposition of $\ModuleAtLevelOf{2}{DE_{10}}$ into irreducible
$\whD_8$ representations is a bit more complicated.
It can be shown that $\ModuleAtLevelOf{2}{DE_{10}}$ can be
represented as a certain coset of the anti-symmetric tensor product
$\wedge^2\ModuleAtLevelOf{1}{DE_{10}}=\wedge^2 L(\wLambda_0+2\wdelta).$
Specifically,
$$
\ModuleAtLevelOf{2}{DE_{10}} = 
  \wedge^2 L(\wLambda_0+2\wdelta)/L(\wLambda_1+3\wdelta)
$$
The tensor product $\wedge^2 L(\wLambda_0+2\wdelta)$
can be decomposed into irreducible $\whD_8$ representations
using the {\it coset construction}.
The result is
\bear
\ModuleAtLevelOf{2}{DE_{10}} &\simeq &
\bigoplus_{n=0}^\infty \coDE{1}{n} L(\wLambda_1+(3-n)\wdelta)\oplus
\bigoplus_{n=0}^\infty \coDE{5}{n} L(\wLambda_5+(1-n)\wdelta)
\label{eqn:DE10Level2}
\eear
where the integer multiplicities 
$\coDE{1}{n}$ and $\coDE{5}{n}$ are given by the generating functions
\bear
\sum_{n=0}^\infty \coDE{1}{n} q^n &=& \frac{1}{\vph(q)}-1
= q + 2 q^2 + 3 q^3 + 5 q^4 + 7 q^5 + 11 q^6 + 15 q^7+\cdots,
\nn\\
\sum_{n=0}^\infty \coDE{5}{n} q^n &=& \frac{1}{\vph(q)}
= 1+ q + 2 q^2 + 3 q^3 + 5 q^4 + 7 q^5 + 11 q^6 + 15 q^7 + \cdots,
\nn
\eear
where
$$
\vph(q)\defineas \prod_{m=1}^\infty (1-q^m)
=\sum_{n=1}^\infty
(-1)^n \bigl(q^{\frac{1}{2}n(3n-1)}+q^{\frac{1}{2}n(3n+1)}\bigr)
=1-q-q^2+q^5+q^7-q^{12}-q^{15}+\cdots,
$$
The generating function $1/\vph(q)$
is the character of a Virasoro representation with
central charge $c=1$ (a free boson).
It appears in this context because the coset construction
$\Affso(2n)_1\otimes\Affso(2n)_1/\Affso(2n)_2$
has central charge $c=1.$
We have also used a theorem from \cite{KW} which states
that for an affine Lie algebra of type ADE with ``odd exponents''
(see \S 14.1 of \cite{KacBook} for a definition),
if a highest-weight representation $L(M)$
appears in $\wedge^2 L(\Lambda),$ for $\Lambda\in\PosWgts$ of level-$1,$
then $\hgt (2\Lambda-M)$ must be odd
(where $\hgt$ is the ``height'' of the root,
i.e. the sum of the coefficients in the root's decomposition
into simple roots). Note that this theorem applies to both $E_9$
and $\whD_8.$
For more details on the 
relevant $\Affso(2n)_1\otimes\Affso(2n)_1/\Affso(2n)_2$
coset construction, see 
\cite{Bardakci:1970nb}\cite{KW}\cite{Halpern:1971ay}-\cite{Ishikawa:2003xh}.

To compare $DE_{10}$ to $\AlgInv\subset E_{10},$ we first define
\belabel{eqn:deltaE10}
\delta \defineas \a_0 +2\a_1 +3\a_2 +4\a_3 +5\a_4 +6\a_5 +4\a_6 +2\a_7 +3\a_8
\in\RtsOf{E_{10}},
\ee
and
\belabel{eqn:ModuleE10}
\ModuleAtLevelOf{k}{E_{10}}\defineas
\bigoplus_{\kform{\a}{\delta}=k}\AlgG(E_{10})_\a.
\ee
We then define
$$
\ModuleAtLevelOf{k}{\AlgInv}\defineas\AlgInv\cap\ModuleAtLevelOf{k}{E_{10}}.
$$
Since $\fDEE(\g_{-1})=\a_{-1}$ [see \eqref{eqn:DeffDEE}]
and $\fDEE(\wdelta)=\delta,$
every level-$k$ root of $DE_{10}$ (with respect to $\wdelta$)
is also a level-$k$ root of $E_{10}$ with respect to $\delta$.
We therefore have
$$
\wfDEE(\ModuleAtLevelOf{k}{DE_{10}})\subseteq\ModuleAtLevelOf{k}{\AlgInv},
$$
where $\wfDEE$ was defined in \eqref{eqn:vDE10}.

$E_{10}$ has an $E_9$ (affine $\widehat{E}_8$) subalgebra
that corresponds to the nodes $\a_0,\dots,\a_8$ in
\figref{fig:DynkinE10}, and
$\ModuleAtLevelOf{k}{E_{10}}$ is an $E_9$-representation.
$\ModuleAtLevelOf{k}{\AlgInv},$ on the other hand, is not an $E_9$-representation,
since $E_9$ contains $\Z_2$-odd generators.
However, the direct sum of the $\Z_2$-even root spaces of $E_9$ is isomorphic to $\whD_8.$
Thus, $\ModuleAtLevelOf{k}{\AlgInv}$ is a $\whD_8$-representation.
In order to complete the comparison of $DE_{10}$ to $\AlgInv$ at low levels,
we need to know how $\ModuleAtLevelOf{k}{\AlgInv}$ decomposes into irreducible
representations of $\whD_8.$
We can do this by decomposing $\ModuleAtLevelOf{k}{E_{10}}$ into
irreducible representations of $E_9,$ 
then decomposing these representations under $\wfDEE(\whD_8)\subset E_9,$
and then intersecting with $\AlgInv.$

The embedding $\whD_8\simeq\wfDEE(\whD_8)\subset E_9$ is easy to describe.
It is the extension of the minimal embedding $\Algso(16)\subset E_8$
[under which the adjoint representation $\rep{248}$ of $E_8$ decomposes
as the sum of the adjoint $\rep{120}$  and one of the spinor $\rep{128}$
representations of $\Algso(16)$] to affine algebras.

The decomposition of $\ModuleAtLevelOf{k}{E_{10}}$ at level $k=1,2$
was calculated in 
\cite{KMW}-\cite{FF}
(and was extended to $k=3$
in \cite{Bauer:1996ca}, but we will not use this result here).
It is given by
\belabel{eqn:E10Level1}
\ModuleAtLevelOf{1}{E_{10}}\simeq L(-\a_{-1}) \equiv L(\Lambda_0+2\delta),
\ee
and
\belabel{eqn:E10Level2}
\ModuleAtLevelOf{2}{E_{10}}\simeq  
\bigoplus_{n=0}^\infty c_n L(\Lambda_1+(3-n)\delta)
\ee
with the generating function
$$
\sum_{n=0}^\infty c_n q^n = 
\prod_{n=1}^\infty (1+q^n)-1
=\frac{\vph(q^2)}{\vph(q)}-1.
$$
To compare with \eqref{eqn:DE10Level1} and \eqref{eqn:DE10Level2},
we need to
decompose \eqref{eqn:E10Level1} and
\eqref{eqn:E10Level2} under $\whD_8\subset E_9.$
For level-$1$ we have
\belabel{eqn:E9Level1Decom}
L(\Lambda_0+2\delta) \simeq
L(\wLambda_1+2\wdelta)\oplus L(\wLambda_6+2\wdelta).
\ee
Note that on the left-hand side 
we have an irreducible representation of $E_9,$
and on the right-hand side we have irreducible representations of $\whD_8.$
To complete the calculation of $\ModuleAtLevelOf{1}{\AlgInv}$
we must intersect \eqref{eqn:E9Level1Decom} with $\AlgInv.$

{}From \eqref{eqn:ADEinverse} we find
\bear
-\wLambda_1 &=&
2\g_{-1}+4\g_0+6\g_1+6\g_2+6\g_3+6\g_4+6\g_5+3\g_6+3\g_7+3\g_8,
\nn\\ 
-\wLambda_5 &=&
2\g_{-1}+4\g_0+6\g_1+5\g_2+4\g_3+3\g_4+2\g_5+\g_6+3\g_7+\g_8,
\nn\\ 
-\wLambda_6 &=&
\g_{-1}+2\g_0 +3\g_1 +\tfrac{5}{2}\g_2 +2\g_3 +\tfrac{3}{2}\g_4 +\g_5 +\tfrac{3}{2}\g_7+\tfrac{1}{2}\g_8.
\nn
\eear
(We have also listed $\wLambda_5,$ which will be needed later on.)
Since
$$
-\fDEE(\wLambda_6) = 
\a_{-1}+2\a_0+3\a_1+4\a_2+5\a_3+5\a_4+6\a_5+4\a_6+2\a_7+3\a_8
$$
is a $\Z_2$-odd level-$1$ root,
all the weights of the module $L(\wLambda_6+2\wdelta)$
map, under $\fDEE,$ to $\Z_2$-odd roots of $E_{10}.$
the intersection $\AlgInv\cap L(\Lambda_0+2\delta)$ therefore
leaves only $L(\wLambda_1+2\wdelta),$ and thus
$$
\ModuleAtLevelOf{1}{DE_{10}} = \ModuleAtLevelOf{1}{\AlgInv}.
$$
There is another way to arrive at this result.
Recall that $L(\Lambda_0+2\delta)$ can be realized as the Hilbert
space of $16$ 1+1D chiral Majorana fermions with either all periodic
or all anti-periodic boundary conditions \cite{Goddard:1985xp}.
According to \cite{Bardakci:1970nb}, $16$ Majorana fermions
also realize $\Affso(16).$
The representation $L(\wLambda_1+2\wdelta)$ corresponds to
the sector with anti-periodic boundary conditions and odd fermion number,
and the representation $L(\wLambda_6 +2\wdelta)$ corresponds
to the sector with periodic boundary conditions.
It is not hard to check that the $\Z_2$-charge
is $+$ or $-$ according to whether
the boundary conditions are anti-periodic or periodic.
Thus,
$$
\ModuleAtLevelOf{1}{\AlgInv}\simeq L(\wLambda_1+2\wdelta)
\simeq \ModuleAtLevelOf{1}{DE_{10}}.
$$
At level-$2$ such an equality will no longer hold.

The decomposition under $\whD_8$ of the $\Z_2$-invariant
part of the level-$2$ $E_9$ representations
that appear in \eqref{eqn:E10Level2} is given by
\belabel{eqn:E10Level2Decom}
L(\Lambda_1+3\delta)^{(\textit{inv})} \simeq
\bigoplus_{m=0}^\infty \coE{1}{m} L(\wLambda_1+(3-m)\wdelta)\oplus
\bigoplus_{m=0}^\infty \coE{5}{m} L(\wLambda_5+(3-m)\wdelta)
\ee
with
$$
\sum_{n=0}^\infty \coE{1}{n}q^n = 
\prod_{n=1}^\infty (1+q^n) = \frac{\vph(q^2)}{\vph(q)},
\qquad
\sum_{n=0}^\infty \coE{5}{n}q^n = 
q \prod_{n=1}^\infty (1+q^n) = \frac{q\vph(q^2)}{\vph(q)}.
$$
These expressions can be derived from the coset construction
$(E_9)_{k=2}/(\whD_8)_{k=2}$ which has central charge $c=\tfrac{1}{2}.$

Combining \eqref{eqn:E10Level2} with \eqref{eqn:E10Level2Decom}
we obtain
\belabel{eqn:AlgInvLevel2}
\ModuleAtLevelOf{2}{\AlgInv} \simeq
\bigoplus_{n=0}^\infty \coEXD{1}{n} L(\wLambda_1+(3-n)\wdelta)\oplus
\bigoplus_{n=0}^\infty \coEXD{5}{n} L(\wLambda_5+(3-n)\wdelta)
\ee
with
\bear
\sum_{n=0}^\infty \coEXD{1}{n} q^n &=&
\prod_{n=1}^\infty (1+q^n)^2 -\prod_{n=1}^\infty(1+q^n) 
\nn\\ &=&
q + 2 q^2 + 4 q^3 + 7 q^4 + 11 q^5 + 18 q^6 + 27 q^7 +\cdots
\nn\\
\sum_{n=0}^\infty \coEXD{5}{n} q^n &=& 
q\Bigl\lbrack\prod_{n=1}^\infty (1+q^n)^2 -\prod_{n=1}^\infty(1+q^n) 
\Bigr\rbrack
\nn\\
&=& q^2 + 2 q^3 + 4 q^4 + 7 q^5 + 11 q^6 + 18 q^7 + 27 q^8 +\cdots
\nn
\eear
Now we can compare \eqref{eqn:AlgInvLevel2} to \eqref{eqn:DE10Level2}.
We have
$$
\ModuleAtLevelOf{2}{\AlgInv}/
\ModuleAtLevelOf{2}{DE_{10}}
\simeq
\bigoplus_{n=0}^\infty \DcoEXDE{1}{n} L(\wLambda_1+(3-n)\wdelta)\oplus
\bigoplus_{n=0}^\infty \DcoEXDE{5}{n} L(\wLambda_5+(2-n)\wdelta)
$$
with
\bear
\sum_{n=0}^\infty \DcoEXDE{1}{n} q^n &=&
\prod_{n=1}^\infty (1+q^n)^2 -\prod_{n=1}^\infty(1+q^n) 
-\prod_{n=1}^\infty(1-q^n)^{-1}+1
\nn\\
&=&
q^3 + 2 q^4 + 4 q^5 + 7 q^6 + 12 q^7 + \cdots
\nn\\
\sum_{n=0}^\infty \DcoEXDE{5}{n} q^n &=& 
\prod_{n=1}^\infty (1+q^n)^2 -\prod_{n=1}^\infty(1+q^n) 
-q\prod_{n=1}^\infty(1-q^n)^{-1}
\nn\\ &=&
q^2 + 2 q^3 + 4 q^4 + 6 q^5 + 11 q^6 + 16 q^7 + 25 q^8 + \cdots 
\nn
\eear
We see that $\AlgInv$ is strictly bigger than $DE_{10}$ at level $2.$
For example, the $DE_{10}$ and $\AlgInv$ multiplicities of 
the $E_{10}$ root
\bear
-\fDEE(\wLambda_1)
&=&
2\a_{-1}+4\a_0+6\a_1 +9\a_2 +12\a_3 +15\a_4 +18\a_5 +12\a_6 +6\a_7 +9\a_8
\nn\\ &&
\rightarrow (2,2,2,3,3,3,3,3,3,3)
\nn
\eear
are $711$ and $727,$ respectively.
To calculate these multiplicities,
one needs to know the multiplicities of 
$\wLambda_1$ in $L(\wLambda_1+(3-n)\wdelta)$ for $n=1,2,3,$
and also the multiplicities 
of $\wLambda_1$ in $L(\wLambda_5 + (2-n)\wdelta)$ for
$n=0,1,2.$
These can be found for example in \cite{KMPS}
and are given by
\bear
\lefteqn{
\sum \mult (\text{$\wLambda_1+(3-n)\wdelta$ in $L(\wLambda_1+3\wdelta)$})q^n
} \nn\\ && \qquad
= 1+27q+362q^2+3377q^3+24831q^4+ 153635q^5 + \cdots
\nn\\
\lefteqn{
\sum \mult (\text{$\wLambda_1+(3-n)\wdelta$ in $L(\wLambda_5+\wdelta)$})q^n
} \nn\\ && \qquad
= 15q^2+277q^3+2917q^4 +22740q^5+\cdots
\nn
\eear
At the next order, the $DE_{10}$ and $\AlgInv$ multiplicities of 
the $E_{10}$ root
\bear
\fDEE(\delta-\wLambda_1)
&=&
2\a_{-1}+5\a_0+8\a_1 +12\a_2 +16\a_3 +20\a_4 +24\a_5 +16\a_6 +8\a_7 +12\a_8
\nn\\ &&
\rightarrow (2,3,3,4,4,4,4,4,4,4)
\nn
\eear
are $7411$ and $7747,$ respectively.

\section{The Kac-Moody Algebra $DE_{18(10)}$}\label{sec:DE18}
\paragraph{}
In \secref{sec:Untwisted} we studied the ``untwisted sectors'' of the
$E_{10}$ $\Z_2$-orbifolds. We defined $\AlgInv$ 
and discovered that the hyperbolic Kac-Moody 
algebra $DE_{10}$ is contained in it.
On physical grounds, we know that there should also be ``twisted sectors''
associated with the $16$ additional objects (exceptional branes) that are present
in the M-theory compactification on $T^{10}/\Z_2.$
In this section we will see that both twisted and untwisted sectors
are unified inside a bigger structure -- a coset 
of the Kac-Moody algebra $DE_{18(10)}.$

This algebra, that will be defined in \secref{subsec:RealDE18} below,
contains a natural $\Algso(16)$ subalgebra.
We will see that this number $16$ is directly related to the existence
of $16$ exceptional branes.
We will propose that the decomposition of $DE_{18(10)}$ into irreducible
representations of $\Algso(16)$ has a physical meaning 
in terms of the $\Z_2$-orbifold of M-theory on $T^{10}$ as follows.
The entire ``untwisted'' sector
is related to elements of $DE_{18(10)}$
that are in singlet representations $\rep{1}$ of $\Algso(16)$
[i.e. commute with $\Algso(16)$].
Physical modes that are related to a single exceptional brane correspond
to elements of $DE_{18(10)}$
that fall into the fundamental representation $\rep{16}$ of $\Algso(16).$
Physical modes that are related to pairs of exceptional branes
correspond to elements of $DE_{18(10)}$
that fall into either the anti-symmetric tensor  representation
$\rep{120}$ or the traceless symmetric tensor representation $\rep{135},$
and so on.

\subsection{Basic Properties of $DE_{18}$}
\paragraph{}
The $\Z_2$-orbifolds that we are studying are dual to heterotic string
theory compactified on $T^9.$
As argued in \cite{Sen:1994wr}, $\Algso(24,8,\R)$ is the relevant Lie algebra 
for heterotic string theory on $T^7.$
The Dynkin diagram of the complexified Lie algebra  is $D_{16}.$
The extension to heterotic string theory
on $T^8$ is the affine $\Affso(24,8,\R)$
(
\cite{Nicolai:1987vy}-\cite{Schwarz:1995td}
and references therein).
The extension to heterotic string theory
on $T^9$ is naturally achieved by adding 
one more node. There are, of course,
many ways to add a node to a given Dynkin diagram,
but it turns out that the right one is the Dynkin diagram $DE_{18}.$
The relation between these three Dynkin diagrams is depicted in
\figref{fig:DynkinDE18AndD16}.

\vskip 10pt
\begin{figure}[h]
\begin{picture}(370,230)
%
%

\put(0,140){\begin{picture}(370,80)
\thicklines
\multiput(20,20)(20,0){16}{\circle{6}}
\multiput(23,20)(20,0){15}{\line(1,0){14}}

\put(60,23){\line(0,1){14}}
\put(60,40){\circle{6}}
\put(300,23){\line(0,1){14}}
\put(300,40){\circle{6}}

\put( 14,7){$\b_{-1}$}
\put( 36,7){$\b_0$}
\put( 56,7){$\b_1$}
\put( 76,7){$\b_2$}
\put( 96,7){$\b_3$}
\put(116,7){$\b_4$}
\put(136,7){$\b_5$}
\put(156,7){$\b_6$}
\put(176,7){$\b_7$}
\put(196,7){$\b_9$}
\put(216,7){$\b_{10}$}
\put(236,7){$\b_{11}$}
\put(256,7){$\b_{12}$}
\put(276,7){$\b_{13}$}
\put(296,7){$\b_{14}$}
\put(316,7){$\b_{15}$}

\put(64,38){$\b_8$}
\put(304,38){$\b_{16}$}
\put(20,50){$DE_{18}$}
\end{picture}} 

\put(0,70){\begin{picture}(370,80)
\thicklines
\multiput(40,20)(20,0){15}{\circle{6}}
\multiput(43,20)(20,0){14}{\line(1,0){14}}

\put(60,23){\line(0,1){14}}
\put(60,40){\circle{6}}
\put(300,23){\line(0,1){14}}
\put(300,40){\circle{6}}

\put( 36,7){$\b_0$}
\put( 56,7){$\b_1$}
\put( 76,7){$\b_2$}
\put( 96,7){$\b_3$}
\put(116,7){$\b_4$}
\put(136,7){$\b_5$}
\put(156,7){$\b_6$}
\put(176,7){$\b_7$}
\put(196,7){$\b_{9}$}
\put(216,7){$\b_{10}$}
\put(236,7){$\b_{11}$}
\put(256,7){$\b_{12}$}
\put(276,7){$\b_{13}$}
\put(296,7){$\b_{14}$}
\put(316,7){$\b_{15}$}

\put(64,38){$\b_8$}
\put(304,38){$\b_{16}$}
\put(20,50){$\whD_{16}\subset DE_{18}$}
\end{picture}} 

\put(0,0){\begin{picture}(370,80)
\thicklines
\multiput(60,20)(20,0){14}{\circle{6}}
\multiput(63,20)(20,0){13}{\line(1,0){14}}

\put(60,23){\line(0,1){14}}
\put(60,40){\circle{6}}
\put(300,23){\line(0,1){14}}
\put(300,40){\circle{6}}

\put( 56,7){$\b_1$}
\put( 76,7){$\b_2$}
\put( 96,7){$\b_3$}
\put(116,7){$\b_4$}
\put(136,7){$\b_5$}
\put(156,7){$\b_6$}
\put(176,7){$\b_7$}
\put(196,7){$\b_{9}$}
\put(216,7){$\b_{10}$}
\put(236,7){$\b_{11}$}
\put(256,7){$\b_{12}$}
\put(276,7){$\b_{13}$}
\put(296,7){$\b_{14}$}
\put(316,7){$\b_{15}$}

\put(64,38){$\b_8$}
\put(304,38){$\b_{16}$}
\put(20,50){$D_{16}\subset \whD_{16}$}
\end{picture}} 

\end{picture}
\caption{The Dynkin diagram of  $DE_{18}$ and its subdiagrams
$\hat{D}_{16}$ and $D_{16}.$}
\label{fig:DynkinDE18AndD16}
\end{figure}
\vskip 10pt

\subsection{The Cartan Matrix}
\paragraph{}
The Cartan matrix for $DE_{18}$ with respect to the 
basis $\b_{-1},\dots,\b_{16}$
is given by 
\be\label{eqn:CartanDE18}
A_{18}=
 \left( \begin{array}{cccccccccccccccccc}
   2 & -1 &  &  &  &  &   &  &  &  &  &  &  &   &  &  &  &  \\
   -1 & 2 & -1 &  &  &  &   &  &  &  &  &  &  &   &  &  &  &  \\
   & -1 & 2 & -1 &  &  &  &  &  & -1 &  &  &  &  &   &  &  &  \\  
   &  & -1 & 2 & -1 &  &  &  &  &   &  &  &  &  &  &  &   &  \\     
   &  &  & -1 & 2 & -1 &  &  &  &  &   &  &  &  &  &  &  &   \\    
   &  &  &  & -1 & 2 & -1 &  &  &  &  &   &  &  &  &  &  &   \\     
   &  &  &  &  & -1 & 2 & -1 &  &  &  &  &   &  &  &  &  &   \\     
   &  &  &  &  &  & -1 & 2 & -1 &  &  &  &  &   &  &  &  &   \\     
   &  &  &  &  &  &  & -1 & 2 &  & -1 &  &  &  &   &  &  &   \\
   &  & -1 &  &  &  &  &  &  & 2 &  &  &  &  &  &   &  &   \\      
   &  &  &  &  &  &  &  & -1 &  & 2 & -1 &  &  &  &  &   &   \\
   &  &  &  &  &  &  &  &  &  & -1 & 2 & -1 &  &  &  &  &    \\
   &  &  &  &  &  &  &  &  &  &  & -1 & 2 & -1 &  &  &  &   \\
   &  &  &  &  &  &  &  &  &  &  &  & -1 & 2 & -1 &  &  &   \\      
   &  &  &  &  &  &  &  &  &  &  &  &  & -1 & 2 & -1 &  &   \\      
   &  &  &  &  &  &  &  &  &  &  &  &  &  & -1 & 2 & -1 & -1  \\ 
   &  &  &  &  &  &  &  &  &  &  &  &  &  &  & -1 & 2 &    \\
   &  &  &  &  &  &  &  &  &  &  &  &  &  &  & -1 &  & 2\end{array} 
\right)
\ee
The determinant is
$$
\det A_{18} = -4,
$$
and the inverse Cartan matrix $A_{18}^{-1}$ is given by,
$$
 \left( \begin{array}{rrrrrrrrrrrrrrrrrr}
 0  & -1 & -2 & -2   & -2 & -2   & -2 & -2  & -2 & -1   
      & -2 & -2 & -2  & -2 & -2  & -2 & -1   & -1   \\ 
 -1 & -2 & -4 & -4   & -4 & -4   & -4 & -4  & -4 & -2   
      & -4 & -4 & -4  & -4 & -4  & -4 & -2   & -2   \\
 -2 & -4 & -6 & -6   & -6 & -6   & -6 & -6  & -6 & -3   
      & -6 & -6 & -6  & -6 & -6  & -6 & -3   & -3   \\  
 -2 & -4 & -6 & -5   & -5 & -5   & -5 & -5  & -5 & -3   
      & -5 & -5 & -5  & -5 & -5  & -5 & -\tfrac{5}{2} & -\tfrac{5}{2} \\
 -2 & -4 & -6 & -5   & -4 & -4   & -4 & -4  & -4 & -3   
      & -4 & -4 & -4  & -4 & -4  & -4 & -2   & -2   \\ 
 -2 & -4 & -6 & -5   & -4 & -3   & -3 & -3  & -3 & -3   
      & -3 & -3 & -3  & -3 & -3  & -3 & -\tfrac{3}{2} & -\tfrac{3}{2} \\
 -2 & -4 & -6 & -5   & -4 & -3   & -2 & -2  & -2 & -3   
      & -2 & -2 & -2  & -2 & -2  & -2 & -1   & -1   \\
 -2 & -4 & -6 & -5   & -4 & -3   & -2 & -1  & -1 & -3   
      & -1 & -1 & -1  & -1 & -1  & -1 & -\tfrac{1}{2}  & -\tfrac{1}{2}  \\
 -2 & -4 & -6 & -5   & -4 & -3   & -2 & -1  & 0  & -3   
      & 0  & 0  & 0   & 0  & 0   & 0  & 0    & 0    \\
 -1 & -2 & -3 & -3   & -3 & -3   & -3 & -3  & -3 & -1   
      & -3 & -3 & -3  & -3 & -3  & -3 & -\tfrac{3}{2} & -\tfrac{3}{2} \\
 -2 & -4 & -6 & -5   & -4 & -3   & -2 & -1  & 0  & -3   
      & 1  & 1  & 1   & 1  & 1   & 1  & \tfrac{1}{2}   & \tfrac{1}{2}   \\
 -2 & -4 & -6 & -5   & -4 & -3   & -2 & -1  & 0  & -3   
      & 1  & 2  & 2   & 2  & 2   & 2  & 1    & 1    \\
 -2 & -4 & -6 & -5   & -4 & -3   & -2 & -1  & 0  & -3   
      & 1  & 2  & 3   & 3  & 3   & 3  & \tfrac{3}{2}  & \tfrac{3}{2}  \\
 -2 & -4 & -6 & -5   & -4 & -3   & -2 & -1  & 0  & -3   
      & 1  & 2  & 3   & 4  & 4   & 4  & 2    & 2    \\
 -2 & -4 & -6 & -5   & -4 & -3   & -2 & -1  & 0  & -3   
      & 1  & 2  & 3   & 4  & 5   & 5  & \tfrac{5}{2}  & \tfrac{5}{2}  \\
 -2 & -4 & -6 & -5   & -4 & -3   & -2 & -1  & 0  & -3   
      & 1  & 2  & 3   & 4  & 5   & 6  & 3    & 3    \\  
 -1 & -2 & -3 & -\tfrac{5}{2} & -2 & -\tfrac{3}{2} & -1 
      & -\tfrac{1}{2} & 0  & -\tfrac{3}{2} & \tfrac{1}{2} & 1  
          & \tfrac{3}{2} & 2  & \tfrac{5}{2} & 3  & 2    & \tfrac{3}{2}  \\
 -1 & -2 & -3 & -\tfrac{5}{2} & -2 & -\tfrac{3}{2} & -1 
      & -\tfrac{1}{2} & 0  & -\tfrac{3}{2} & \tfrac{1}{2} & 1  
          & \tfrac{3}{2} & 2  & \tfrac{5}{2} & 3  & \tfrac{3}{2}  & 2
\end{array} \right)
$$
The rows of $A_{18}^{-1}$ will be useful to us as follows.
Define the {\it fundamental weights} $\DENFwgt_i\in\DualCarOf{DE_{18}}$
($i=-1,\dots 16$) to satsify
$$
\inner{\DENFwgt_i}{\b_j} = \delta_{ij},\qquad i,j=-1,\dots,16.
$$
Then $\DENFwgt_i$ can be explicitly written as a linear combination
of the simple roots $\b_j$ ($j=-1,\dots,16$) with coefficents given
by the $i^{th}$ row of $(A_{18})^{-1}.$

\subsection{The Real Form $DE_{18(10)}$}\label{subsec:RealDE18}
\paragraph{}
We need a certain {\it real form} of the complexified Lie algebra $DE_{18}.$
The real form is defined by the 
Tits-Satake diagram --  a choice of ``black'' and ``white'' nodes
on the Dynkin diagram, as depicted in \figref{fig:DynkinRealDE18}.

\vskip 10pt
\begin{figure}[h]
\begin{picture}(370,80)
%
%
\thicklines
\multiput(20,20)(20,0){9}{\circle{6}}
\multiput(200,20)(20,0){7}{\circle*{6}}
\multiput(23,20)(20,0){15}{\line(1,0){14}}

\put(60,23){\line(0,1){14}}
\put(60,40){\circle{6}}
\put(300,23){\line(0,1){14}}
\put(300,40){\circle*{6}}
\put( 14,7){$\b_{-1}$}
\put( 36,7){$\b_0$}
\put( 56,7){$\b_1$}
\put( 76,7){$\b_2$}
\put( 96,7){$\b_3$}
\put(116,7){$\b_4$}
\put(136,7){$\b_5$}
\put(156,7){$\b_6$}
\put(176,7){$\b_7$}
\put(196,7){$\b_{9}$}
\put(216,7){$\b_{10}$}
\put(236,7){$\b_{11}$}
\put(256,7){$\b_{12}$}
\put(276,7){$\b_{13}$}
\put(296,7){$\b_{14}$}
\put(316,7){$\b_{15}$}

\put(64,38){$\b_8$}
\put(304,38){$\b_{16}$}
\put(20,60){$DE_{18(10)}$}
\end{picture}
\caption{The Dynkin diagram of the real form $DE_{18(10)}$ of $DE_{18}.$ 
The black and white nodes determine the real form.}
\label{fig:DynkinRealDE18}
\end{figure}
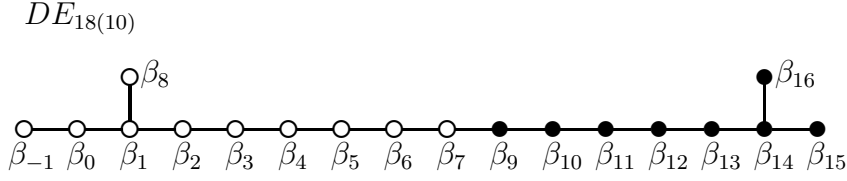
\vskip 10pt

It corresponds to a choice of anti-linear involution on the complexified
Lie algebra $DE_{18},$ given by its action on the Chevalley generators
that can be constructed as follows \cite{PP}
(see also \cite{Keurentjes:2002rc} for a comprehensive discussion).

First, we construct a linear involution $\omega^*$ on $\CarOf{DE_{18}}^*$
that preserves the Killing form $\kform{\cdot}{\cdot}.$
It is defined by its action on the simple roots
\bear
\omega^*(\b_i) &=& \b_i
\quad\text{for $i=-1,\dots 6$ and $i=8$}\nn\\
\omega^*(\b_7) &=& \b_7 + 2\bigr(\sum_{i=1}^6\b_{8+i}\bigl)+\b_{15}+\b_{16},\nn\\
\omega^*(\b_i) &=& -\b_i
\quad\text{for $i=9,\dots 16$}\nn
\eear
In general, $\omega^*$ is determined by the requirement that:
\begin{itemize}
\item
$\omega^*$ preserves the bilinear form $\kform{\cdot}{\cdot}$;
\item
$\omega^*(\b_i)=-\b_i$ for every ``black'' simple root $\b_i$;
\item
For every ``white'' simple root $\b_j,$
we require $\omega^*(\b_j)-\b_j$ to be a linear combination
of the ``black'' simple roots only.
\end{itemize}
Now that we have $\omega^*$ we can construct
an anti-linear involution $\omega$ on the Lie algebra $DE_{18}$ itself.
We pick a set of Chevalley generators $h_i, e_i, f_i$
($i=-1,\dots, 16$) such that $h_i\in\CarOf{DE_{18}},$
$e_i$ generates the root space of $\a_i,$
and $f_i$ generates the root space of $-\a_i,$
and \cite{KacBook},
$$
[e_i, f_i] = h_i,
\qquad
[h_i, e_i] = 2 e_i, 
\qquad
[h_i, f_i] = -2 f_i.
$$
Now note that $\omega^*(\b_i)$ is a real root for $i=-1,\dots,16.$
Its root space is therefore 1-dimensional, and 
a generator of this root space is uniquely determined,
up to a multiplicative constant.
We can now define the {\it anti-linear} involution $\omega$
in such a way that
$$
\omega(e_i)\in \AlgG(DE_{18})_{\omega^*(\b_i)},
$$
In particular, we can take $\omega(e_i)=-f_i$ and $\omega(f_i)=-e_i$
for $i=9,\dots,16.$
For $i=-1,\dots,6,8$ we can 
take $\omega(e_i)=e_i$ and $\omega(f_i)=f_i,$
and $\omega(e_7)$ has to be chosen as a generator
of the root space of $\omega^*(\b_7).$
This determines the action of $\omega$ on $\CarOf{DE_{18}},$
and the requirement that $[\omega(e_7),\omega(f_7)]=\omega(h_7)$
then determines the normalization of $\omega(f_7).$
The real Lie algebra $DE_{18(10)}$ is then defined as the invariant subalgebra
$$
DE_{18(10)} \defineas \{x\in DE_{18}\suchthat \omega(x)=x\}.
$$
Note that
$$
h_{-1},\dots,h_6, h_8, 
i h_9,\dots, i h_{16}
\in DE_{18(10)}.
$$

\subsection{The Subalgebras $DE_{10}$ and $\Algso(16)$}\label{subsec:subalgDE18}
\paragraph{}
We will now define two more subalgebras of $DE_{18(10)}.$
First note that the ``black'' nodes in the $DE_{18(10)}$ Dynkin diagram 
(\figref{fig:DynkinRealDE18}) define a subalgebra isomorphic to $\Algso(16)$
and the ``white'' nodes define a subalgebra isomorphic to $E_{10}.$
This is depicted in \figref{fig:DynkinDE10andD8}.
\vskip 10pt
\begin{figure}[h]
\begin{picture}(370,80)
%
%

\put(0,0){\begin{picture}(220,80)
\thicklines
\multiput(20,20)(20,0){9}{\circle{6}}
\multiput(23,20)(20,0){8}{\line(1,0){14}}

\put(60,23){\line(0,1){14}}
\put(60,40){\circle{6}}
\put( 14,7){$\b_{-1}$}
\put( 36,7){$\b_0$}
\put( 56,7){$\b_1$}
\put( 76,7){$\b_2$}
\put( 96,7){$\b_3$}
\put(116,7){$\b_4$}
\put(136,7){$\b_5$}
\put(156,7){$\b_6$}
\put(176,7){$\b_7$}

\put(64,38){$\b_8$}
\put(20,50){$E_{10}\subset DE_{18}$}
\end{picture}}

\put(20,0){\begin{picture}(220,80)
\thicklines
\multiput(200,20)(20,0){7}{\circle*{6}}
\multiput(203,20)(20,0){6}{\line(1,0){14}}
\put(300,23){\line(0,1){14}}
\put(300,40){\circle*{6}}
\put(196,7){$\b_{9}$}
\put(216,7){$\b_{10}$}
\put(236,7){$\b_{11}$}
\put(256,7){$\b_{12}$}
\put(276,7){$\b_{13}$}
\put(296,7){$\b_{14}$}
\put(316,7){$\b_{15}$}
\put(304,38){$\b_{16}$}
\put(220,50){$D_8\subset DE_{18(10)}$}
\end{picture}}

\end{picture}
\caption{We split the Dynkin diagram of
\figref{fig:DynkinRealDE18} into two sub-diagrams.
The Dynkin sub-diagrams are $E_{10}$ and $D_8.$}
\label{fig:DynkinDE10andD8}
\end{figure}
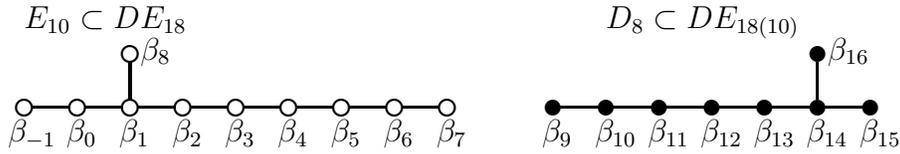
\vskip 10pt
Note that $E_{10}$ is only a subalgebra of the complexified $DE_{18},$
but $\Algso(16)\subset DE_{18(10)}.$
These two subalgebras intersect trivially, but they do not commute with
each other, because the node $\b_7$ of $E_{10}$ is connected to the
node $\b_9$ of $\Algso(16).$
We denote the commutant of $\Algso(16)$ by
$$
\Commutant = \{x\in DE_{18(10)}\suchthat 
   [x,y]=0\quad \forall y\in\Algso(16)\subset DE_{18(10)}\}.
$$

\begin{prop}\label{prop:Commutant}
$\Commutant$ contains a $DE_{10}$ subalgebra.
\end{prop}
\begin{proof}
We begin with the complexified algebra.
Recall the root-space decomposition 
$$
DE_{18} = \bigoplus_{\a\in\RtsOf{DE_{18}}} \AlgG(DE_{18})_\a.
$$
Note that the subspaces $\AlgG(DE_{18})_{\pm\a},$ 
for $\a=\b_{-1},\dots,\b_6,\b_8,$ commute with $\Algso(16).$
Now define
\belabel{eqn:DefYrt}
\Yrt = 
(\b_6+2\b_7)
+(2\b_9 + 2\b_{10}+2\b_{11}+2\b_{12}+2\b_{13}+2\b_{14}+\b_{15}+\b_{16}).
\ee
Note that the involution $\omega^*$ defined in \secref{subsec:RealDE18}
satisfies $\omega^*(\Yrt)=\Yrt.$ This implies
that the coroot $\Yrtv$ is in the real $DE_{18(10)}.$

Since $\Yrt$ is a root of a finite $D_{10}$ subalgebra
(corresponding to the Dynkin
sub-diagram that includes only the nodes 
$\b_6,\b_7,\b_9,\cdots,\b_{16}$),
$\dim \AlgG(DE_{18})_{\pm \Yrt} = 1.$
It is not hard to see that the elements of $\AlgG(DE_{18})_{\pm \Yrt}$
commute with $\Algso(16)$
[they can be embedded in $\Algso(20)\supset \Algso(16)$].
Consider the set of positive real roots $\b_{-1},\dots,\b_6, \b_8, \Yrt.$
These roots all square to $2,$ and their intersection matrix is
encoded in the Dynkin diagram of \figref{fig:DynkinAlgGpr}.
This is again the Dynkin diagram of $DE_{10}.$
The rest of the proof is technical and can be found in \appref{app:ProofDE18}.
\end{proof}

\vskip 10pt
\begin{figure}[h]
\begin{picture}(370,80)
%
%
\thicklines
\multiput(20,20)(20,0){8}{\circle{6}}
\multiput(23,20)(20,0){7}{\line(1,0){14}}

\put(60,23){\line(0,1){14}}
\put(60,40){\circle{6}}
\put(140,23){\line(0,1){14}}
\put(140,40){\circle{6}}
\put( 14,7){$\b_{-1}$}
\put( 36,7){$\b_0$}
\put( 56,7){$\b_1$}
\put( 76,7){$\b_2$}
\put( 96,7){$\b_3$}
\put(116,7){$\b_4$}
\put(136,7){$\b_5$}
\put(156,7){$\b_6$}

\put(64,38){$\b_8$}
\put(144,38){$\Yrt$}
\put(20,60){$\AlgG'\simeq DE_{10}$}
\end{picture}
\caption{The Dynkin diagram of the subalgebra $\AlgG'\subset DE_{18}.$}
\label{fig:DynkinAlgGpr}
\end{figure}
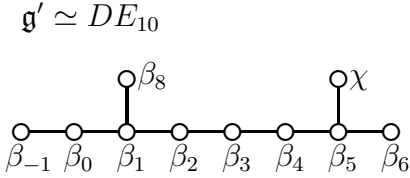
\vskip 10pt

By now, we have defined three different algebras that contain $DE_{10}$:
the $\Z_2$-invariant subalgebra $\AlgInv\subset E_{10},$
the $\Algso(16)$-commutant subalgebra $\Commutant\subset DE_{18},$
and the abstract $DE_{10}$ algebra.
We need a way to relate them to each other.
The embedding of $DE_{10}$ in $E_{10}$ sends the Cartan subalgebra
$\CarOf{DE_{10}}$ to the Cartan subalgebra $\CarOf{E_{10}}.$
It thus induces a linear isomorphism between the Cartan subalgebra $\CarOf{DE_{10}}$
and $\CarOf{E_{10}}.$ 
Similarly, the embedding of $DE_{10}$ in $DE_{18}$
induces a linear map from $\CarOf{DE_{10}}$ to $\CarOf{DE_{18}}.$
Since these maps are going to be important later on, 
we will now describe them in more detail.

Comparing the Dynkin diagram of \figref{fig:DynkinDE10}
to that of \figref{fig:DynkinAlgGpr}, we define a map $\fOH$
from $\CarOf{DE_{10}}^*$ to 
$\CarOf{DE_{18}}^*$ that sends the simple roots of 
$DE_{10}$ to roots of $DE_{18}$ and preserves their inner products.
It acts on the generators as follows:
\bear
&&\fOH(\g_{-1}) = \b_{-1},\quad
\fOH(\g_0) = \b_0,\quad
\fOH(\g_1) = \b_1,\quad
\fOH(\g_2) = \b_2,\quad
\fOH(\g_3) = \b_3,
\nn\\
&&\fOH(\g_4) = \b_4,\quad
\fOH(\g_5) = \b_5,\quad
\fOH(\g_6) = \b_6,\quad
\fOH(\g_7) = \b_8,\quad
\fOH(\g_8) = \Yrt.
\label{eqn:DeffOH}
\eear
$\fOH$ is the unique injective map that is compatible with
the bilinear forms $\kform{\cdot}{\cdot}$ on $DE_{10}$ and $DE_{18}$
and satisfies
$$
0=\kform{\fOH(\a)}{\b},\qquad
\text{for
$\a\in\CarOf{DE_{10}}^*$ and 
$\b\in\sum_{j=9}^{16}\C\b_j\equiv\CarOf{D_{16}}^*$}.
$$
Now we will define a map from $\CarOf{DE_{18}}^*$ to 
$\CarOf{DE_{10}}^*.$
Given $\b\in\CarOf{DE_{18}}^*$ we first take its projection
to the image of $\fOH,$ and, using $\fOH^{-1},$ we identify this projection
with an element of $\CarOf{DE_{10}}^*.$
Thus, we define a linear map $\pHO':\CarOf{DE_{18}}^*\rightarrow\CarOf{DE_{10}}^*,$
that satisfies
$$
\pHO'\circ\fOH=\Id,\qquad 
\pHO'(\b)=0\quad
\text{for
$\b\in\sum_{j=9}^{16}\C\b_j\equiv\CarOf{D_{16}}^*$}.
$$
Explicitly, $\pHO'$ can be defined by
\begin{align}
&\pHO'(\b_{-1})=\g_{-1},\quad
\pHO'(\b_0)=\g_0,\quad
\pHO'(\b_1)=\g_1,\quad
\pHO'(\b_2)=\g_2,\quad
\pHO'(\b_3)=\g_3,
\nn\\
&\pHO'(\b_4)=\g_4,\quad
\pHO'(\b_5)=\g_5,\quad
\pHO'(\b_6)=\g_6,\quad
\pHO'(\Yrt)=\g_8,\quad
\pHO'(\b_8)=\g_7,
\nn\\
&\pHO'(\b_9)=\pHO'(\b_{10})=\cdots=\pHO'(\b_{16})=0.
\label{eqn:ExplicitOmegaPr}
\end{align}

Finally, we need a map from $\CarOf{DE_{18}}^*$ back to $\CarOf{E_{10}}^*.$
Using \eqref{eqn:DeffDEE}, we define
$\pHO:\CarOf{DE_{18}}^*\rightarrow\CarOf{E_{10}}^*$ by
$$
\pHO\defineas \fDEE\circ\pHO'.
$$
Explicitly, $\pHO$ can be defined by
\bear
&&\pHO(\b_{-1})=\a_{-1},\quad
\pHO(\b_0)=\a_0,\quad
\pHO(\b_1)=\a_1,\quad
\pHO(\b_2)=\Xrt,\quad
\pHO(\b_3)=\a_7,
\nn\\
&&\pHO(\b_4)=\a_6,\quad
\pHO(\b_5)=\a_5,\quad
\pHO(\b_6)=\a_4,\quad
\pHO(\Yrt)=\a_8,\quad
\pHO(\b_8)=\a_2,
\nn\\
&&\pHO(\b_9)=\pHO(\b_{10})=\cdots=\pHO(\b_{16})=0.
\label{eqn:ExplicitOmega}
\eear
Using \eqref{eqn:ExplicitOmega} and 
\eqref{eqn:DefYrt}, we get
\belabel{eqn:pHObseven}
\pHO(\b_7) = \frac{1}{2}(\pHO(\Yrt)-\pHO(\b_6))
=\frac{1}{2}(\a_8-\a_4)
\ee
Given a root $\b\in\RtsOf{DE_{18}}\subset\CarOf{DE_{18}}^*$
we define its {\it physical action} by,
\belabel{eqn:PhysAcb}
\Action(\b)\defineas 2\pi e^{\inner{\pHO(\b)}{\vh}},
\ee
where $\vh\in\CarOf{E_{10}}$ is a linear function of the $\log$'s of 
the compactification radii, as defined in \eqref{eqn:vhRadii}.
We will also denote by 
\belabel{eqn:eSODE}
\eSODE:\CarOf{D_8}^*\rightarrow\CarOf{DE_{18}}^*
\ee
the linear map that sends the simple roots $\b_9,\dots,\b_{16}$
of $D_8$ to the simple roots $\b_9,\dots,\b_{16}$ of $DE_{18}.$

The linear maps that we have defined so far are collected in the following
line
$$
\CarOf{D_8}^*\xrightarrow{\eSODE}\CarOf{DE_{18}}^*
\xrightarrow{\pHO'}\CarOf{DE_{10}}^*
\xrightarrow{\fDEE}\CarOf{E_{10}}^*
\xrightarrow{\fDEE^{-1}}
\CarOf{DE_{10}}^*\xrightarrow{\fOH}
\CarOf{DE_{18}}^*
\qquad
\pHO= \fDEE\circ\pHO'.
$$
(Note that this is not an exact sequence!)

We note the analogous statements for M-theory on $T^8.$
If we replace M-theory on $T^{10}$
with M-theory on $T^8,$ we have to replace 
$DE_{18(10)}$ with $\Algso(24,8),$
and $DE_{10}$ with $\Algso(8,8).$
The $E_{10}\subset DE_{18}$ would be analogous to $A_8=sl(9,\R),$
which is the subalgebra corresponding to the ``white'' nodes of the
Dynkin diagram of $D_{16(8)}$.


It is interesting to compare $\AlgInv$ to $\Commutant,$
since both contain a $DE_{10}$ subalgebra,
and both have $\CarOf{DE_{10}}$ as a maximal abelian subalgebra.
Both $\AlgInv$ and $\Commutant$ have a root-space decomposition
with respect to $\CarOf{DE_{10}}$ and the root-spaces of both algebras
are identical to $\RtsOf{DE_{10}}.$

It turns out that $\Commutant$ is strictly bigger than $DE_{10},$
and this can be seen already at level-$1,$
namely, some multiplicities of level-$1$ roots of $\AlgInv$ are smaller
than the corresponding $\Commutant$ multiplicities.
The question of whether $\Commutant$ contains $\AlgInv$ or not will be studied
in another paper.

\subsection{The Maximal Compact Subalgebra $\AlgKomOf{DE_{18(10)}}$}
\paragraph{}
We need to define a ``maximal compact subalgebra''
 $\AlgKomOf{DE_{18(10)}}\subset DE_{18(10)}.$
[If we replace $T^{10}$ with $T^8,$ for comparison, $\AlgKom$ will become
$\Algso(24)\times \Algso(8).$]
To define the maximal compact subalgebra, we first
define the {\it compact involution} $\omega_c$ on $DE_{18(10)}$ as
follows.
We define it on the Chevalley generators (see \secref{subsec:RealDE18})
by
$$
\omega_c(h_i) = -h_i,\quad
\omega_c(e_i) = f_i,\quad
\omega_c(f_i) = e_i,
\qquad
\text{for $i=-1\dots 16.$}
$$
We then define
$$
\AlgKomOf{DE_{18(10)}} = \{x\in DE_{18(10)}\suchthat \omega_c(x)=x\}.
$$
Note that 
$$
i h_j, e_j + f_j, i(e_j - f_j)
\in \AlgKomOf{DE_{18(10)}}
\qquad 
\text{for $j=9,\dots,16.$}
$$
These generators generate a real orthogonal subalgebra 
$\Algso(16)\subset \AlgKomOf{DE_{18(10)}}.$
In general, little is known about such ``maximal compact'' subalgebras
of infinite dimensional Kac-Moody algebras
(but see \cite{Nicolai:2004nv} for a recent study of $\AlgKomOf{E_9}$).
The physical degrees of freedom of M-theory on $T^{10}/\Z_2$
are presumed to be related to the coset
of Lie algebras \textit{$DE_{18(10)}/\AlgKomOf{DE_{18(10)}}$}.

\section{The Twisted Sector}\label{sec:Twisted}
\paragraph{}
We are now ready to discuss the twisted sector of the orbifold.
As in \secref{sec:Untwisted}, we restrict attention to the orbifold
$T^5\times (T^5/\Z_2)$ from \secref{subsec:T5Z2}.
The twisted sector consists of $16$ M5-branes which
support fluxes and particle charges.
Boundaries of M2-branes attached to the M5-branes are the origin
for these charges.
Experience with $E_{10}$ suggests that fluxes are related to
real roots and particle charges are related to imaginary
(isotropic) roots of the Kac-Moody  algebra.
We would like to extend this type of relation to the twisted sectors.

We have seen in \secref{subsec:subalgDE18}
that $DE_{18}$ has a $D_8\simeq\Algso(16)$ subalgebra.
This number $16$ that appears in $\Algso(16)$ is directly related
to the 16 M5-branes as follows.
The Lie algebra $DE_{18}$ can be decomposed into (an infinite number of)
irreducible representations of $\Algso(16)\subset DE_{18}.$
We will propose below that the twisted sector fluxes and charges
can be matched with such irreducible $\Algso(16)$ representations.

\subsection{$\Algso(16)$-packs}
\paragraph{}
In \secref{subsec:subalgDE18} we identified a 
$DE_{10}\subset DE_{18}$ subalgebra.
The Cartan subalgebra $\CarOf{DE_{10}}$ is identified with 
$$
\C\bv_{-1}+\C\bv_0+\C\bv_1+\C\bv_2+\C\bv_3+\C\bv_4+\C\bv_5+\C\bv_6+\C\bv_8
+\C\Yrtv
$$
[where $\bv_i$ ($i=-1\dots 16$) are the simple coroots, and $\Yrtv$ 
is the coroot dual to $\Yrt$].
It commutes with $\Algso(16)$ and is isomorphic to $\C^{10}.$
We can therefore decompose $DE_{18}$ under its $\CarOf{DE_{10}}\oplus\Algso(16)$
subalgebra.
The irreducible representations of $\CarOf{DE_{10}}\oplus\Algso(16)$
that appear in this decomposition will be called $\Algso(16)$-{\it packs}.

It is slightly more convenient to work with $\CarOf{E_{10}}$ rather
than $\CarOf{DE_{10}}.$ 
We switch to $\CarOf{E_{10}}$ using the map
$\fDEE:\CarOf{DE_{10}}^*\rightarrow\CarOf{E_{10}}^*$ defined in \eqref{eqn:DeffDEE}.
We now have the decomposition
\belabel{eqn:DErootpackDecom}
DE_{18}=\bigoplus_\rtlam\rtAlgG_\rtlam,\qquad
\text{where $\rtlam$ labels irreps of $\Algso(16)\oplus\pHO^*(\CarOf{E_{10}}).$}
\ee
Here $\pHO$ is the map from \eqref{eqn:ExplicitOmega}
that allows us to translate an $E_{10}$ root to a $DE_{18}$ root.

An irreducible representation of $\pHO^*(\CarOf{E_{10}})\simeq\CarOf{E_{10}}$
is 1-dimensional and is labeled by an element $\lam\in\CarOf{E_{10}}^*.$
Moreover, if $\b\in\RtsOf{DE_{18}}$ then
$\pHO'(\b)\in\WgtLatOf{DE_{10}},$
where $\WgtLatOf{DE_{10}}$ was defined in \eqref{eqn:DefWLDE},
and $\pHO'$ is the map from \eqref{eqn:ExplicitOmegaPr}.
The irreducible representations of $\Algso(16)\oplus\pHO^*(\CarOf{E_{10}})$
that appear in \eqref{eqn:DErootpackDecom}
are therefore labeled by a pair
$$
\rtlam = (\lam,\rep{r}),\qquad
\text{where
$\lam\in \fDEE(\WgtLatOf{DE_{10}}),$
and $\rep{r}$ is an irrep of $\Algso(16)$}.
$$
We call such a pair an $\Algso(16)$-{\it pack}.
The weight $\lam$ will eventually describe the dependence
of the physical object associated with $\rtlam$ on the radii $R_1,\dots, R_{10},$
and $\rep{r}$ will determine the object's relation to the $16$ M5-branes.

The representations $\rep{r}$ are all finite.
They are completely determined by their lowest (or highest) $D_8$ weight 
$\wwLambda\in\WgtLatOf{D_8}.$
Let $x\in\rtAlgG_\rtlam\subset DE_{18}$
be a vector with $D_8$ lowest-weight $\wwLambda.$
Then $x$ belongs to a single root-space $(DE_{18})_\b,$
and the root $\b$ of $DE_{18}$ can be calculated as follows:
$$
\b = \fOH(\lam)+\eSODE(\wwLambda).
$$
[See \eqref{eqn:DeffOH} and \eqref{eqn:eSODE} for
the definitions of the linear maps $\fOH$ and $\eSODE.$]
Alternatively, if the $DE_{18}$-root $\b$ is given, then
the $\Algso(16)$-representation of which $x$ is a lowest-weight vector
can be found
by calculating the lowest weight $\wwLambda$:
\belabel{eqn:LambfOH}
\eSODE(\wwLambda) = \b-\fOH(\pHO'(\b)).
\ee
For $\wwLambda$ to be a lowest-weight vector, it must satisfy
$$
0 \ge \kform{\eSODE(\wwLambda)}{\b_j},\qquad j=9,\dots,16.
$$
We denote the fundamental weights of $D_8$ by
$\wwLambda_9,\dots,\wwLambda_{16}\in\WgtLatOf{D_8}.$
This notation is compatible with the simple roots of $D_8$
that we take to be $\b_9,\dots,\b_{16}.$

Now take $\g$ to be a $DE_{18}$-root with support only on the white nodes of
\figref{fig:DynkinRealDE18},
$$
\g = \sum_{i=-1}^8 k_i\b_i.
$$
Take an element $x\in\AlgG(DE_{18})_\g$ and complete it to an $\Algso(16)$
representation by acting on it with all possible combinations
of the Chevalley generators 
$e_9, f_9, h_9,\dots, e_{16}, f_{16}, h_{16}$
(that generate $\Algso(16)\subset DE_{18}$).
This is denoted by $U(\Algso(16))x,$ where $U(\Algso(16))$ is
the universal enveloping algebra of $\Algso(16).$
It is easy to check that $\kform{\g}{\b_j}=\kform{-k_7\wwLambda_9}{\b_j}$
for all the simple roots $\b_j$ ($j=9,\dots,16$) of $\Algso(16).$
Without loss of generality, suppose $k_7>0.$
Then $[f_j,\g]=0$ for $j=9,\dots,16$ and therefore $U(\Algso(16))x$
is a lowest-weight representation with lowest-weight $-k_7\wwLambda_9.$

\begin{prop}\label{prop:EasyRootPacks}
Let \textit{$\g = \sum_{i=-1}^8 k_i\b_i$} be a root of $DE_{18}$ and
let $0\neq x\in\AlgG_\g.$
Then $U(\Algso(16))x\simeq L_{\Algso(16)}(\abs{k_7}\wwLambda_9)$
where $L_{\Algso(16)}(\wwLambda)$
is the irreducible $\Algso(16)$-representation with highest-weight $\wwLambda.$
In particular,
\begin{description}

\item[a.]
If $k_7=0,$ then $U(\Algso(16))x$ is equivalent to the $1$-dimensional singlet
representation~$\rep{1}.$
\item[b.]
If $k_7=\pm 1,$ then $U(\Algso(16))x$ 
is equivalent to the $16$-dimensional vector representation~$\rep{16}.$
\item[c.]
If $k_7=\pm 2,$ then $U(\Algso(16))x$ is equivalent 
to the traceless symmetric tensors~$\rep{135}.$
\end{description}
\end{prop}
The proof can be found in \appref{app:ProofProp}.

Of course, not all $\Algso(16)$-packs are of the form described in
\propref{prop:EasyRootPacks}. The support of the root $\Yrt$ 
[from \eqref{eqn:DefYrt}], for example,
is not restricted to the ``white'' nodes, and it is a highest (lowest)
weight of a singlet $\Algso(16)$-representation $\rep{1}.$

\subsection{Examples of $\Algso(16)$-packs}
\paragraph{}
We will now give a few examples of $\Algso(16)$-packs.
Let $\rtlam=(\lam,\rep{r})$ be an $\Algso(16)$-pack.
It is more convenient to present $\fDEE(\lam)$ instead of $\lam,$
and it is convenient to write 
$\fDEE(\lam)\in\CarOf{E_{10}}^*$ in the basis
\eqref{eqn:AlphaExplicit} [$\fDEE$ was defined in \eqref{eqn:DeffDEE}].
For small representations $\rep{r},$ it is convenient to present
the dimension $\dim\rep{r}.$
Thus, we will present an $\Algso(16)$-pack as 
\belabel{eqn:PhysNotationRtlets}
\rtlam = (\tn_1,\dots,\tn_{10};\dim\rep{r}),
\qquad
\tn_j\in\tfrac{1}{2}\Z,\quad
(j=1,\dots, 10).
\ee
Note that from \eqref{eqn:annn} and 
\eqref{eqn:ExplicitOmega}-\eqref{eqn:pHObseven}
it follows that $\tn_j$ ($j=1\dots 10$)
could be integers or half-integers.

Using \propref{prop:EasyRootPacks},
we can make a list of some real $DE_{18}$ roots that are 
$\Algso(16)$-lowest-weights. 
Using
\eqref{eqn:ExplicitOmega}-\eqref{eqn:pHObseven}
and \eqref{eqn:AlphaExplicit}, we can translate
the roots to the physical notation \eqref{eqn:PhysNotationRtlets}.
We get,
\begin{align}
\b_7
  &\rightarrow (0,0,0,0,0,
     -\half,\half,\half,\half,\half;\rep{16}),\nn\\
\sum_{j=6}^7\b_j
  &\rightarrow (0,0,0,0,0,
     \half,-\half,\half,\half,\half;\rep{16}),\nn\\
\sum_{j=5}^7\b_j
  &\rightarrow (0,0,0,0,0,
     \half,\half,-\half,\half,\half;\rep{16}),\nn\\
\sum_{j=4}^7\b_j
  &\rightarrow (0,0,0,0,0,
     \half,\half,\half,-\half,\half;\rep{16}),\nn\\
\sum_{j=3}^7\b_j
  &\rightarrow (0,0,0,0,0,
     \half,\half,\half,\half,-\half;\rep{16}),\nn\\
 & \label{eqn:RealRootpacksI}
\end{align}
In general,
it is obvious that given an $\Algso(16)$-pack in the form
$$
(n_1, n_2, n_3, n_4, n_5, n_6, n_7, n_8, n_9, n_{10}; \rep{r}),
$$
we can generate more $\Algso(16)$-packs by reshuffling
$n_1,\dots, n_5$ and reshuffling $n_6,\dots, n_{10}.$
That is, for every pair of permutations $(\sigma,\sigma')\in S_5\times S_5,$
$$
(
n_{\sigma(1)}, 
n_{\sigma(2)}, 
n_{\sigma(3)}, 
n_{\sigma(4)}, 
n_{\sigma(5)}, 
n_{5+\sigma'(1)}, 
n_{5+\sigma'(2)}, 
n_{5+\sigma'(3)}, 
n_{5+\sigma'(4)}, 
n_{5+\sigma'(5)}; \rep{r}),
$$
is also an $\Algso(16)$-pack.
Physically, $S_5\times S_5,$
which acts by permuting the radii,
is clearly a symmetry of the problem.
Mathematically, $S_5\times S_5$ is a subgroup of the Weyl group of $DE_{18}.$
It is therefore sufficient -- as we will do from now on --
to list only one $\Algso(16)$-pack
from each $S_5\times S_5$ multiplet.

The next \textit{$\binom{5}{2}$} $\Algso(16)$-packs in our examples are 
$S_5\times S_5$ permutations of
\belabel{eqn:RealRootpacksII}
\sum_{j=2}^7\b_j
\rightarrow (0,0,0,1,1,
     \half,\half,\half,\half,\half;\rep{16}).
\ee
Finally, we will need the $5\times 5$ 
permutations of
\belabel{eqn:RealRootpacksIII}
\b_0+2\b_1+2\b_2+\b_3+\b_4+\b_5+\b_6+\b_7+\b_8
\rightarrow
(0,1,1,1,1,\tfrac{1}{2},\tfrac{1}{2},\tfrac{1}{2},\tfrac{1}{2},\tfrac{3}{2};\rep{16})
\ee

\section{Physical Interpretation}\label{sec:PhysicalInt}
\paragraph{}

We will now propose a physical interpretation for $\Algso(16)$-packs.
The vector $\lam=(\tn_1,\dots,\tn_{10})$ that appears in
\eqref{eqn:PhysNotationRtlets}
is an element of $\CarOf{E_{10}}^*$ and it is therefore natural
to examine the inner product $\inner{\lam}{\vh},$
where $\vh$ is the vector of $\log$s of radii, defined
in \eqref{eqn:vhRadii}.
$$
\Action(\lam)\defineas
2\pi e^{\inner{\lam}{\vh}} = 2\pi\prod_{i=1}^{10} (M_p R_i)^{\tn_i},
$$
can be interpreted as an ``action.''
This definition is consistent with \eqref{eqn:PhysAcb}
with $\lam=\pHO(\b)$ for any root $\b$ of the $\Algso(16)$-pack.

The $\Algso(16)$-representation $\rep{r}$
that appears in \eqref{eqn:PhysNotationRtlets}
can tell us the origin of the physical object in relation to the $16$
M5-branes.
For example, $\rep{r}=\rep{16}$ suggests that the object is
related to a single M5-brane.
Below we will study the physical interpretation of $\Algso(16)$-packs
in several examples.
We will begin with $\Algso(16)$-packs that can be interpreted
as fluxes, and we will end with more complicated examples that
can be interpreted as ``charges.''

\subsection{Fluxes of the Twisted Sector}\label{subsec:fluxes}
\paragraph{}
The twisted-sector provides more varieties of fluxes.
Consider, for concreteness, the orbifold $T^5\times (T^5/\Z_2)$
from \secref{subsec:T5Z2}.
Let the radii of the $T^5$ factor be $R_1,\dots,R_5$ and the radii
of the $(T^5/\Z_2)$ factor be $R_6,\dots,R_{10}.$
The twisted sector consists of 16 M5-branes \cite{Witten:1995em}, and
each M5-brane carries a 5+1D tensor multiplet of $N=(2,0)$ supersymmetry.
The bosonic fields of the tensor multiplet are an anti-self-dual tensor field
and 5 scalar fields.
The 5 scalar fields can be interpreted as the coordinates of the M5-brane
in the $T^5/\Z_2$ directions.

We will assume that we are in the asymptotic region of ``moduli-space,''
\eqref{eqn:Asymptotic}.
In this region, the M5-branes are very heavy,
and a nonrelativistic limit is a good approximation in many cases.
For example,
with $k$ units of momentum in, say, the $6^{th}$ direction,
an M5-brane has a kinetic energy which, in the nonrelativistic limit
\eqref{eqn:Asymptotic}, is 
\belabel{eqn:Ewrtt}
\frac{1}{2 M_p^6 R_1\cdots R_5}\left(\frac{k}{R_6}\right)^2.
\ee
We need to measure energy with respect to conformal time, which means
an extra factor of $2\pi M_p^9 R_1\cdots R_{10}.$
Multiplying \eqref{eqn:Ewrtt} by this factor, we get
\belabel{eqn:TensorFluxI}
\Phi =\pi M_p^3 k^2\frac{R_7 R_8 R_9 R_{10}}{R_6}.
\ee
Similarly, we get four more fluxes corresponding to M5-brane momenta
in the $7^{th},\dots,10^{th}$ directions.

The anti-self-dual 3-form field strength $H_{IJK}$ ($I=0,\dots,5$)
of the tensor multiplet will give us 10 more possible fluxes,
with energies proportional to
\belabel{eqn:TensorFluxII}
\pi M_p^9 R_1^2 R_2^2 R_6 R_7 R_8 R_9 R_{10},\dots,
\pi M_p^9 R_4^2 R_5^2 R_6 R_7 R_8 R_9 R_{10},
\ee
in units of conformal time.
We will now relate these fluxes to $\Algso(16)$-packs.

\subsection{$\Algso(16)$-packs for Fluxes}
\paragraph{}
Recall that real roots of $E_{10}$ are related to fluxes,
and the squared action $\exp(2\inner{\a}{\vh})$ of a real root $\a$
is the energy stored in the flux, measured in units conjugate to conformal time
\cite{Damour:2000hv}\cite{Damour:2002et}
\cite{Pioline:2002qz}\cite{Brown:2004jb}.
In other words,
$$
E \propto N^2\frac{e^{2\inner{\a}{\vh}}}{M_p^9 \prod_{j=1}^{10}R_j}
$$
is the energy stored in the flux, where $N\in\Z$ is the number of flux units.
We will now extend this formula to the twisted-sectors.

The $\Algso(16)$-packs from \eqref{eqn:RealRootpacksI}-\eqref{eqn:RealRootpacksII}
should be associated with the fluxes 
\eqref{eqn:TensorFluxI}-\eqref{eqn:TensorFluxII}.
Using \eqref{eqn:pHObseven}, the flux energy \eqref{eqn:TensorFluxI}
can be written as
\textit{$\pi k^2 e^{2\inner{\pHO(\b_7)}{\vh}}$}.
We therefore associate this flux to $\b_7,$ which is the 
lowest-weight of the first $\Algso(16)$-pack
in \eqref{eqn:RealRootpacksI}.
Similarly, all the other $\Algso(16)$-packs from \eqref{eqn:RealRootpacksI} 
correspond to fluxes of the scalar components of the $16$ tensor multiplets.

The roots from \eqref{eqn:RealRootpacksII} match the fluxes 
of the anti-self-dual tensor field \eqref{eqn:TensorFluxII}.
For example,
$$
N^2 M_p^9 R_4^2 R_5^2 R_6 R_7 R_8 R_9 R_{10}
=
N^2 e^{2\inner{\pHO(\sum_{j=2}^7\b_j)}{\vh}}.
$$

Now consider the $5\times 5$ $\Algso(16)$-packs 
which are $S_5\times S_5$ permutations of \eqref{eqn:RealRootpacksIII}.
The $\Algso(16)$-pack from \eqref{eqn:RealRootpacksIII} has energy
\belabel{eqn:Fluxpxphi}
E= \frac{1}{2} N^2 \frac{M_p^6 R_2 R_3 R_4 R_5 R_{10}^2}{R_1}
\ee
This flux has the following interpretation.

Let $X^I$ ($I=5,\dots,10$) be the $5$ scalars of the tensor-multiplet
on one of the 16 exceptional M5-branes. The indices $I=5,\dots,10$ are chosen
to match the directions of the $T^5/\Z_2.$
Now consider a constant flux $\px{1}X^{10}.$
Geometrically, this can be described as an exceptional M5-brane
that wraps the diagonal of the  torus in the $1^{st}$ and $10^{th}$ directions.
Note that the M5-brane is a point in the $6^{th},\dots, 9^{th}$ directions.
The moduli space for $X^I$ is actually $T^5/\Z_2,$ but
for fixed $X^6,\dots,X^9,$ away from the fixed points of the $\Z_2$-action,
$T^5/\Z_2$ looks like an $S^1$ fibration (corresponding to, say, the $10^{th}$
direction) over $T^4/\Z_2$ (corresponding to the $6^{th},\dots, 9^{th}$ directions).
The fibration does not preserve the orientation of the fiber, and therefore
winding number (of the exceptional M5-brane) in the $10^{th}$ direction is 
not conserved.
This is consistent with the fact that we projected out the imaginary roots
$$
(1,2,2,2,2,1,1,1,1,2;\rep{1}),
$$
which would correspond to an M5-brane wrapping the
$2^{nd}, 3^{rd}, 4^{th}, 5^{th},$ and  $10^{th}$ directions.
However, the constant flux $\px{1}X^{10}$ is still allowed.
The nonconservation of the winding number implies that transitions 
between zero and nonzero $\px{1}X^{10}$ flux are possible.

The energy of a wrapped brane is 
$$
E = M_p^6 R_2 R_3 R_4 R_5 \sqrt{R_1^2 + N^2 R_{10}^2}
$$
In the asymptotic region \eqref{eqn:Asymptotic},
the excess energy stored in the $\px{1}X^{10}$ flux can be written as
\eqref{eqn:Fluxpxphi}, approximately.

The $\Algso(16)$-packs \eqref{eqn:RealRootpacksI}-\eqref{eqn:RealRootpacksIII},
which we discussed so far, correspond to real roots of $DE_{18}.$
Take $\b_7,$ for example. It is a lowest weight of an $\Algso(16)$-pack
($\Algso(16)$ irreducible representation) that can be decomposed into 
$DE_{18}$ weight spaces. 
The $\Algso(16)$-pack is isomorphic to the fundamental
representation $\rep{16}$, and, as is easy to check, all its $DE_{18}$-weights can
be generated by acting on $\b_7$ with simple reflections around
$\b_9,\dots,\b_{16}.$
Thus, all the $DE_{18}$-weights corresponding to 
this $\Algso(16)$-pack square to $2$ and are therefore real roots of $DE_{18}.$
The same statement is true for all the other 
$\Algso(16)$-packs from \eqref{eqn:RealRootpacksI}-\eqref{eqn:RealRootpacksIII}.
Next, we will discuss $\Algso(16)$-packs that contain imaginary roots.

\subsection{Imaginary Roots}\label{subsec:imroots}
\paragraph{}
An isotropic root $\a$ is an imaginary root that squares to zero.
We are interested in prime isotropic roots, i.e. roots that are not 
a nontrivial integer multiple of another root.
Such roots were found in \cite{Brown:2004jb} to correspond
to brane charges.

According to Proposition 5.7 of \cite{KacBook}, 
for any simply-laced Kac-Moody algebra, a root $\a$ is isotropic if
and only if it is $\Weyl$-equivalent to an imaginary root $\b$
such that $\supp\b$ (the set of nodes of the Dynkin diagram
whose corresponding simple roots have a nonzero
coefficient in $\b$) is a subdiagram of affine type.

We will now apply this to $DE_{18}.$
There are two distinct subdiagrams of affine type.
One subdiagram is $E_9$ and the other is $\whD_{16}.$
$\whD_{16}$ is obtained by dropping the node $\b_{-1}$
(see \figref{fig:DynkinDE18AndD16}),
and
$E_9$ is obtained by dropping nodes $\b_7,\b_9,\dots,\b_{16}.$
The positive prime isotropic root with support in $E_9$ is given by
\belabel{eqn:delta1}
\delta_1 = 
2\b_{-1}
+4\b_0
+6\b_1
+5\b_2
+4\b_3
+3\b_4
+2\b_5
+\b_6
+3\b_8.
\ee
Its multiplicity is $\mult\delta_1=8,$ as in $E_9.$
The positive prime isotropic root with support in $\whD_{16}$ is given by
\belabel{eqn:delta2}
\delta_2 = 
\b_0+\b_8+\b_{15}+\b_{16}+2\sum_{i=1}^7\b_i+2\sum_{i=9}^{14}\b_i.
\ee
Its multiplicity is $\mult\delta_2 = 16,$ as in $\whD_{16}.$
It follows that $\delta_1$ and $\delta_2$ are not $\WeylOf{DE_{18}}$-equivalent.
Note that
\belabel{eqn:pHOdelta1}
\pHO(\delta_1) = 
2\a_{-1}
+4\a_0
+6\a_1
+8\a_2
+10\a_3
+11\a_4
+12\a_5
+8\a_6
+4\a_7
+5\a_8
=\OrbRt_{\textit{M5}}
\ee
where we used \eqref{eqn:OrbRtT5Z2}.
Thus, $\delta_1$ corresponds to the charge of an M5-brane
from the twisted sector.
But note that $\delta_1$ is a $\Z_2$-even root
and belongs to the untwisted sector.
(There is of course no contradiction, since $\delta_1$ would correspond to
an extra M5-brane parallel to the already existing $16$ M5-branes.) 
Note also that
\belabel{eqn:pHOdelta2}
\pHO(\delta_2) = 
\a_0+2\a_1+3\a_2+4\a_3+5\a_4+6\a_5+4\a_6+2\a_7+3\a_8
=\delta
\ee
where $\delta$ was defined in \eqref{eqn:deltaE10}
and is identified with the charge of a Kaluza-Klein particle
in the $1^{st}$ direction. It is also a member of the untwisted sector.

In \cite{Brown:2004jb}, a strategy to understand the meaning of 
imaginary roots $\g$ was to express them as a sum of real roots,
$\g=\a+\b,$ and check what happens when both fluxes associated with
$\a$ and $\b$ are present.
We will now employ a similar strategy for imaginary $\Algso(16)$-packs.

The analogs of $\a, \b$ are two real $\Algso(16)$-packs
$$
\rtlam = (\lam,\rep{r}),
\qquad
\rtlam' = (\lam',\rep{r}').
$$
The analog of $\g$ is an $\Algso(16)$-pack
\belabel{eqn:lamsplit}
\rtlam'' = (\lam+\lam';\rep{r}''),
\ee
such that the $\Algso(16)$-representation $\rep{r}''$
appears in the decomposition of the tensor product $\rep{r}\otimes\rep{r}'.$
If $\rtlam$ and $\rtlam'$ can both be associated with fluxes,
$\rtlam''$ should be associated with an anomalous charge
that is present when the two fluxes overlap.
We will now demonstrate these ideas in explicit examples.

\subsubsection*{Wess-Zumino Interactions}
\paragraph{}
Take
\bear
\lefteqn{
\rtlam'' =
(2,1,1,1,1,\tfrac{3}{2},\tfrac{3}{2},\tfrac{3}{2},\tfrac{3}{2},\tfrac{3}{2};\rep{16}),
}\nn\\
&&
(\text{with lowest-weight 
$\g=2\b_{-1}+3\b_0+4\b_1+3\b_2+3\b_3+3\b_4+3\b_5+2\b_6+\b_7+2\b_8+\Yrt$}),
\nn
\eear
and split it, as in \eqref{eqn:lamsplit}, into
\bear
\lefteqn{\rtlam =
(2,1,1,0,0,1,1,1,1,1;\rep{1}),
}\nn\\
&&
(\text{with lowest-weight 
$2\b_{-1}+3\b_0+4\b_1+2\b_2+2\b_3+2\b_4+2\b_5+\b_6+\Yrt+2\b_8$}),
\nn\\
\lefteqn{
\rtlam' =
(0,0,0,1,1,\tfrac{1}{2},\tfrac{1}{2},\tfrac{1}{2},\tfrac{1}{2},\tfrac{1}{2};\rep{16}),
}\nn\\
&&
(\text{with lowest-weight $\b_2+\b_3+\b_4+\b_5+\b_6+\b_7$}).
\nn
\eear
To analyze this configuration of fluxes more conveniently,
we formally reduce from M-theory to type-IIA.
For fixed $6^{th}, 7^{th}, 8^{th}, 9^{th}, 10^{th}$ coordinates,
we reduce to type-IIA with the $1^{st}$ circle taken as the M-theory ``$11^{th}$''
direction;
the exceptional M5-branes become D4-branes;
the $\rtlam$-flux becomes an RR 2-form field-strength $\FRR_{45}$
(keeping the same notation for directions as for M-theory),
and $\rtlam'$ corresponds to a field-strength $F_{23}$ of the gauge
field on the D4-brane.
We need to assume that $M_p R_1\ll 1$ for this reduction,
and therefore relax the condition \eqref{eqn:WeakAsympt}.
Since we are dealing with discrete charges, however, the final conclusion
will be valid in the limit \eqref{eqn:WeakAsympt}.

Suppose there are $N$ units of $\FRR_{45}$-flux and $N'$ units
of $F_{23}$-flux.
The Wess-Zumino interaction $A\wedge F\wedge \FRR$ on the D4-brane world-volume
\cite{Callan:1988wz}\cite{Douglas:1995bn}
now produces, in the presence of the $\rtlam$ and $\rtlam'$ fluxes, a background
charge for the D4-brane gauge field $A.$
This charge must be cancelled by $N N'$ open strings that end on the D4-brane.
Elevating back to M-theory, we see that these strings become
open M2-branes with one end on one of the exceptional M5-branes;
the other end could be at one of the $32$ $\Z_2$ fixed-hyperplanes
(which extend in the $1^{st},\dots,5^{th}$ directions
and run parallel to the 16 M5-branes).
We conclude that $\rtlam''$ is associated with the charge
of open M2-branes that end on the exceptional M5-branes.
Note that the lowest-weight $\g$ is an imaginary root of $DE_{18}$
and it satisfies $\g^2=0.$
In fact,
\belabel{eqn:gbT5Z2}
\g=
2\b_{-1}+3\b_0+4\b_1+3\b_2+3\b_3+3\b_4+3\b_5+3\b_6+3\b_7+2\b_8
+\b_{15}+\b_{16}+2\sum_{j=1}^6\b_{8+j}
\ee
It is not hard to check that $\g$ is $\WeylOf{DE_{18}}$-equivalent to
$\delta_2$ from \eqref{eqn:delta2}.
Its multiplicity, as a $DE_{18}$ root, is therefore
$$
\mult\g = \mult\delta_2 = 16.
$$
We do not know yet what the physical significance of the multiplicity is.

\subsubsection*{Scalar-Field Fluxes}
\paragraph{}
Now take the following $\Algso(16)$-pack
\bear
\rtlam'' &=&
(2,2,2,2,2,\tfrac{1}{2},\tfrac{3}{2},\tfrac{3}{2},\tfrac{3}{2},\tfrac{3}{2};\rep{16})
\nn\\
&&
(\text{with lowest-weight 
$2\b_{-1}+4\b_0+6\b_1+5\b_2+4\b_3+3\b_4+2\b_5+\b_6+\b_7+3\b_8$}),
\nn
\eear
and split it into
\bear
\rtlam &=& 
(2,1,1,1,1,0,1,1,1,0;\rep{1})
\nn\\
&& \qquad 
(\text{with lowest-weight $2\b_{-1}+3\b_0+4\b_1+3\b_2+3\b_3+2\b_4+\b_5+2\b_8$}),
\nn\\
\rtlam' &=&
(0,1,1,1,1,\tfrac{1}{2},\tfrac{1}{2},\tfrac{1}{2},\tfrac{1}{2},\tfrac{3}{2};\rep{16})
\nn\\
&& \qquad 
(\text{with lowest-weight $\b_0+2\b_1+2\b_2+\b_3+\b_4+\b_5+\b_6+\b_7+\b_8$}).
\nn
\eear
The flux associated with $\rtlam$ is a nonzero first Chern class $c_1$
for the fibration of the circle in the $1^{st}$ direction
over the $T^2$ that is made of the circles in the $6^{th}$ and $10^{th}$
directions.
The relevant part of the metric is
$$
R_1^2 (dx_1 -\tfrac{N}{2\pi} x_6 dx_{10})^2 +R_6^2 dx_6^2 + R_{10}^2 dx_{10}^2.
$$
The flux associated with $\rtlam'$ was discussed after
\eqref{eqn:Fluxpxphi} and corresponds to $N'=\px{1}X^{10}.$
Here, $X^{10}$ is the $10^{th}$ coordinate of one of the exceptional M5-branes.

The effect we are looking for comes from a cubic coupling
of the form
\belabel{eqn:NNprPh}
-N M_p^6 R_1 R_2 R_3 R_4 R_5 X^6\px{1}X^{10}
=
-N N' M_p^6 R_1 R_2 R_3 R_4 R_5 X^6,
\ee
on the M5-brane world-volume.
(The term ``cubic'' refers to the powers of $N$ and $X$'s together.)
This term can be inferred by expanding the Nambu-Goto action for an M5-brane,
which in our case reads:
\bear
M_p^6 R_2 R_3 R_4 R_5 
\sqrt{G_{11} + 2 G_{1,10}\px{1}X^{10} + G_{10,10}(\px{1}X^{10})^2}
\approx
M_p^6 R_2 R_3 R_4 R_5 
(\sqrt{G_{11}} + \frac{G_{1,10}}{\sqrt{G_{11}}}\px{1}X^{10})
\nn\\ 
\approx
M_p^6 R_2 R_3 R_4 R_5 
\left(\sqrt{G_{11}} + \left.\frac{G_{1,10}}{\sqrt{G_{11}}}\right|_0\px{1}X^{10} + 
\left.\px{6}\left(\frac{G_{1,10}}{\sqrt{G_{11}}}\right)\right|_0 X^6\px{1}X^{10}
+\cdots\right)
\nn\\
\label{eqn:NG}
\eear
where $G_{IJ}$ ($I,J=1\dots,10$) is the target-space metric.
It is the last term in \eqref{eqn:NG} that is responsible for
the cubic term \eqref{eqn:NNprPh}.
The term \eqref{eqn:NNprPh} is linear in the field $X^6,$
and hence is a source for that field.
We conclude that the $\Algso(16)$-pack $\rtlam''$ is associated  
with a source term for $X^6.$
It would be interesting to look for the implications of other
nonlinear terms in the effective action of the M5-brane
\cite{Aganagic:1997zq} (and see also 
\cite{DeCastro}).

\section{Type IA}\label{sec:TypeIA}
\paragraph{}
For completeness, we will now present some examples of $\Algso(16)$-packs
for type-IA string theory.
The corresponding $E_{10}$ orbifold was discussed in \secref{subsec:typeIA}.
The $\Z_2$-charge is $k_6+k_7.$
A basis for the sublattice of $\RtLatOf{E_{10}}$
generated by $\Z_2$-invariant roots
is given by
$$
\a_{-1},
\a_0, \a_1, \a_2, \a_3, \a_4, \a_5, \a_8, \Xrt', \Xrt''
$$
where
\bear
\Xrt' &=& \a_6 + \a_7
\rightarrow (0,0,0,0,0,0,0,1,0,-1),
\nn\\
\Xrt'' &=& \a_2 + 2\a_3 + 3\a_4 + 4\a_5 + 3\a_6 + \a_7 + 2\a_8
\rightarrow (0,0,0,1,1,1,1,1,0,1).
\nn
\eear
The inner products are given by
\vskip 10pt
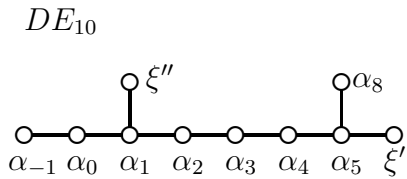
\begin{figure}[h]
\begin{picture}(370,80)
%
%
\thicklines
\multiput(20,20)(20,0){8}{\circle{6}}
\multiput(23,20)(20,0){7}{\line(1,0){14}}

\put(60,23){\line(0,1){14}}
\put(60,40){\circle{6}}
\put(140,23){\line(0,1){14}}
\put(140,40){\circle{6}}
\put( 14,7){$\a_{-1}$}
\put( 36,7){$\a_0$}
\put( 56,7){$\a_1$}
\put( 76,7){$\a_2$}
\put( 96,7){$\a_3$}
\put(116,7){$\a_4$}
\put(136,7){$\a_5$}
\put(156,7){$\Xrt'$}

\put(66,38){$\Xrt''$}
\put(144,38){$\a_8$}
\put(20,60){$DE_{10}$}
\end{picture}
\caption{The Dynkin diagram of the 
$\Z_2$-invariant subalgebra $DE_{10}\subset E_{10}$
that is associated with the type-IA orbifold.}
\label{fig:DynkinDE10IA}
\end{figure}
\vskip 10pt
Instead of \eqref{eqn:DeffDEE}, we define a linear map
$\fDEE_{\textit{IA}}:\CarOf{DE_{10}}^*\rightarrow\CarOf{E_{10}}^*$ by
\begin{align}
&\fDEE_{\textit{IA}}(\g_{-1})=\a_{-1},\quad
\fDEE_{\textit{IA}}(\g_0)=\a_0,\quad
\fDEE_{\textit{IA}}(\g_1)=\a_1,\quad
\fDEE_{\textit{IA}}(\g_2)=\a_2,\quad
\fDEE_{\textit{IA}}(\g_3)=\a_3,
\nn\\
&\fDEE_{\textit{IA}}(\g_4)=\a_4,\quad
\fDEE_{\textit{IA}}(\g_5)=\a_5,\quad
\fDEE_{\textit{IA}}(\g_6)=\Xrt',\quad
\fDEE_{\textit{IA}}(\g_7)=\Xrt'',\quad
\fDEE_{\textit{IA}}(\g_8)=\a_8.
\label{eqn:DeffDEEIA}
\end{align}
Instead of \eqref{eqn:ExplicitOmega}, we define
$$
\pHO_{\textit{IA}}\defineas \fDEE_{\textit{IA}}\circ\pHO',
$$
which gives, using \eqref{eqn:DeffDEEIA} and \eqref{eqn:ExplicitOmegaPr},
\bear
&&
\pHO_{\textit{IA}}(\b_{-1})=\a_{-1},\quad
\pHO_{\textit{IA}}(\b_0)=\a_0,\quad
\pHO_{\textit{IA}}(\b_1)=\a_1,\quad
\pHO_{\textit{IA}}(\b_2)=\a_2,\quad
\pHO_{\textit{IA}}(\b_3)=\a_3,
\nn\\
&&
\pHO_{\textit{IA}}(\b_4)=\a_4,\quad
\pHO_{\textit{IA}}(\b_5)=\a_5,\quad
\pHO_{\textit{IA}}(\b_6)=\Xrt',\quad
\pHO_{\textit{IA}}(\Yrt)=\a_8,\quad
\pHO_{\textit{IA}}(\b_8)=\Xrt'',
\nn\\
&&
\pHO_{\textit{IA}}(\b_9)=
\pHO_{\textit{IA}}(\b_{10})=\cdots=
\pHO_{\textit{IA}}(\b_{16})=0.
\label{eqn:ExplicitOmegaIA}
\nn
\eear
Using \eqref{eqn:DefYrt}, we get
$$
\pHO_{\textit{IA}}(\b_7) = 
\frac{1}{2}(\pHO_{\textit{IA}}(\Yrt)-\pHO_{\textit{IA}}(\b_6))=
\frac{1}{2}(\a_8-\a_6-\a_7)
$$
In type-IA the twisted sectors are D8-branes.
We will now analyze the Wess-Zumino interactions
on D8-branes,
in a fashion similar to \secref{subsec:imroots}.

Take for example the Wess-Zumino interaction
$A_0\wedge F_{12}\wedge \FRR_{345678},$
where again $\FRR$ is a Ramond-Ramond 6-form field-strength,
$A$ is the 1-form Yang-Mills gauge field on a D8-brane,
and $F=dA$ is its field-strength.
The flux $F_{12}$ is associated to
the $\Algso(16)$-pack
\belabel{eqn:rtlamprIA}
\rtlam' = (\lam',\rep{16}),\quad\lam'=
(0,0,1,1,1,1,1,1,\tfrac{1}{2},1)
=
\a_1+2\a_2+3\a_3+4\a_4+5\a_5
+\tfrac{7}{2}\a_6
+\tfrac{3}{2}\a_7
+\tfrac{5}{2}\a_8
\ee
This $\Algso(16)$-pack has lowest $DE_{18}$ weight
\belabel{eqn:gprIA}
\g' = \b_1+\b_2+\b_3+\b_4+\b_5+\b_6+\b_7+\b_8,
\qquad
\pHO_{\textit{IA}}(\g')=\lam'.
\ee
Equation \eqref{eqn:rtlamprIA}
is easily checked by calculating the energy stored in the flux,
$$
e^{2\inner{\lam'}{\vh}}
=
\frac{1}{g_s}\frac{M_s^5 R_3 \cdots R_8}{R_1 R_2}
M_p^9 R_1 \cdots R_{10}
=
M_p^{15} R_3^2 \cdots R_8^2 R_9 R_{10}^2
$$
where $M_s= M_p^{3/2} R_{10}^{1/2}$ is the string scale, and 
$g_s= (M_p R_{10})^{3/2}$ is the string coupling constant.

The flux $\FRR_{345678}$ is associated to
$$
\rtlam'' = (\lam'',\rep{1}),\qquad\lam''=
(1,1,0,0,0,0,0,0,1,0).
$$
This singlet $\Algso(16)$-pack has $DE_{18}$ weight
\belabel{eqn:gprprIA}
\g'' = \b_{-1}+2\b_0+2\b_1+2\b_2+2\b_3+2\b_4+2\b_5+2\b_6+2\b_7
+2\Bigl(\sum_{j=1}^{6}\b_{8+j}\Bigr)+\b_{15}+\b_{16},
\ee
(So that $\pHO_{\textit{IA}}(\g'')=\lam''.$)
We conclude that the $\Algso(16)$-pack
$$
\rtlam = (\lam'+\lam'',\rep{16})=
(1,1,1,1,1,1,1,1,\tfrac{3}{2},1; \rep{16})
$$
With lowest weight
\belabel{eqn:gbtypeIA}
\g =\b_{-1} +2\b_0
+3\b_1+3\b_2+3\b_3+3\b_4+3\b_5+3\b_6+3\b_7+\b_8
+2\Bigl(\sum_{j=1}^{6}\b_{8+j}\Bigr)+\b_{15}+\b_{16},
\ee
corresponds to the charge that couples to $A_0$ -- the 
open-string end-point charge.
Note that $\g^2=0$ and that, in fact, $\g$ is $\WeylOf{DE_{18}}$-equivalent
to $\delta_2$ from \eqref{eqn:delta2},
and therefore its multiplicity in $DE_{18}$ is
$$
\mult\g=\mult\delta_2=16.
$$
All 16 weights of $\rtlam$ can be obtained from $\g$ by Weyl
reflections around $\b_9,\dots,\b_{16},$ and therefore 
square to $0$ like $\g.$
As we mentioned in \secref{subsec:imroots},
we do not know the physical meaning of the multiplicity.

\section{The Orbifold $(T^4/\Z_2)\times (T^4/\Z_2)$}\label{sec:T4T4}
\paragraph{}
We will briefly remark on a more complicated orbifold -- 
$(T^4/\Z_2)\times (T^4/\Z_2).$

First, let us recall a few of the physical properties of M-theory
on $(T^4/\Z_2)\times (T^4/\Z_2).$
We take the supersymmetric case, without discrete torsion.
The $T^4/\Z_2$ singular spaces can be deformed into smooth $K_3$ spaces.
We then get M-theory on $K_3\times K_3\times T^2.$
According to \cite{Sethi:1996es} there 
must also be $24$ M2-branes in this compactification which are points
on $K_3\times K_3.$
Alternatively, one can argue \cite{Sethi:1996es} that M-theory on $K_3$ is dual to 
heterotic string theory on $T^3$ \cite{Witten:1995ex}.
The compactification of M-theory on $K_3\times K_3\times T^2$
is then dual to heterotic string theory on $K_3\times T^5.$
If the heterotic gauge-bundle is trivial on $K_3,$ then $24$ NS5-branes
are required to cancel anomalies.

Let us analyze the $\Z_2\times\Z_2$ action on $E_{10}.$
The $\Z_2\times\Z_2$ charges are determined by
the values of $k_0+k_8$ and $k_4+k_8.$
They can be expressed in terms of the two imaginary roots
$$
\OrbRt_1 = (2,2,2,2,2,2,1,1,1,3),\qquad
\OrbRt_2 = (2,2,1,1,1,3,2,2,2,2).
$$
The subset of roots
\belabel{eqn:T4T4def}
\RtsInv = \{\a\suchthat
 \kform{\OrbRt_1}{\a}\equiv\kform{\OrbRt_2}{\a}\equiv 0\pmod{2}\},
\ee
defines the untwisted sector of M-theory on $(T^4/\Z_2)\times (T^4/\Z_2).$
The 24 exceptional M2-branes
that are required by \cite{Sethi:1996es} correspond to the imaginary root
$$
\OrbRt_{\textit{M2}} = (2,2,1,1,1,1,1,1,1,1).
$$
It is now interesting to note that
$$
\kform{\OrbRt_{\textit{M2}}}{\a}\equiv 0 \pmod{2},\qquad\forall\a\in\RtsInv.
$$
In fact,
$$
\OrbRt_{\textit{M2}} \equiv\OrbRt_1 + \OrbRt_2\pmod{2}.
$$
Let us find the analog of the $DE_{10}\subset E_{10}$ subgroup
from \secref{sec:Untwisted}.
The steps are similar to \secref{subsec:subsetDE10}.
We set
\bear
\Xrt^{(1)} &=& \a_0 +\a_1 +\a_2 +\a_3 +\a_4 +\a_5 +\a_8,
\nn\\
\Xrt^{(2)} &=& \a_3 +2\a_4 +3\a_5 +2\a_6 +\a_7 +2\a_8,
\nn\\
\Yrt &=& \Xrt^{(1)} +\a_3 +2\a_4 +2\a_5 +2\a_6 +\a_7,
\nn
\eear
and take $\a_{-1},$ $\a_1,$ $\a_3,$ $\a_5,$ $\a_6,$ $\a_7,$ $\Yrt,$ $\Xrt^{(1)},$ $\Xrt^{(2)}$
as a basis.
The resulting Dynkin diagram is depicted in \figref{fig:DynkinT4Z2T4Z2}.
\vskip 10pt
\begin{figure}[h]
\begin{picture}(370,90)
\thicklines
\multiput(30,40)(20,0){6}{\circle{6}}
\multiput(33,40)(20,0){5}{\line(1,0){14}}

\put(50,43){\line(0,1){14}}\put(50,60){\circle{6}}
\put(50,37){\line(0,-1){14}}\put(50,20){\circle{6}}

\put(132,42){\line(1,1){10}}\put(145,55){\circle{6}}
\put(132,38){\line(1,-1){10}}\put(145,25){\circle{6}}
\put( 25,47){$\Xrt^{(2)}$}
\put( 46,8){$\a_1$}
\put( 46,66){$\a_3$}
\put( 51,31){$\a_2$}
\put( 68,47){$\Yrt$}
\put( 86,31){$\a_{-1}$}
\put( 105,47){$\Xrt^{(1)}$}
\put( 120,31){$\a_6$}
\put( 150,53){$\a_7$}
\put( 150,23){$\a_5$}
\end{picture}
\caption{The Dynkin diagram of the 
$(\Z_2\times\Z_2)$-invariant subalgebra of $E_{10}$
that is associated with the $(T^4/\Z_2)\times (T^4/\Z_2)$ orbifold.}
\label{fig:DynkinT4Z2T4Z2}
\end{figure}
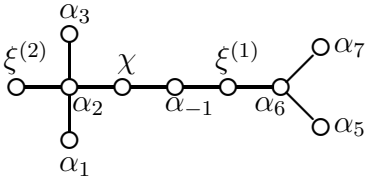
\vskip 10pt

Exploring  possible analogs of $DE_{18}$ will be left for future work.
We have seen in \secref{sec:Twisted} that
the number $8$ of extra nodes needed to extend 
the Dynkin diagram of $DE_{10}$ to $DE_{18}$
was the rank of $\Algso(16),$ and $16$ was the number
of exceptional branes.
By the same token,
we expect
the Kac-Moody algebra that describes 
the $(T^4/\Z_2)\times (T^4/\Z_2)$ orbifold to be of rank
$10+(16+16+24)/2=38,$ 
since there are $2\times 16$ exceptional Kaluza-Klein monopoles
(corresponding to each separate $T^4/\Z_2$ as in \secref{subsec:T4Z2})
and $24$ exceptional M2-branes.
It should correspond to
a Dynkin diagram with black and white nodes such that
the black nodes generate an $\Algso(16)+\Algso(16)+\Algso(24)$
subalgebra, corresponding to the twisted sectors,
and the commutant of $\Algso(16)+\Algso(16)+\Algso(24)$
should contain the Kac-Moody algebra depicted in \figref{fig:DynkinT4Z2T4Z2}.

Heterotic string theory on $K_3\times T^2$ is dual to
M-theory on $X\times S^1,$
where $X$ is a certain $K_3$-fibered Calabi-Yau manifold
\cite{Kachru:1995wm}\cite{Ferrara:1995yx}.
It follows that if we could understand fully the $\Z_2\times\Z_2$ 
orbifold of $E_{10},$ we might be able to give a Lie-algebraic
description for M-theory on certain Calabi-Yau manifolds.

\section{Conclusions and Discussion}\label{sec:concl}
\paragraph{}

In simple $\Z_2$ orbifolds of M-theory on $T^{10},$
the $\Z_2$ action on spacetime defines a $\Z_2$ action on $E_{10}.$
The spacetime fields and branes of the untwisted sector of the orbifold
correspond to roots of the $\Z_2$-invariant subalgebra of $E_{10}$ --
the algebra we called $\AlgInv.$
We have seen that this algebra contains a
Kac-Moody subalgebra $DE_{10},$
but $\AlgInv$ is bigger.

The twisted sectors of the orbifold comprise of branes that
correspond to imaginary roots that 
are orthogonal mod $2$ to all the $\Z_2$ invariant roots.
Both twisted and untwisted sectors appear to be encoded in a larger
structure -- the Kac-Moody algebra $DE_{18}.$
This algebra contains an $\Algso(16)$ subalgebra whose commutant
$\Commutant$ contains $DE_{10}.$
The infinite dimensional Lie algebra $\Commutant$ is also larger
than $DE_{10},$ and it would be very interesting to compare
it to $\AlgInv.$

$DE_{18}$ has two distinct Weyl equivalence classes of prime isotropic roots
(roots that square to zero and are not a nontrivial multiple
of any other root). 
Roots in the first class have multiplicity $8,$
while roots in the other class have multiplicity $16.$
We have given several examples of brane-charges that correspond to
roots from each class
[see equations 
\eqref{eqn:pHOdelta1},
\eqref{eqn:pHOdelta2},
\eqref{eqn:gbT5Z2},
\eqref{eqn:gbtypeIA}].

Finally, we constructed a Kac-Moody algebra that describes the
untwisted sector of a $(T^4/\Z_2)\times (T^4/\Z_2)$ orbifold.
It would be interesting to extend this algebra to include twisted sectors.

Another direction for further study is a possible connection with K-theory.
K-theory classifies
Ramond-Ramond charges and fluxes in perturbative string theory
(\cite{Witten:2000cn} and references therein),
while real roots of $E_{10}$ and $DE_{18}$ appear to classify 
fluxes nonperturbatively, and prime isotropic imaginary roots 
extend the notion of charge.
In particular, $\Z_2$ orientifolds \cite{Horava:1989ga}
have been discussed in the K-theory context in
\cite{Bergman:2001ax}\cite{Bergman:2001rp}.
It would be interesting to understand the various
discrete types of orientifolds in terms of $DE_{18}$;
their M-theory interpretation is described in 
\cite{Gimon:1998be}-\cite{deBoer:2001px}.
It would also be interesting to study the physical
interpretations of other Kac-Moody algebras, such as the ones
appearing in \cite{deBuyl:2004md}.

Finally, let us discuss our results in the context of
the ``mysterious duality'' between toroidally compactified
M-theory and del-Pezzo surfaces \cite{Iqbal:2001ye}.
Iqbal, Neitzke and Vafa have established 
an intriguing correspondence
between brane charges of M-theory on $T^k$ 
($k=0,\dots,8$) and cohomology classes
of rational curves in the del-Pezzo surface $\dPS_k,$
which can be described as $\CP^2$ blown up at  $k$ points.
(See also \cite{Abounasr:2004fn} for a recent discussion.)
The connection with $E_{10}$ appears to be that
the root lattice of
$E_k$ is a sublattice of the $H^2(\Z)$ cohomology lattice of $\dPS_k.$
In fact, it appears that the ``mysterious duality'' can be extended
to $k=10.$ The del-Pezzo surface would be replaced by a noncompact
Calabi-Yau surface which can be constructed by
starting with a $\CP^2,$ 
picking two cubic varieties on it and
blowing-up their 9 intersection points,
and then removing one cubic (hence the noncompactness).
This gives a ``$\tfrac{1}{2}K_3$'' surface and the $H^2(\Z)$ cohomology lattice
is  $\RtLatOf{E_{10}}.$ 
(Infinite dimensional Lie algebras have been introduced
in the context of this del-Pezzo surface in
\cite{DeWolfe:1998eu}\cite{Hauer:2000xy}.)

Instantons in M-theory on $T^{10}$ are connected with real roots
in $\RtLatOf{E_{10}},$ which in turn are related to cohomology classes
of $\tfrac{1}{2}K_3$ which square to $2.$
The duality between brane charges and del-Pezzo surfaces with smaller
$k$ can be obtained from $\tfrac{1}{2}K_3$ by blowing down,
in a way similar to the one described in \cite{Iqbal:2001ye}
for lower $k$'s.

It would be very interesting to study imaginary roots of $E_{10}$ 
in the context of the generalized Iqbal-Neitzke-Vafa duality.
They would correspond to curves of genus $1.$
Also, the orbifolds of $T^{10}$ that we discussed in \secref{sec:Orbifolds}
are described by assigning a $\Z_2$-gradation on 
$\RtLatOf{E_{10}}.$ On the del-Pezzo side of the duality, this corresponds
to an element of $H^2(\Z_2).$ Perhaps this could be further
interpreted as a Stiefel-Whitney class of a real vector bundle.

\vskip 20pt
\acknowledgments\nobreak
We are indebted to Richard Borcherds and Marty Halpern
for helpful explanations and references.
We also wish to thank 
Aaron Bergman, Oren Bergman, Eric Gimon and Alexandre Givental 
for helpful discussions,
and Axel Kleinschmidt and Herman Nicolai 
for correspondence and for sending us a list of multiplicities of $E_{10}$
and $DE_{10}$ roots.
The work of OJG was supported in part by the Director, Office of Science,
Office of High Energy and Nuclear Physics, of the U.S. Department of
Energy under Contract DE-AC03-76SF00098, and in part by
the NSF under grant PHY-0098840.

\newpage
\appendix
\setcounter{equation}{0}
\renewcommand{\theequation}{\thesection-\arabic{equation}}

\section{Proof that $DE_{10}\subset\AlgInv$}\label{app:proof}
\paragraph{}
We will now complete the details of the proof from 
\secref{subsec:subsetDE10}.
Using the map $\fDEE,$ defined in \eqref{eqn:DeffDEE}, we can construct an
injective homomorphism of  Lie algebras,
$$
\wfDEE:DE_{10}\rightarrow\AlgInv\subset E_{10}
$$
such that

\begin{description}

\item[a.]
For any $\a\in\RtsOf{DE_{10}}$ we have
$\wfDEE(\AlgG(DE_{10})_\a)\subset\AlgG_{\psi(\a)},$
and the restriction of $\wfDEE$ to $\AlgG(DE_{10})_\a$ is an injection.
[Here $\AlgG(E_{10})_\a$ and $\AlgG(DE_{10})_\a$ are the root spaces
of $DE_{10}$ and $E_{10},$ as defined in \eqref{eqn:RtSpaces}.]
\item[b.]
$\wfDEE$ is an isomorphism between the Cartan subalgebras
$\CarOf{DE_{10}}$ and $\CarOf{E_{10}}.$
\item[c.]
If $\a\in\ReRootsOf{DE_{10}}$ (a real root) then
$\wfDEE$ is an isomorphism between the root spaces 
$\AlgG(DE_{10})_\a$ and $\AlgG_{\psi(\a)}.$
\end{description}
\begin{proof}
$\wfDEE$ can be defined naturally on the Cartan subalgebra $\CarOf{DE_{10}}.$
Pick Chevalley generators $e'_i\in\AlgG(DE_{10})_{\g_i}$ ($i=-1\dots 8$),
and pick nonzero elements $x_i\in\AlgG(E_{10})_{\fDEE(\g_i)}.$
Define $\wfDEE(e'_i)=x_i.$
The Serre relations among the $e'_i$'s are satisfied by the $x_i$'s.
We can see this by using the Weyl-group $\WeylOf{E_{10}}$ to turn pairs of
$x_i$'s into simple roots.
Using the invariant bilinear forms on $DE_{10}$ and on $E_{10}$
we can find $y_i\in\AlgG(E_{10})_{-\fDEE(\g_i)}$ 
such that $\kform{x_i}{y_j}=\delta_{ij}.$
Pick Chevalley generators $f'_i\in\AlgG(DE_{10})_{-\g_i}$ and set
$\wfDEE(f'_i)=y_i.$

The map $\wfDEE$ is well-defined.
To see that it is an injection, note that
the kernel $\Ker\wfDEE$ is an ideal of $DE_{10}$ that intersects
$\CarOf{DE_{10}}$ trivially. Therefore, according to \cite{KacBook}, 
$\Ker\wfDEE=0.$

Parts (b) and (c) follow immediately from (a).
\end{proof}

\section{Proof that $DE_{10}\subset \Commutant$}\label{app:ProofDE18}
\paragraph{}
We now complete the missing details of \propref{prop:Commutant}.
Set $x^{\pm}$ to be nonzero generators of the root spaces 
$\AlgG(DE_{18})_{\pm \Yrt},$ and let $\AlgG'\subset DE_{18}$ 
be the smallest subalgebra that contains the set
\belabel{eqn:GeneratorsOfAlgGpr}
\{e_{-1}, f_{-1}, \cdots, e_6, f_6, 
e_8, f_8, x^{+}, x^{-}\}\subset DE_{18(10)}.
\ee
We will now show that $\AlgG'\simeq DE_{10}.$
Consider the set of positive real roots $\b_{-1},\dots,\b_6, \b_8, \Yrt.$
These roots all square to $2,$ and their intersection matrix is
encoded in the Dynkin diagram of \figref{fig:DynkinAlgGpr}, which, as we saw,
is the Dynkin diagram of $DE_{10}.$
We can therefore construct a surjective map $\phi:DE_{10}\rightarrow\AlgG'$
that maps the Chevalley generators of $DE_{10}$ to
the corresponding elements \eqref{eqn:GeneratorsOfAlgGpr}.
We need to prove that $\phi$ is injective, i.e. that there are no extra
relations among the elements of \eqref{eqn:GeneratorsOfAlgGpr} in addition
to those of the Kac-Moody algebra $DE_{10}.$
To see this note that $\{\phi^{-1}(h_{-1}),\dots,\phi^{-1}(h_8)\}$
generate the Cartan subalgebra $\CarOf{DE_{10}},$ and therefore
the kernel of $\phi$ intersects $\CarOf{DE_{10}}$ trivially.
But the kernel of $\phi$ is an ideal of $DE_{10}$
and a Kac-Moody algebra has no nontrivial ideals that intersect the Cartan
subalgebra trivially. (This follows from the construction
in \S1 of \cite{KacBook}.)
It follows that $\phi$ is an isomorphism of algebras and $DE_{10}\simeq\AlgG'.$

\section{Proof of \propref{prop:EasyRootPacks}}\label{app:ProofProp}
\propref{prop:EasyRootPacks} states that for
$$
0\neq x\in\AlgG_\g,
\qquad
\g = \sum_{i=-1}^8 k_i\b_i\in\RtsOf{DE_{18}},
$$
$U(\Algso(16))x\simeq L_{\Algso(16)}(\abs{k_7}\wwLambda_9).$
\begin{proof}
Let $e_i, f_i, h_i$ ($i=-1,\dots,16$) be Chevalley generators for $DE_{18}.$
Suppose, without loss of generality, that $k_7<0.$
The element $x$ is a linear combination of multiple commutators of 
$f_{-1},\dots,f_8$ and $f_7$ appears $\abs{k_7}$ times.
Consider a particular commutator
\belabel{eqn:zmultiplecom}
z \defineas 
\underbrace{[\cdots [f_7,\cdots [f_7,\cdots \cdots ]]]}_{\textit{$\abs{k_7}$ times}},
\ee
where the other generators that appear in $\cdots$ are from the list
$f_{-1},\dots,f_6, f_8.$
$\Algso(16)$ commutes with all these generators, and
$$
V\defineas U(\Algso(16))f_7\simeq L(\wwLambda_9)
$$
is isomorphic to the fundamental
representation of $\Algso(16)$ 
[since $e_7, f_7, h_7, e_9,\dots, h_{16}$ generate a finite $\Algso(18)$
Lie algebra].
Note that $f_7\in V$ is a lowest-weight vector
(a generator of the weight-space of $\wwLambda_9$).
Let $V^{\otimes\abs{k_7}}$ be the tensor product of $\abs{k_7}$
copies of the fundamental representation $\rep{16}.$
There is a surjective map
$g: V^{\otimes\abs{k_7}}\rightarrow U(\Algso(16))z$
generated by
$$
v_1\otimes v_2\otimes\cdots\otimes v_{\abs{k_7}}
\mapsto
[\cdots [v_1,\cdots [v_2,\cdots \cdots ]]],
$$
where the various $(\cdots)$'s are the same series of commutators as appear
in the corresponding expression \eqref{eqn:zmultiplecom}.
This map sends $f_7\otimes\cdots\otimes f_7$ to $z.$
Now set
$$
W\defineas U(\Algso(16))(\underbrace{f_7\otimes\cdots\otimes 
 f_7}_{\textit{$\abs{k_7}$ times}})\simeq L(\abs{k_7}\wwLambda_9).
$$
$W$ is isomorphic to the irreducible representation of 
rank-$\abs{k_7}$ traceless symmetric tensors.
$g$ induces a map $g':W\rightarrow U(\Algso(16))z,$
since $W\subset V^{\otimes\abs{k_7}}.$ 
Also, $f_7\otimes\cdots\otimes f_7\in W,$
so $g'$ is surjective. Since $W$ is irreducible, $g'$ is an isomorphism.
This proves that $U(\Algso(16))z$ is isomorphic to $L_{\Algso(16)}(\abs{k_7}\wwLambda_9).$
It is easy to extend this proof to $x,$ which is a linear combination 
of expressions like \eqref{eqn:zmultiplecom}.
\end{proof}

\section{Equivalence of Root lattices}
\label{app:EquivRt}
In this section we prove that:
\begin{enumerate}
\item
The $\AlgInv$ root lattice is isomorphic to the 
$DE_{10}$ root lattice. 
\item
The $\Commutant$ root lattice is isomorphic to the  
$DE_{10}$ root lattice.
\end{enumerate}
\begin{proof}
We already established an embedding of the $DE_{10}$ root lattices,
in \secref{subsec:subsetDE10} using the isometry $\fDEE$, 
and in \secref{subsec:subalgDE18} using the isometry $\fOH$.   
We will now show that both $\fDEE$ and $\fOH$ are actually surjective on positive roots; 
the surjection on negative roots then follows automatically.  
Thus we will have shown that $\fDEE$ and $\fOH$ are
surjective embeddings, i.e. isomorphims.
\begin{enumerate}
\item
Let $\a\in\PosRootsOf{\AlgInv}$.  
Then 
$$
\a=\sum_{i=-1}^2 a_i\a_i +2k\alpha_3+\sum_{j=4}^8 a_j\a_j;
\qquad
k,a_i,a_j\in\N.
$$
Using the $E_{10}$ Cartan matrix corresponding to the $E_{10}$ 
Dynkin diagram (\figref{fig:DynkinE10}) we find 
$$
\kform{\a}{\a}_{E_{10}}=
2\bigl(\sum_{i=-1}^2 a_i^2+4k^2+\sum_{j=4}^8 a_j^2\bigr)
-2\bigl(\sum_{i=-1}^{1}a_i a_{i+1}
 +2ka_2+2k a_4+a_5 a_8+\sum_{j=4}^6 a_j a_{j+1}\bigr).
$$
Note that $\a\in\PosRootsOf{\AlgInv}$ implies that $\kform{\a}{\a}_{E_{10}}\in 2\Z$
and is less than or equal to $2.$

We rewrite $\a$ in terms of $\wfDEE(DE_{10})\subset \AlgInv$ 
simple roots, and denote by $\b$ the resulting element of $\fDEE(\RtLatOf{DE_{10}})\subset\RtLatOf{\AlgInv}$. 
We find
$$
\b=\sum_{i=-1}^1 a_i\a_i+k\xi+(a_2-k)\a_2
+(a_4-2k)\a_4+(a_5-2k)\a_5
+(a_6-k)\a_6+a_7\a_7+(a_8-k)\a_8.
$$
Next we calculate $\kform{\b}{\b}_{DE_{10}}$,
using the $DE_{10}$ Cartan matrix from 
\secref{subsec:DE10}, and obtain 
$$
\kform{\b}{\b}_{DE_{10}}=
2\bigl(\sum_{i=-1}^{2}a_i^2+4k^2
+\sum_{j=4}^8 a_j^2\bigr)
-2\bigl(\sum_{i=-1}^1 a_i a_{i+1}+2ka_2+2ka_4
+a_5 a_8 +\sum_{j=4}^6 a_j a_{j+1}\bigr),
$$
so $\kform{\b}{\b}_{DE_{10}}=\kform{\a}{\a}_{E_{10}}$.
Since $\kform{\a}{\a}_{E_{10}}\leq 2$ and is even, 
and since $DE_{10}$ is hyperbolic, Proposition 5.10 of \cite{KacBook}
implies that $\b\in \RtsOf{DE_{10}}$, and $\fDEE(\b)=\a$ clearly.
We want to show that $\b$ is also a positive root: 
if $k>0$ we are finished, since if one simple root coefficient of a root is positive, then all simple root coefficients must be non-negative.
  If $k=0$, it is equally clear that $\b$ is a positive root, by assumption on $\a$.  Thus we have proven that $\fDEE:\PosRootsOf{DE_{10}}\rightarrow \PosRootsOf{\AlgInv}$ is a surjective map.

\item
Let $\a\in \PosRootsOf{\Commutant}$.
Then
$$
\a=\sum_{i=-1}^{6}a_i\b_i+a_8\b_8+2k\b_7
+\sum_{j=9}^{14}2k\b_j+k\b_{15}+k\b_{16};
\qquad
k, a_i\in\N.
$$ 
Using the $DE_{18}$ Cartan matrix from \eqref{eqn:CartanDE18},
we find 
$$
\kform{\a}{\a}_{DE_{18}}=
2(\sum_{i=-1}^6 a_i^2+a_8^2+2k^2)
-2(\sum_{j=-1}^5 a_j a_{j+1}+2ka_6 +a_1 a_8).
$$ 
We rewrite $\a$ in terms of the $\phi(DE_{10})\subset \Commutant$ 
simple roots, and denote by $\b$ the resulting element of 
$\phi(\RtLatOf{DE_{10}})\subset\RtLatOf{\Commutant}$.  We find,
$$
\b=\sum_{i=-1}^5 a_i\b_i
+a_8\b_8+k\chi+(a_6-k)\b_6.
$$ 
Using the $\AlgG'\cong DE_{10}$ 
Dynkin diagram (\figref{fig:DynkinAlgGpr}), 
we calculate 
$$
\kform{\b}{\b}_{DE_{10}}=
2(\sum_{i=-1}^{6}a_i^2+a_8^2+2k^2)
-2(\sum_{j=-1}^5 a_j a_{j+1}+2ka_6 +a_1 a_8),
$$
so $\kform{\b}{\b}_{DE_{10}}=\kform{\a}{\a}_{DE_{18}}$.
 We remark here that since $\a\in\PosRootsOf{\Commutant}$, 
then $\kform{\a}{\a}_{DE_{18}}\leq 2$ (see Proposition 5.2c of \cite{KacBook}); 
however, the converse statement no longer holds, because
$DE_{18}$ is not hyperbolic.
But we do not need the converse statement.
Since $DE_{10}$ is hyperbolic, Proposition 5.10 of \cite{KacBook}
implies that $\b\in\RtsOf{DE_{10}}$, and $\fOH(\b)=\a$ clearly.
The same reasoning as in part (1)
now shows that $\b\in\PosRootsOf{DE_{10}}$, 
so $\fOH:\PosRootsOf{DE_{10}}\rightarrow\PosRootsOf{\Commutant}$ 
is a surjective map.
\end{enumerate}
\end{proof}

\section{Denominator Formula for $\AlgInv$}\label{app:DenomFormula}
\paragraph{}
In this section we will present a ``denominator formula''
that captures the multiplicities of $\AlgInv$ roots.
Recall the {\it denominator identity} (formula (10.4.4) of \cite{KacBook}),
$$
\prod_{\a\in\PosRoots} (1 - e^{-\a})^{\mult \a}
=\sum_{w\in\Weyl}(\sgn w) e^{w(\rho)-\rho},
$$
Here, as usual in character formulas, we expand each side in a formal
power series in the formal variables \textit{$e^{-\a_i}$}.
The multiplication is according to the rule \textit{$e^{-\a} e^{-\b} = e^{-\a-\b}$}.
The sum on the right-hand side is over
all Weyl-group elements $w,$ and $\sgn w$ is the ``signature'' of $w$ --
$(+1)$ if $w$ is a product of an even number of simple reflections
and $(-1)$ otherwise.
The weight $\rho$ is chosen such that $\kform{\rho}{\a}=1$ for all
simple roots $\a.$
For $E_{10},$ with the assignment of simple roots as in
the \figref{fig:DynkinE10}, we have
$$
\rho = 
-30\a_{-1}
-61\a_0
-93\a_1
-126\a_2
-160\a_3
-195\a_4
-231\a_5
-153\a_6
-76\a_7
-115\a_8.
$$

We now calculate
$$
\prod_{\a\in\PosRoots} (1 - e^{-t \a})^{\mult \a}
=\sum_{w\in\Weyl}(\sgn w) e^{t w(\rho)-t \rho}
$$
and
$$
\prod_{\a\in\PosRoots} (1 - (-1)^{\kform{\a}{\OrbRt}}e^{-\a})^{\mult \a}
=\sum_{w\in\Weyl}(\sgn w) (-1)^{\kform{\OrbRt}{w(\rho)-\rho}} e^{w(\rho)-\rho}
$$
It follows that
$$
\prod_{\a\in\PosRtsInv} (1 - e^{-\a})^{\mult \a}
\prod_{\a\in\PosRoots\setminus\PosRtsInv} (1 + e^{-\a})^{\mult \a}
=\sum_{w\in\Weyl}(\sgn w) (-1)^{\kform{\OrbRt}{w(\rho)-\rho}} e^{w(\rho)-\rho}
$$
Multiplying by the original denominator formula, we get 
\bear
\lefteqn{
\prod_{\a\in\PosRtsInv} (1 - e^{-\a})^{2\mult \a}
\prod_{\a\in\PosRoots\setminus\PosRtsInv} (1 -e^{-2\a})^{\mult \a}
}
\nn\\
&& =\left(\sum_{w\in\Weyl}(\sgn w) (-1)^{\kform{\OrbRt}{w(\rho)-\rho}} e^{w(\rho)-\rho}\right)
\left(\sum_{w\in\Weyl}(\sgn w) e^{w(\rho)-\rho}\right).
\nn
\eear
Finally, dividing by
$$
\prod_{\a\in\PosRoots} (1 - e^{-2 \a})^{\mult \a}
=\sum_{w\in\Weyl}(\sgn w) e^{2 w(\rho)-2 \rho}
$$
we obtain the requisite formula
\belabel{eqn:CharacterInv}
\prod_{\a\in\PosRtsInv}(\tanh\frac{\a}{2})^{\mult \a}
=
\frac{\left(\sum_{w\in\Weyl}(\sgn w)
   (-1)^{\kform{\OrbRt}{w(\rho)-\rho}} e^{w(\rho)-\rho}\right)
\left(\sum_{w\in\Weyl}(\sgn w) e^{w(\rho)-\rho}\right)}{
\sum_{w\in\Weyl}(\sgn w) e^{2 w(\rho)-2 \rho}}
\ee
where we used
$$
\prod_{\a\in\PosRtsInv}(\tanh\frac{\a}{2})^{\mult \a}
=
\prod_{\a\in\PosRtsInv} \left(\frac{1 - e^{-\a}}{1+e^{-\a}}\right)^{\mult\a}
=
\prod_{\a\in\PosRtsInv} \frac{(1 - e^{-\a})^{2\mult \a}}{(1-e^{-2\a})^{\mult\a}}
$$
Now we can compare \eqref{eqn:CharacterInv} to a similar expression for $DE_{10}.$ 
Using similar manipulations we find
\belabel{eqn:CharacterDE10}
\prod_{\a\in\PosRtsInv}(\tanh\frac{\a}{2})^{\mult' \a}
=
\frac{\left(\sum_{w'\in\WeylOf{DE_{10}}}(\sgn w') e^{w'(\rho')-\rho'}\right)^2}{
\sum_{w'\in\WeylOf{DE_{10}}}(\sgn w') e^{2 w'(\rho')-2 \rho'}},
\qquad
\rho'\equiv\rho(DE_{10}),
\ee
where $\mult'$ denotes $DE_{10}$ multiplicities, and we used the identification $\PosRtsInv=\PosRootsOf{DE_{10}},$
where $\CarOf{DE_{10}}^*$ is implicitly identified with 
$\CarOf{E_{10}}^*$ using \eqref{eqn:DeffDEE}.

Using \eqref{eqn:rhoE10} and \eqref{eqn:rhoDE10} we can write
$$
\rho' = \rho + 8\OrbRt',
$$
where $\OrbRt'$ is some root such that
$$
\OrbRt-\OrbRt'\in 2\RtLatOf{E_{10}}
$$
so that
$$
(-1)^{\kform{\a}{\OrbRt}} = 
(-1)^{\kform{\a}{\OrbRt'}}
\qquad
\text{for all $\a\in\RtLatOf{E_{10}}$}.
$$
Using \eqref{eqn:CharacterDE10} we get
\begin{align}
\prod_{\a\in\PosRtsInv}(\tanh\frac{\a}{2})^{\mult' \a}
&=
\frac{\left(\sum_{w'\in\WeylOf{DE_{10}}}
     (\sgn w') e^{w'(\rho)-\rho +8 w'(\OrbRt')-8 \OrbRt'}\right)^2}{
\sum_{w'\in\WeylOf{DE_{10}}}(\sgn w') e^{2 w'(\rho)-2 \rho + 16 w'(\OrbRt')-16\OrbRt'}}
\nn\\
\prod_{\a\in\PosRtsInv}(\tanh\frac{\a}{2})^{\mult \a}
&=
\frac{\left(\sum_{w\in\WeylOf{E_{10}}}(\sgn w) 
    (-1)^{\inner{\OrbRt'}{w(\rho)-\rho}} e^{w(\rho)-\rho}\right)
\left(\sum_{w\in\WeylOf{E_{10}}}(\sgn w) e^{w(\rho)-\rho}\right)}{
\sum_{w\in\WeylOf{E_{10}}}(\sgn w) e^{2 w(\rho)-2 \rho}}
\nn
\end{align}

Let us now discuss the relation between the Weyl groups
$\WeylOf{DE_{10}}$ and $\WeylOf{E_{10}}.$
We can identify both 
$\WeylOf{DE_{10}}$ and $\WeylOf{E_{10}}$ as subgroups of the isometry group
of the dual of the Cartan subalgebra
$\CarOf{DE_{10}}^*\simeq\CarOf{E_{10}}^*.$

\begin{lem}
We have:
\begin{description}

\item[a.]
$\WeylOf{DE_{10}}\subset\WeylOf{E_{10}}$;
\item[b.]
$\WeylOf{DE_{10}} \simeq
\{ w\in\WeylOf{E_{10}}\suchthat w(\OrbRt)-\OrbRt\in 2\RtLatOf{E_{10}}\}$;
\item[c.]
The coset $\WeylOf{DE_{10}}/\WeylOf{E_{10}}$ is finite and
has $527$ elements.
\end{description}
\end{lem}
\begin{proof}
$\WeylOf{DE_{10}}$ is generated by reflections around
$\g_{-1},\dots,\g_8.$
Using \eqref{eqn:DeffDEE},
we can thus identify $\WeylOf{DE_{10}}$ with the subset of
$\WeylOf{E_{10}}$ that is generated by reflections
around the real roots $\fDEE(\g_{-1}),\dots,\fDEE(\g_8).$
This proves part (a).

To prove (b), consider a fundamental Weyl reflection $r_i\in\WeylOf{DE_{10}}$ around 
the simple root $\g_i.$ This maps to a reflection 
$r_{\fDEE(\g_i)}\in \WeylOf{E_{10}}.$
We now calculate
$$
r_{\fDEE(\g_i)}(\OrbRt) -\OrbRt = \kform{\OrbRt}{\fDEE(\g_i)}\fDEE(\g_i)\in 2\RtLatOf{E_{10}},
$$
since $\kform{\OrbRt}{\fDEE(\g_i)}$ is even.
$\WeylOf{E_{10}}$ is generated by the fundamental reflections $r_i,$
and it therefore follows that
$\WeylOf{DE_{10}} \subset
\{ w\in\WeylOf{E_{10}}\suchthat w(\OrbRt)-\OrbRt\in 2\RtLatOf{E_{10}}\}.$

To prove $\supset,$ take 
$w\in\WeylOf{E_{10}}$ such that 
$$
w(\OrbRt)-\OrbRt = 2\b
\qquad
\text{for some $\b\in \RtLatOf{E_{10}}$}.
$$
note first that
$$
\kform{\OrbRt}{\OrbRt}=\OrbRt^2=0\equiv 0\pmod{2}
$$
and therefore  $\OrbRt\in\RtLatInv\simeq\RtLatOf{DE_{10}}.$
According to Proposition 5.10b of \cite{KacBook},
since both $E_{10}$ and $DE_{10}$ are hyperbolic,
the Weyl groups $\WeylOf{DE_{10}}$ and $\WeylOf{E_{10}}$
are equivalent to a $\Z/2\Z$\footnote{Here
we use the notation $\Z/2\Z\simeq\Z_2$ to avoid any confusion
with the $\Z_2$ orbifold group.}
quotient of the
group of automorphisms of the respective root lattices
$\RtLatOf{DE_{10}}$ and $\RtLatOf{E_{10}}.$
The $\Z/2\Z$ quotient is the identification of
the automorphisms $\phi, -\phi\in\Aut(\RtLat).$
{}From part (a) it follows that every automorphism of
$\RtLatOf{DE_{10}}$ can be extended to an automorphism 
of $\RtLatOf{E_{10}}.$
Thus
\begin{multline}
\OrbChg{w(\a)}\equiv
\kform{w(\a)}{\OrbRt}=
\kform{\a}{w^{-1}(\OrbRt)}=
\kform{\a}{\OrbRt-2w^{-1}(\b)}
\equiv\kform{\a}{\OrbRt}\equiv\OrbChg{\a}\nn\\
\text{for any
$\a\in\RtLatOf{DE_{10}}\simeq\RtLatInv\subset\RtLatOf{E_{10}}$}.
\nn
\end{multline}
This proves that $w$ preserves $\RtLatInv\simeq\RtLatOf{DE_{10}}$
and therefore
$$
\WeylOf{DE_{10}} \supset
\{ w\in\WeylOf{E_{10}}\suchthat w(\OrbRt)-\OrbRt\in 2\RtLatOf{E_{10}}\}.
$$

To prove (c), note that if $w_1, w_2\in\WeylOf{E_{10}}$
then $w_1(\OrbRt)-w_2(\OrbRt)\in 2\RtLatOf{E_{10}},$
if and only if $w_1^{-1} w_2(\OrbRt)-\OrbRt\in 2\RtLatOf{E_{10}},$
and so, by (b), $w_1^{-1}w_2\in\WeylOf{DE_{10}}.$
Therefore, the map
$t: \WeylOf{E_{10}}/\WeylOf{DE_{10}}\rightarrow \RtLatOf{E_{10}}/2\RtLatOf{E_{10}}$
that sends the equivalence class of $w\in\WeylOf{E_{10}}$
to the equivalence class of $w(\OrbRt)-\OrbRt$ is an injection.
This proves the finiteness of $\WeylOf{E_{10}}/\WeylOf{DE_{10}}.$

To count the size of $\WeylOf{E_{10}}/\WeylOf{DE_{10}}$ we need to calculate
the size of the image of $t.$
This is most conveniently done in the basis \eqref{eqn:AlphaExplicit}.
Define the set
$$
S\defineas
\{(n_1,\dots, n_{10})\suchthat
n_i\in\Z,\quad
\sum_{i=1}^{10}n_i\in 3\Z,\quad
\sum_{i=1}^{10}(-1)^{n_i}\in\{-8,-6,0,2,8\}\}
$$
Using \eqref{eqn:RootLatRep}, we see that $S$ can be identified with a certain
a subset of $\RtLatOf{E_{10}}.$
Using \eqref{eqn:AlphaExplicit}, it is easy to check that the
fundamental Weyl reflections preserve $S\subset\RtLatOf{E_{10}}.$
$S$ is therefore $\WeylOf{E_{10}}$-invariant.
Using the explicit expressions found in \secref{sec:Orbifolds},
we see that $\OrbRt\in S.$
The coset $S/2\RtLatOf{E_{10}}$ [where we use the explicit
representation of $\RtLatOf{E_{10}}$ as in \eqref{eqn:AlphaExplicit}]
contains
$$
\binom{10}{9}
+\binom{10}{8}
+\binom{10}{5}
+\binom{10}{4}
+\binom{10}{1}
=527 
\,\text{elements}.
$$
Therefore, the image in $\RtLatOf{E_{10}}/2\RtLatOf{E_{10}}$ of the map $t$
contains at most $527$ elements.
To see that it contains exactly $527$ elements,
recall that by Proposition 5.7 of \cite{KacBook} all positive prime 
isotropic roots of $E_{10}$ are $\WeylOf{E_{10}}$-equivalent to $\OrbRt.$
Below is a list of such roots $\a=w(\OrbRt)$ [in the basis \eqref{eqn:AlphaExplicit}]
such that $t(w)=[\a-\OrbRt]$ (the equivalence class of $\a-\OrbRt$
in $\RtLatOf{E_{10}}/2\RtLatOf{E_{10}}$) exhaust all $527$ possibilities.
\begin{align}
\text{$\binom{10}{9}=10$ distinct permutations of}\quad
& (0,1,1,1,1,1,1,1,1,1),
\nn\\
\text{$\binom{10}{8}=45$ distinct permutations of}\quad
& (2,2,1,1,1,1,1,1,1,1)\equiv (0,0,1,1,1,1,1,1,1,1),
\nn\\
\text{$\binom{10}{5}=252$ distinct permutations of}\quad
& (2,2,2,2,2,1,1,1,1,1)\equiv (0,0,0,0,0,1,1,1,1,1),
\nn\\
\text{$\binom{10}{4}=210$ distinct permutations of}\quad
& (2,2,2,2,2,2,1,1,1,3)\equiv (0,0,0,0,0,0,1,1,1,1),
\nn\\
\text{$\binom{10}{1}=10$ distinct permutations of}\quad
& (2,2,2,2,2,2,2,2,4,1)\equiv (0,0,0,0,0,0,0,0,0,1).
\nn
\end{align}
The $\equiv$'s are $\pmod{2}.$
\end{proof}


\newpage
\begin{thebibliography}{1}

\bibitem{Brown:2004jb}
J.~Brown, O.~J.~Ganor and C.~Helfgott,
{``M-theory and $E_{10}$: Billiards, branes, and imaginary roots,''}
arXiv:hep-th/0401053.

\bibitem{Julia}
B.~Julia,
{``On Infinite Dimensional Symmetry Groups in Physics,''}
Published in {\it Copenhagen Bohr Symp.}(1985)0215

\bibitem{Nicolai:kx}
H.~Nicolai,
{``A Hyperbolic Lie Algebra From Supergravity,''}
Phys.\ Lett.\ B {\bf 276}, 333 (1992).

\bibitem{Dixon:1985jw}
L.~J.~Dixon, J.~A.~Harvey, C.~Vafa and E.~Witten,
{``Strings On Orbifolds,''}
Nucl.\ Phys.\ B {\bf 261}, 678 (1985);
L.~J.~Dixon, J.~A.~Harvey, C.~Vafa and E.~Witten,
{``Strings On Orbifolds. 2,''}
Nucl.\ Phys.\ B {\bf 274}, 285 (1986).

\bibitem{Halpern:2004ud}
M.~B.~Halpern and C.~Helfgott,
{``The general twisted open WZW string,''}
arXiv:hep-th/0406003;
{``A basic class of twisted open WZW strings,''}
arXiv:hep-th/0402108;
{``On the target-space geometry of open-string orientation-orbifold sectors,''}
Annals Phys.\  {\bf 310}, 302 (2004)
[arXiv:hep-th/0309101].
\bibitem{Borisov:1997nc}
L.~Borisov, M.~B.~Halpern and C.~Schweigert,
{``Systematic approach to cyclic orbifolds,''}
Int.\ J.\ Mod.\ Phys.\ A {\bf 13}, 125 (1998)
[arXiv:hep-th/9701061].


\bibitem{Horava:1995qa}
P.~\Horava and E.~Witten,
{``Heterotic and type I string dynamics from eleven dimensions,''}
Nucl.\ Phys.\ B {\bf 460}, 506 (1996)
[arXiv:hep-th/9510209].

\bibitem{Polchinski:1995df}
J.~Polchinski and E.~Witten,
{``Evidence for Heterotic - Type I String Duality,''}
Nucl.\ Phys.\ B {\bf 460}, 525 (1996)
[arXiv:hep-th/9510169].

\bibitem{Dasgupta:1995zm}
K.~Dasgupta and S.~Mukhi,
{``Orbifolds of M-theory,''}
Nucl.\ Phys.\ B {\bf 465}, 399 (1996)
[arXiv:hep-th/9512196].

\bibitem{Witten:1995em}
E.~Witten,
{``Five-branes and M-theory on an orbifold,''}
Nucl.\ Phys.\ B {\bf 463}, 383 (1996)
[arXiv:hep-th/9512219].

\bibitem{Sen:1996na}
A.~Sen,
{``Duality and Orbifolds,''}
Nucl.\ Phys.\ B {\bf 474}, 361 (1996)
[arXiv:hep-th/9604070];

\bibitem{Sen:1996zq}
A.~Sen,
{``Orbifolds of M-Theory and String Theory,''}
Mod.\ Phys.\ Lett.\ A {\bf 11}, 1339 (1996)
[arXiv:hep-th/9603113];

\bibitem{Dasgupta:1996yh}
K.~Dasgupta and S.~Mukhi,
{``A note on low-dimensional string compactifications,''}
Phys.\ Lett.\ B {\bf 398}, 285 (1997)
[arXiv:hep-th/9612188].

\bibitem{Dasgupta:1997cd}
K.~Dasgupta, D.~P.~Jatkar and S.~Mukhi,
{``Gravitational couplings and $\Z_2$ orientifolds,''}
Nucl.\ Phys.\ B {\bf 523}, 465 (1998)
[arXiv:hep-th/9707224].

\bibitem{Kumar:1996mj}
A.~Kumar and K.~Ray,
{``Compactification of $M$-Theory to Two Dimensions,''}
Phys.\ Lett.\ B {\bf 383}, 160 (1996)
[arXiv:hep-th/9604164].

\bibitem{Motl:1999cy}
L.~Motl and T.~Banks,
{``On the hyperbolic structure of moduli spaces with 16 SUSYs,''}
JHEP {\bf 9905}, 015 (1999)
[arXiv:hep-th/9904008].

\bibitem{Nicolai:1987vy}
H.~Nicolai,
{``Two-Dimensional Supergravities, Hidden Symmetries And Integrable Systems,''}
KA-THEP-87/10
{\it Lectures given at 1987 Cargese Summer School on Particle Physics, Cargese, France, Aug 3-21, 1987}

\bibitem{Sen:1995qk}
A.~Sen,
{``Duality symmetry group of two-dimensional heterotic string theory,''}
Nucl.\ Phys.\ B {\bf 447}, 62 (1995)
[arXiv:hep-th/9503057].

\bibitem{Schwarz:1995td}
J.~H.~Schwarz,
{``Classical symmetries of some two-dimensional models,''}
Nucl.\ Phys.\ B {\bf 447}, 137 (1995)
[arXiv:hep-th/9503078];
{``Classical duality symmetries in two dimensions,''}
arXiv:hep-th/9505170.

\bibitem{Damour:2000hv}
T.~Damour and M.~Henneaux,
{``$E_{10},$ $BE_{10}$ and arithmetical chaos in superstring cosmology,''}
Phys.\ Rev.\ Lett.\  {\bf 86}, 4749 (2001)
[arXiv:hep-th/0012172].

\bibitem{Kleinschmidt:2004dy}
A.~Kleinschmidt and H.~Nicolai,
{``$E_{10}$ and $SO(9,9)$ invariant supergravity,''}
JHEP {\bf 0407}, 041 (2004)
[arXiv:hep-th/0407101].

\bibitem{Mkrtchyan:2004ah}
H.~Mkrtchyan and R.~Mkrtchyan,
{``$E_{11}, K_{11}$ and $EE_{11}$,''}
arXiv:hep-th/0407148.

\bibitem{Chaudhuri:2004zh}
S.~Chaudhuri,
{``Hidden symmetry unmasked: Matrix theory and $E_{11} = E_8^3$,''}
arXiv:hep-th/0404235;
{``Spacetime reduction of locally supersymmetric theories: A nonperturbative
proposal for M theory,''}
arXiv:hep-th/0408057.

\bibitem{Sethi:1996es}
S.~Sethi, C.~Vafa and E.~Witten,
{``Constraints on low-dimensional string compactifications,''}
Nucl.\ Phys.\ B {\bf 480}, 213 (1996)
[arXiv:hep-th/9606122].

\bibitem{KacBook}
V.G.~Kac,
{\it Infinite dimensional Lie algebras,}
$3^{rd}$ ed., Cambridge University Press, 1990.

\bibitem{Obers:1998fb}
N.~A.~Obers and B.~Pioline,
{``U-duality and M-theory,''}
Phys.\ Rept.\  {\bf 318}, 113 (1999)
[arXiv:hep-th/9809039].

\bibitem{Elitzur:1997zn}
S.~Elitzur, A.~Giveon, D.~Kutasov and E.~Rabinovici,
{``Algebraic aspects of matrix theory on $T^d,$''}
Nucl.\ Phys.\ B {\bf 509}, 122 (1998)
[arXiv:hep-th/9707217].

\bibitem{Banks:1998vs}
T.~Banks, W.~Fischler and L.~Motl,
{``Dualities versus singularities,''}
JHEP {\bf 9901}, 019 (1999)
[arXiv:hep-th/9811194].

\bibitem{Hull:1995mz}
C.~M.~Hull and P.~K.~Townsend,
{``Enhanced gauge symmetries in superstring theories,''}
Nucl.\ Phys.\ B {\bf 451}, 525 (1995)
[arXiv:hep-th/9505073].

\bibitem{Ganor:1999ui}
O.~J.~Ganor,
{``Two conjectures on gauge theories, gravity, and infinite dimensional
Kac-Moody groups,''}
arXiv:hep-th/9903110.

\bibitem{Damour:2002cu}
T.~Damour, M.~Henneaux and H.~Nicolai,
{``$E_{10}$ and a 'small tension expansion' of M theory,''}
Phys.\ Rev.\ Lett.\  {\bf 89}, 221601 (2002)
[arXiv:hep-th/0207267].

\bibitem{West:2001as}
P.~C.~West,
{``$E_{11}$ and M theory,''}
Class.\ Quant.\ Grav.\  {\bf 18}, 4443 (2001)
[arXiv:hep-th/0104081].

\bibitem{Englert:2004it}
F.~Englert and L.~Houart,
{``G+++ invariant formulation of gravity and M-theories: Exact intersecting
brane solutions,''}
JHEP {\bf 0405}, 059 (2004)
[arXiv:hep-th/0405082].

\bibitem{Keurentjes:2004bv}
A.~Keurentjes,
{``$E_{11}$: Sign of the times,''}
arXiv:hep-th/0402090.

\bibitem{Miemiec:2004iv}
A.~Miemiec and I.~Schnakenburg,
{``Killing spinor equations from nonlinear realisations,''}
arXiv:hep-th/0404191.

\bibitem{HoravaE9}
P.~\Horava, unpublished.

\bibitem{Adams:2002rx}
A.~Adams and J.~Evslin,
{``The loop group of $E_8$ and K-theory from 11d,''}
JHEP {\bf 0302}, 029 (2003)
[arXiv:hep-th/0203218].

\bibitem{Bergman:2004ne}
A.~Bergman and U.~Varadarajan,
{``Loop groups, Kaluza-Klein reduction and M-theory,''}
arXiv:hep-th/0406218.


\bibitem{Horava:1989ga}
P.~\Horava,
{``Background Duality Of Open String Models,''}
Phys.\ Lett.\ B {\bf 231}, 251 (1989);
{``Strings On World Sheet Orbifolds,''}
Nucl.\ Phys.\ B {\bf 327}, 461 (1989);
{``Properties Of Strings On World Sheet Orbifolds,''}
PRA-HEP-90-1


\bibitem{Bardakci:1970nb}
K.~Bardakci and M.~B.~Halpern,
{``New Dual Quark Models,''}
Phys.\ Rev.\ D {\bf 3}, 2493 (1971);

\bibitem{KMW}
V.~G.~Kac, R.~V.~Moody and M.~Wakimoto,
{``On $E_{10},$''}
preprint 1988.

\bibitem{KW}
V.G.~Kac and M.~Wakimoto,
{``Unitarizable highest weight representations of the Virasoro,
Neveu-Schwarz and Ramond algebras,''}
1986 Lect. Notes in Phys. 261

\bibitem{FF}
A.~Feingold and I.~Frenkel,
{``A Hyperbolic Kac-Moody Algebra and the Theory of Siegel Modular Forms of Genus 2,''}
Math.\ Ann.\ {\bf 263} (1983) 87.

\bibitem{Bauer:1996ca}
M.~Bauer and D.~Bernard,
{``On root multiplicities of some hyperbolic Kac-Moody algebras,''}
Lett.\ Math.\ Phys.\  {\bf 42}, 153 (1997)
[arXiv:hep-th/9612210].

\bibitem{Halpern:1971ay}
M.~B.~Halpern,
{``The Two Faces Of A Dual Pion - Quark Model,''}
Phys.\ Rev.\ D {\bf 4} (1971) 2398.

\bibitem{Goddard:1984vk}
P.~Goddard, A.~Kent and D.~I.~Olive,
{``Virasoro Algebras And Coset Space Models,''}
Phys.\ Lett.\ B {\bf 152}, 88 (1985).

\bibitem{Schellekens:1990xy}
A.~N.~Schellekens and S.~Yankielowicz,
{``Simple Currents, Modular Invariants And Fixed Points,''}
Int.\ J.\ Mod.\ Phys.\ A {\bf 5}, 2903 (1990).

\bibitem{Ishikawa:2003xh}
H.~Ishikawa and A.~Yamaguchi,
{``Twisted boundary states in $c = 1$ coset conformal field theories,''}
JHEP {\bf 0304}, 026 (2003)
[arXiv:hep-th/0301040].

\bibitem{Goddard:1985xp}
P.~Goddard, D.~I.~Olive and A.~Schwimmer,
{``The Heterotic String And A Fermionic Construction Of The $E_8$ Kac-Moody
Algebra,''}
Phys.\ Lett.\ B {\bf 157}, 393 (1985).

\bibitem{KMPS}
S.~Kass, R.V.~Moody, J.~Patera and R.~Slansky,
{``Affine Lie algebras, weight multiplicities,
and branching rules,''}
Vol.1,2 University of California Press, 1990.


\bibitem{Sen:1994wr}
A.~Sen,
{``Strong - weak coupling duality in three-dimensional string theory,''}
Nucl.\ Phys.\ B {\bf 434}, 179 (1995)
[arXiv:hep-th/9408083].

\bibitem{PP}
K.C.~Pati and D.~Parashar,
{``Iwasawa decomposition of affine Kac-Moody algebras using Satake diagrams,''}
1998 J. Math. Phys. {\bf 39}(9) 5015;
{``Involutive automorphisms and Iwasawa decomposition
of some hyperbolic Kac-Moody Algebras,''} 1999 J. Math. Phys. {\bf 40}(1) 501

\bibitem{Keurentjes:2002rc}
A.~Keurentjes,
{``The group theory of oxidation. II: Cosets of non-split groups,''}
Nucl.\ Phys.\ B {\bf 658}, 348 (2003)
[arXiv:hep-th/0212024].

\bibitem{Nicolai:2004nv}
H.~Nicolai and H.~Samtleben,
{``On $K(E_9)$,''}
arXiv:hep-th/0407055.


\bibitem{Damour:2002et}
T.~Damour, M.~Henneaux and H.~Nicolai,
{``Cosmological billiards,''}
Class.\ Quant.\ Grav.\  {\bf 20}, R145 (2003)
[arXiv:hep-th/0212256].

\bibitem{Pioline:2002qz}
B.~Pioline and A.~Waldron,
{``Quantum cosmology and conformal invariance,''}
Phys.\ Rev.\ Lett.\  {\bf 90}, 031302 (2003)
[arXiv:hep-th/0209044].


\bibitem{Callan:1988wz}
C.~G.~.~Callan, C.~Lovelace, C.~R.~Nappi and S.~A.~Yost,
{``Loop Corrections To Superstring Equations Of Motion,''}
Nucl.\ Phys.\ B {\bf 308}, 221 (1988).

\bibitem{Douglas:1995bn}
M.~R.~Douglas,
{``Branes within branes,''}
arXiv:hep-th/9512077.

\bibitem{Aganagic:1997zq}
M.~Aganagic, J.~Park, C.~Popescu and J.~H.~Schwarz,
{``World-volume action of the M-theory five-brane,''}
Nucl.\ Phys.\ B {\bf 496}, 191 (1997)
[arXiv:hep-th/9701166].

\bibitem{DeCastro}
A.~De Castro and A.~Restuccia,
{``Master canonical action and BRST charge of 
the M theory bosonic five brane,''}
Nucl.\ Phys.\ B {\bf 617}, 215 (2001)
[arXiv:hep-th/0103123];
{``Super five brane Hamiltonian 
and the chiral degrees of freedom,''}
Phys.\ Rev.\ D {\bf 66}, 024037 (2002)
[arXiv:hep-th/0204052];
{``On the super five brane Hamiltonian,''}
Rev.\ Mex.\ Fis.\  {\bf 49S3}, 95 (2003)
[arXiv:hep-th/0406081].



\bibitem{Witten:1995ex}
E.~Witten,
{``String theory dynamics in various dimensions,''}
Nucl.\ Phys.\ B {\bf 443}, 85 (1995)
[arXiv:hep-th/9503124].

\bibitem{Kachru:1995wm}
S.~Kachru and C.~Vafa,
{``Exact results for N=2 compactifications of heterotic strings,''}
Nucl.\ Phys.\ B {\bf 450}, 69 (1995)
[arXiv:hep-th/9505105].

\bibitem{Ferrara:1995yx}
S.~Ferrara, J.~A.~Harvey, A.~Strominger and C.~Vafa,
{``Second quantized mirror symmetry,''}
Phys.\ Lett.\ B {\bf 361}, 59 (1995)
[arXiv:hep-th/9505162].


\bibitem{Witten:2000cn}
E.~Witten,
{``Overview of K-theory applied to strings,''}
Int.\ J.\ Mod.\ Phys.\ A {\bf 16}, 693 (2001)
[arXiv:hep-th/0007175].

\bibitem{Bergman:2001ax}
O.~Bergman, E.~G.~Gimon and B.~Kol,
{``Strings on orbifold lines,''}
JHEP {\bf 0105}, 019 (2001)
[arXiv:hep-th/0102095].

\bibitem{Bergman:2001rp}
O.~Bergman, E.~G.~Gimon and S.~Sugimoto,
{``Orientifolds, RR torsion, and K-theory,''}
JHEP {\bf 0105}, 047 (2001)
[arXiv:hep-th/0103183].

\bibitem{Gimon:1998be}
E.~G.~Gimon,
{``On the M-theory interpretation of orientifold planes,''}
arXiv:hep-th/9806226.

\bibitem{Hanany:1999jy}
A.~Hanany, B.~Kol and A.~Rajaraman,
``Orientifold points in M theory,''
JHEP {\bf 9910}, 027 (1999)
[arXiv:hep-th/9909028];
A.~Hanany and B.~Kol,
``On orientifolds, discrete torsion, branes and M theory,''
JHEP {\bf 0006}, 013 (2000)
[arXiv:hep-th/0003025].

\bibitem{deBoer:2001px}
J.~de Boer, R.~Dijkgraaf, K.~Hori, A.~Keurentjes, J.~Morgan, D.~R.~Morrison and S.~Sethi,
``Triples, fluxes, and strings,''
Adv.\ Theor.\ Math.\ Phys.\  {\bf 4}, 995 (2002)
[arXiv:hep-th/0103170].

\bibitem{deBuyl:2004md}
S.~de Buyl and C.~Schomblond,
{``Hyperbolic Kac Moody algebras and Einstein billiards,''}
arXiv:hep-th/0403285.

\bibitem{Iqbal:2001ye}
A.~Iqbal, A.~Neitzke and C.~Vafa,
{``A mysterious duality,''}
Adv.\ Theor.\ Math.\ Phys.\  {\bf 5}, 769 (2002)
[arXiv:hep-th/0111068].

\bibitem{Abounasr:2004fn}
R.~Abounasr, M.~Ait Ben Haddou, A.~El Rhalami and E.~H.~Saidi,
{``Algebraic geometry realization of quantum Hall soliton,''}
arXiv:hep-th/0406036.

\bibitem{DeWolfe:1998eu}
O.~DeWolfe, T.~Hauer, A.~Iqbal and B.~Zwiebach,
{``Uncovering the symmetries on (p,q) 7-branes: Beyond the Kodaira
classification,''}
Adv.\ Theor.\ Math.\ Phys.\  {\bf 3}, 1785 (1999)
[arXiv:hep-th/9812028].

\bibitem{Hauer:2000xy}
T.~Hauer, A.~Iqbal and B.~Zwiebach,
{``Duality and Weyl symmetry of 7-brane configurations,''}
JHEP {\bf 0009}, 042 (2000)
[arXiv:hep-th/0002127].


\end{thebibliography}
\end{document}